\documentclass[aps,pra,amsmath,amssymb,reprint,floatfix,superscriptaddress,showpacs]{revtex4-1}
\usepackage{graphicx}
\usepackage[bookmarks=false,colorlinks=true,linkcolor=blue,filecolor=blue,citecolor=blue,urlcolor=blue]{hyperref}
\usepackage{bm}
\usepackage{soul}

\newcommand{\algg}{\mathfrak{g}}

\newcommand{\tphi}{\tilde{\phi}}
\newcommand{\ttheta}{\tilde{\theta}}

\newcommand{\bfone}{{\bm 1}}
\newcommand{\bfn}{\mathbf{n}}
\newcommand{\bfr}{\mathbf{r}}
\newcommand{\bfR}{\mathbf{R}}
\newcommand{\bfe}{\mathbf{e}}
\newcommand{\bfJ}{\mathbf{J}}
\newcommand{\bfS}{\mathbf{S}}
\newcommand{\bfdelta}{{\bm \delta}}

\newcommand{\bJ}{\bar{J}}
\newcommand{\bbfJ}{\bar{\mathbf{J}}}

\newcommand{\calS}{\mathcal{S}}
\newcommand{\calM}{\mathcal{M}}
\newcommand{\calZ}{\mathcal{Z}}
\newcommand{\calV}{\mathcal{V}}

\begin{document}

\title{From coupled wires to coupled layers: Model with three-dimensional fractional excitations}

\author{Yohei Fuji}
\affiliation{Condensed Matter Theory Laboratory, RIKEN CPR, Wako, Saitama 351-0198, Japan}

\author{Akira Furusaki}
\affiliation{Condensed Matter Theory Laboratory, RIKEN CPR, Wako, Saitama 351-0198, Japan}
\affiliation{RIKEN Center for Emergent Matter Science, Wako, Saitama 351-0198, Japan}

\date{\today}

\begin{abstract}
We propose a systematic approach to constructing microscopic models with fractional excitations in three-dimensional (3D) space. 
Building blocks are quantum wires described by the (1+1)-dimensional conformal field theory (CFT) associated with a current algebra $\mathfrak{g}$. 
The wires are coupled with each other to form a 3D network through the current-current interactions of $\mathfrak{g}_1$ and $\mathfrak{g}_2$ CFTs that are related to the $\mathfrak{g}$ CFT by a nontrivial conformal embedding $\mathfrak{g} \supset \mathfrak{g}_1 \times \mathfrak{g}_2$. 
The resulting model can be viewed as a layer construction of a 3D topologically ordered state, in which the conformal embedding in each wire implements the anyon condensation between adjacent layers. 
Local operators acting on the ground state create point-like or loop-like deconfined excitations depending on the branching rule. 
We demonstrate our construction for a simple solvable model based on the conformal embedding $SU(2)_1 \times SU(2)_1 \supset U(1)_4 \times U(1)_4$. 
We show that the model possesses extensively degenerate ground states on a torus with deconfined quasiparticles, and that appropriate local perturbations lift the degeneracy and yield a 3D $Z_2$ gauge theory with a fermionic $Z_2$ charge.
\end{abstract}

\maketitle

\emph{Introduction.---}Two-dimensional (2D) topologically ordered phases, such as fractional quantum Hall states \cite{Tsui82, Laughlin83} and the toric codes \cite{Kitaev03}, harbor deconfined quasiparticle excitations obeying nontrivial braiding statistics \cite{Nayak08}. 
In three-dimensional (3D) space, topologically ordered phases can have two types of deconfined quasiparticles: point-like or loop-like. 
While the statistics between point-like quasiparticles can only be bosonic or fermionic in 3D space, there are possibilities of nontrivial point-loop, loop-loop, three-loop, and loop-loop-point braiding statistics \cite{ChenjieWang14, SJiang14, CMJian14, JCWang15, CHLin15, Putrov17, AtMaChan18}.
Along with the development of mathematical frameworks for classifying topologically ordered phases \cite{Wen16, ZCGu15, TLan16, TLan18, TLan19}, construction of microscopic Hamiltonians is also desired for their realizations in the real world. 
For certain 2D topological orders, a systematic construction scheme of exactly solvable Hamiltonians has been proposed by Levin and Wen in their string-net models \cite{Levin05}. 
Although our understanding of the 3D topological orders is much more limited, there have been several proposed schemes to write down exactly solvable Hamiltonians such as the Dijkgraaf-Witten model \cite{Dijkgraaf90, YidunWan15}, the Walker-Wang model \cite{Walker12, vonKeyserlingk13, ZitaoWang17}, and their relatives \cite{Williamson17}. 

In this Rapid Communication, we propose yet another way to construct microscopic Hamiltonians for 3D topologically ordered phases. 
Our approach is based on two key ingredients. 
The first one is coupled-wire construction of 2D topological phases originally developed by Kane and co-workers \cite{Kane02, Teo14}.
This construction uses a hybrid of continuum and lattice descriptions: One spatial direction is a discrete lattice, while the other direction is continuum and described by $(1+1)$-dimensional conformal field theory (CFT) \cite{dFMS}.
It has been successfully applied to various 2D topological phases \cite{YMLu12, Mong14, Oreg14, Neupert14, Sagi14, Klinovaja14, Vaezi14, PHHuang16, Fuji17, Kane17, Kane18}, and there have been several applications to the surface \cite{Mross15, Sahoo16, FLu17, SHong17, Volpez17, MCheng18, BHan18} or bulk \cite{Sagi15, Meng15, Iadecola16, Iadecola17, MJPark18, Raza19} of 3D topological phases. 
The second ingredient is the coupled-layer construction of 3D topological phases \cite{CMJian14, ChongWang13, Fidkowski13, BenZion16}. 
In this construction, one starts from stacked layers of 2D topologically ordered states and then induces anyon condensation \cite{Bais09} between adjacent layers. 
This approach is conceptually appealing and insightful, but its microscopic implementation is not straightforward, since a quantum phase transition from stacked layers of a 2D topological phase to a genuine 3D topological phase must be achieved by controlling interactions between adjacent layers.

Our construction brings these two ideas together and yields fully tractable 3D models with deconfined quasiparticles. 
We begin with a 3D system built out of quantum wires.
They constitute stacked layers of 2D topologically ordered states, each of which is described as a 2D coupled-wire model. 
A crucial step is to use conformal embedding \cite{dFMS} in each wire, which imposes nontrivial constraints on local operators and implements an effect like the anyon condensation. 
Depending on the choice of embeddings, the constructed models possess intrinsically point-like or loop-like quasiparticles deconfined in the full 3D space, in contrast to the previous 3D coupled-wire models \cite{Sagi15, Meng15, Iadecola16, Iadecola17, MJPark18, Raza19}.
However, the ground-state (GS) degeneracy turns out to be extensive due to the presence of quasiparticles confined in individual wires.
We argue that such a degeneracy can be lifted by local perturbations to give rise to a pure 3D topological order. 
We explain the construction of our model in more detail below.

\begin{figure}
\includegraphics[clip,width=0.45\textwidth]{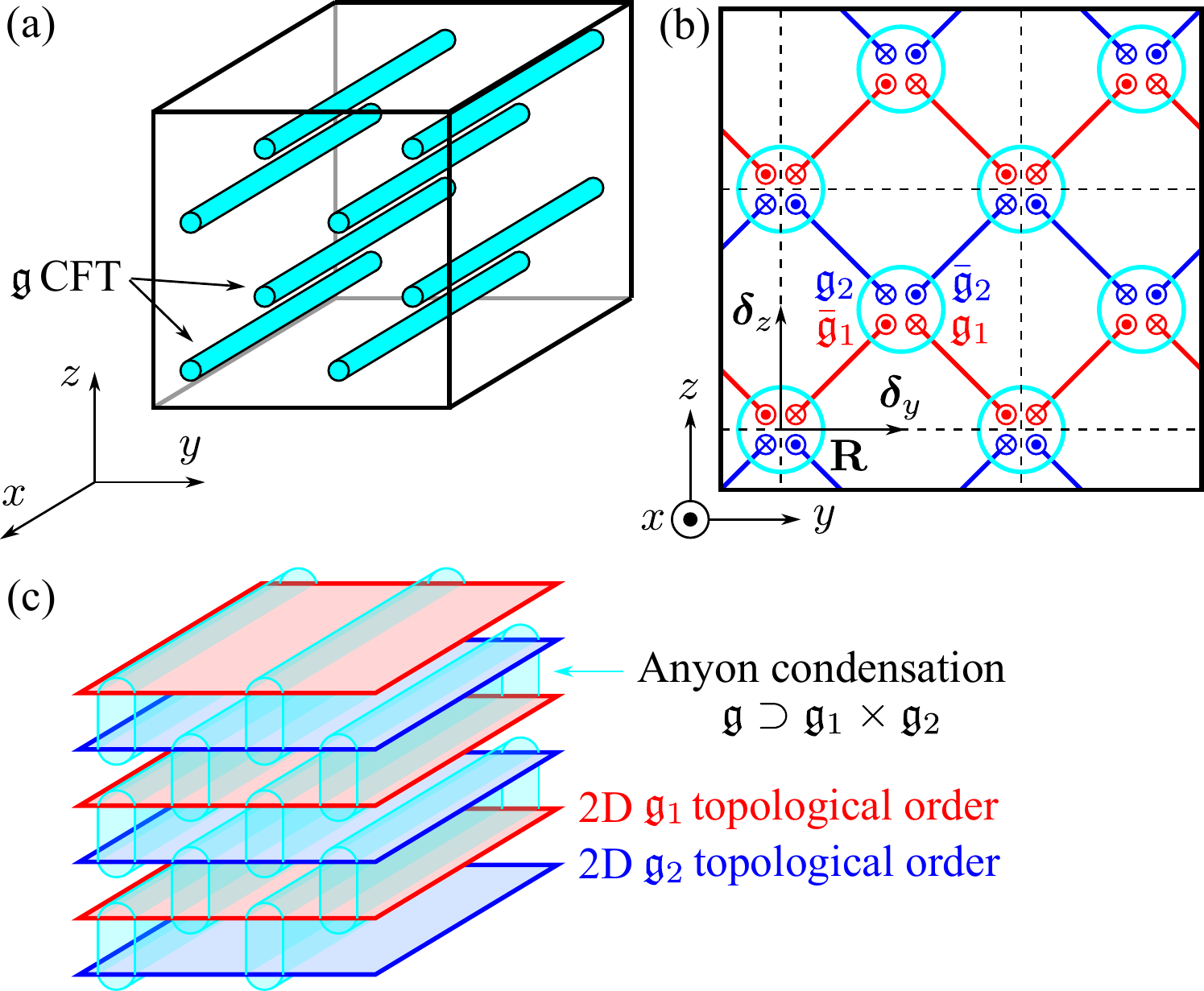}
\caption{(a) Schematic picture of our model composed of quantum wires described by the $\algg$ CFT and aligned along the $x$ axis. 
(b) The model projected onto the $yz$ plane. 
The dotted (crossed) circles denote the right-going (left-going) modes. 
The red (blue) colors are associated to the $\algg_1$ ($\algg_2$) CFTs. 
The plaquettes drawn by dashed lines represent the unit cells containing two wires of the $\algg$ CFT. 
(c) The model may be viewed as the layer construction of a 3D topologically ordered state.}
\label{fig:Model}
\end{figure}

\emph{Model.---}Our model consists of quantum wires described by the CFT associated with a current algebra $\mathfrak{g}$ \cite{Schoutens16}, whose primary fields are generated by the $\algg$ currents $\bfJ^\algg$ and $\bbfJ^\algg$ in the left- and right-going sectors, respectively, as local operators with respect to the currents. 
The wires are aligned in parallel to the $x$ axis and form a square lattice in the $yz$ plane as depicted in Fig.~\ref{fig:Model}~(a). 
We choose the unit cell to contain two wires, one from each sublattice of the square lattice as in Fig.~\ref{fig:Model}~(b). 
We then write the decoupled-wire Hamiltonian as 
\begin{align}
H_0 = \sum_\bfR (H_{0,\bfR} +H_{0,\bfR+\bfdelta_y +\bfdelta_z}),
\end{align}
where $\bfR=(y,z) \in (\mathbb{Z},\mathbb{Z})$ specifies the unit cell, $\bfdelta_y=(\frac{1}{2},0)$, $\bfdelta_z=(0,\frac{1}{2})$, and $H_{0,\bfr}$ is the Hamiltonian for a single wire on $\bfr$ standing for $\bfR$ or $\bfR+\bfdelta_y+\bfdelta_z$ . 
If $\algg$ is the affine Lie algebra of a simple Lie group $G$ with level $k$, $H_{0,\bfr}$ is given in the Sugawara form \cite{dFMS}, 
\begin{align}
H_{0,\bfr} = \frac{v}{4\pi (k+g)} \int dx \ \bigl( \mathrel{:\!\bfJ^{G_k}_{\bfr} \cdot \bfJ^{G_k}_{\bfr} \!:} +\mathrel{:\! \bbfJ^{G_k}_{\bfr} \cdot \bbfJ^{G_k}_{\bfr} \!:} \bigr), 
\end{align}
where $g$ is the dual Coxeter number of $G$ and $v$ is a velocity. 

We now assume that $\algg$ has some nontrivial conformal embedding \cite{dFMS} $\algg \supset \algg_1 \times \algg_2$, where $\algg_{1,2}$ are also current algebras generated by the currents $\bfJ^{\algg_1}_\bfr$ and $\bfJ^{\algg_2}_\bfr$, respectively.
A physical meaning of this embedding is that the primary fields of individual $\algg_1$ or $\algg_2$ do not constitute local operators of the theory, but their products do when they form a representation of $\algg$.
If there exists such an embedding, the single-wire Hamiltonian is split into the sum of the Hamiltonians for $\algg_{1,2}$.
We then consider the following interactions between adjacent wires: 
\begin{align} \label{eq:InterwireInt}
H_1 &= \gamma \sum_\bfR \int dx \ \bigl( \bfJ^{\algg_1}_\bfR \cdot \bbfJ^{\algg_1}_{\bfR+\bfdelta_y +\bfdelta_z} +\bbfJ^{\algg_1}_\bfR \cdot \bfJ^{\algg_1}_{\bfR -\bfdelta_y +\bfdelta_z} \nonumber \\
&\qquad\qquad
+\bbfJ^{\algg_2}_\bfR \cdot \bfJ^{\algg_2}_{\bfR +\bfdelta_y -\bfdelta_z} +\bfJ^{\algg_2}_\bfR \cdot \bbfJ^{\algg_2}_{\bfR -\bfdelta_y -\bfdelta_z} \bigr).
\end{align}
If $\algg_{1,2}$ are affine Lie algebras, the one-loop renormalization group analysis shows that this interaction flows to the strong-coupling limit for $\gamma>0$. 
For some $\algg_1$ and $\algg_2$, the model is exactly solvable and known to be gapped \cite{[{See, e.g., }][{ and references therein.}]James18}. 
This Hamiltonian is a straightforward extension of previous coupled-wire models based on the current algebra \cite{Teo14, PHHuang16, Sahoo16, Fuji17, BHan18}. 
The model may appear to be alternatively stacked layers of the 2D $\algg_1$ and $\algg_2$ topologically ordered states as in Fig.~\ref{fig:Model}~(c). 
However, the conformal embedding implements the anyon condensation \cite{Bais09} at which the wires are placed. 
This makes quasiparticles of the model essentially deconfined in the full 3D space in contrast to the stacked 2D topologically ordered states.
In order to see this, we first focus on a simple solvable model and then argue physical properties of the model for general $\algg$.

\emph{Example.---}We here consider the case of $\algg=SU(2)_1 \times SU(2)_1$ and $\algg_1 = \algg_2 = U(1)_4$. 
The latter is described by the free boson CFT with the compactification radius $2$, which is related to the $\nu=\frac{1}{4}$ Laughlin state \cite{Nayak08}. 
Each quantum wire consists of two critical spin-$\frac{1}{2}$ chains, and each spin chain is described by the $SU(2)_1$ CFT and conveniently represented by bosonic fields \cite{GNT, Fradkin}. 
We thus introduce pairs of dual bosonic fields $\varphi^l_{\bfr}, \theta^l_{\bfr}$ with the chain index $l=1,2$, which obey the commutation relations $[\theta^l_{\bfr}(x), \varphi^{l'}_{\bfr'}(x')] = i\pi \delta_{\bfr, \bfr'} \delta_{l,l'} \Theta (x-x')$ with $\Theta(x)$ being the Heaviside step function. 
We also define chiral bosonic fields by $\phi^l_{R/L,\bfr} \equiv \varphi^l_\bfr \pm \theta^l_\bfr$. 
In terms of these bosonic fields, the $SU(2)_1$ currents $\bfJ^\algg_\bfr = (\bfJ^1_\bfr, \bfJ^2_\bfr)$ are represented as $(J^l_\bfr)^\pm \propto e^{\pm i \sqrt{2} \phi^l_{L,\bfr}}$ and $(J^l_\bfr)^z \propto \partial_x \phi^l_{L,\bfr}$ and similarly for $\bbfJ^l_\bfr$ with $\phi^l_{R,\bfr}$. 
These correspond to the uniform parts of lattice spin operators $\bfS_{\bfr,l}(x)$. 
The Hamiltonian for each wire is given by a free boson theory,
\begin{align}
H_{0,\bfr} = \frac{v}{4\pi} \int dx \sum_{l=1,2} \Bigl[ (\partial_x \phi^l_{R,\bfr})^2 +(\partial_x \phi^l_{L,\bfr})^2 \Bigr].
\end{align}
Local operators of the theory are written in terms of left-right products of the spin-$\frac{1}{2}$ primary fields $e^{(i/\sqrt{2})\phi^l_{R/L,\bfr}}$ with the conformal weight $h=1/4$ for individual $l$'s, such as $e^{i\sqrt{2} \varphi^l_\bfr}$ and $e^{i\sqrt{2} \theta^l_\bfr}$. 
Physically, these operators are related to the N\'eel or dimer order parameters of the spin chain $l$: $\bfn_{\bfr,l}(x) = (-1)^x \bfS_{\bfr,l}(x)$, $\epsilon_{\bfr,l}(x) = (-1)^x \bfS_{\bfr,l}(x) \cdot \bfS_{\bfr,l}(x+1)$. 

Let us introduce the symmetric and antisymmetric combinations of the bosonic fields, $\tphi^{S/A}_{P,\bfr} \equiv (\phi^1_{P,\bfr} \pm \phi^2_{P,\bfr})/\sqrt{2}$ for $P=R,L$, as is usually done in the analysis of two coupled spin chains \cite{Schulz86, Strong92, Shelton96}. 
We can then represent the $U(1)_4$ currents $(\bfJ^{\algg_1}_\bfr, \bfJ^{\algg_2}_\bfr) = (\bfJ^S_\bfr,\bfJ^A_\bfr)$  by $(J^S_\bfr)^\pm \propto e^{\pm i2 \tphi^S_{L,\bfr}}$, $(J^A_\bfr)^\pm \propto e^{\pm i2 \tphi^A_{L,\bfr}}$, which are products of the $SU(2)_1$ currents. 
The $U(1)_4$ CFT has three primary fields $e^{i\tphi^\rho_{P,\bfr}}$ and $e^{\pm (i/2) \tphi^\rho_{P,\bfr}}$ with $h=1/2$ and $1/8$, respectively, for each of $\rho=S,A$. 
The currents of the $SU(2)_1 \times SU(2)_1$ CFT are now represented by bilinears of the primary fields with $h=1/2$ of the $U(1)_4 \times U(1)_4$ CFT; this establishes the desired conformal embedding.
Plugging the above expressions into Eq.~\eqref{eq:InterwireInt}, we consider the interwire interactions,
\begin{align} \label{eq:InterwireSpin}
H_1 &= \gamma \sum_\bfR \int dx \ \Bigl[ e^{i2(\tphi^S_{L,\bfR} -\tphi^S_{R,\bfR+\bfdelta_y+\bfdelta_z})} \nonumber \\
&\ \ \ +e^{i2(\tphi^S_{R,\bfR} -\tphi^S_{L,\bfR-\bfdelta_y+\bfdelta_z})} +e^{i2(\tphi^A_{L,\bfR} -\tphi^A_{R,\bfR-\bfdelta_y-\bfdelta_z})} \nonumber \\
&\ \ \ +e^{i2(\tphi^A_{R,\bfR} -\tphi^A_{L,\bfR+\bfdelta_y-\bfdelta_z})} +\textrm{H.c.} \Bigr],
\end{align}
which may be generated by four-spin interactions, such as $S^+_{\bfr,1} S^+_{\bfr,2} S^-_{\bfr',1} S^-_{\bfr',2}$.
However, we remark that these interactions are irrelevant for the $SU(2)$-symmetric spin chains, in contrast to similar microscopic constructions for the 2D case where interactions are marginal \cite{Gorohovsky15, Lecheminant17, JHChen17}, since $\bfJ^l_\bfr$ has the weight $h=1$ and thus the interactions have the scaling dimension $4$. 
In principle, they can be made relevant by adding appropriate forward scattering interactions to $H_0$ \cite{Teo14}, but we do not pursue their detailed microscopic forms at this stage. 
The Hamiltonian $H_0+H_1$ is a sine-Gordon model with many cosine terms. 
On a torus with linear sizes $L_x \times L_y \times L_z$, there are $2L_y L_z$ wires and thus $4L_y L_z$ pairs of the bosonic fields. 
Since the linearly independent set of the $4L_y L_z$ fields on the links satisfies the Haldane's null-vector condition \cite{Haldane95}, the interaction can open a gap. 
Furthermore, we can solve the model exactly in the limit of $\gamma \to \infty$, where the link fields are pinned at potential minima. 
In this limit, we can adopt the formal recipe developed by Ganeshan and Levin \cite{Ganeshan16} to obtain the ground state. 
As we show below, the model has \emph{extensive} GS degeneracy $4 \cdot 2^{2L_y L_z}$ \footnote[1]{See Supplemental Material, which includes Refs.~\cite{Newman, Ganeshan17, Walton89, Altschuler90}, for a detailed discussion on the ground state and full forms of the branching rules.}.

\begin{figure}
\includegraphics[clip,width=0.48\textwidth]{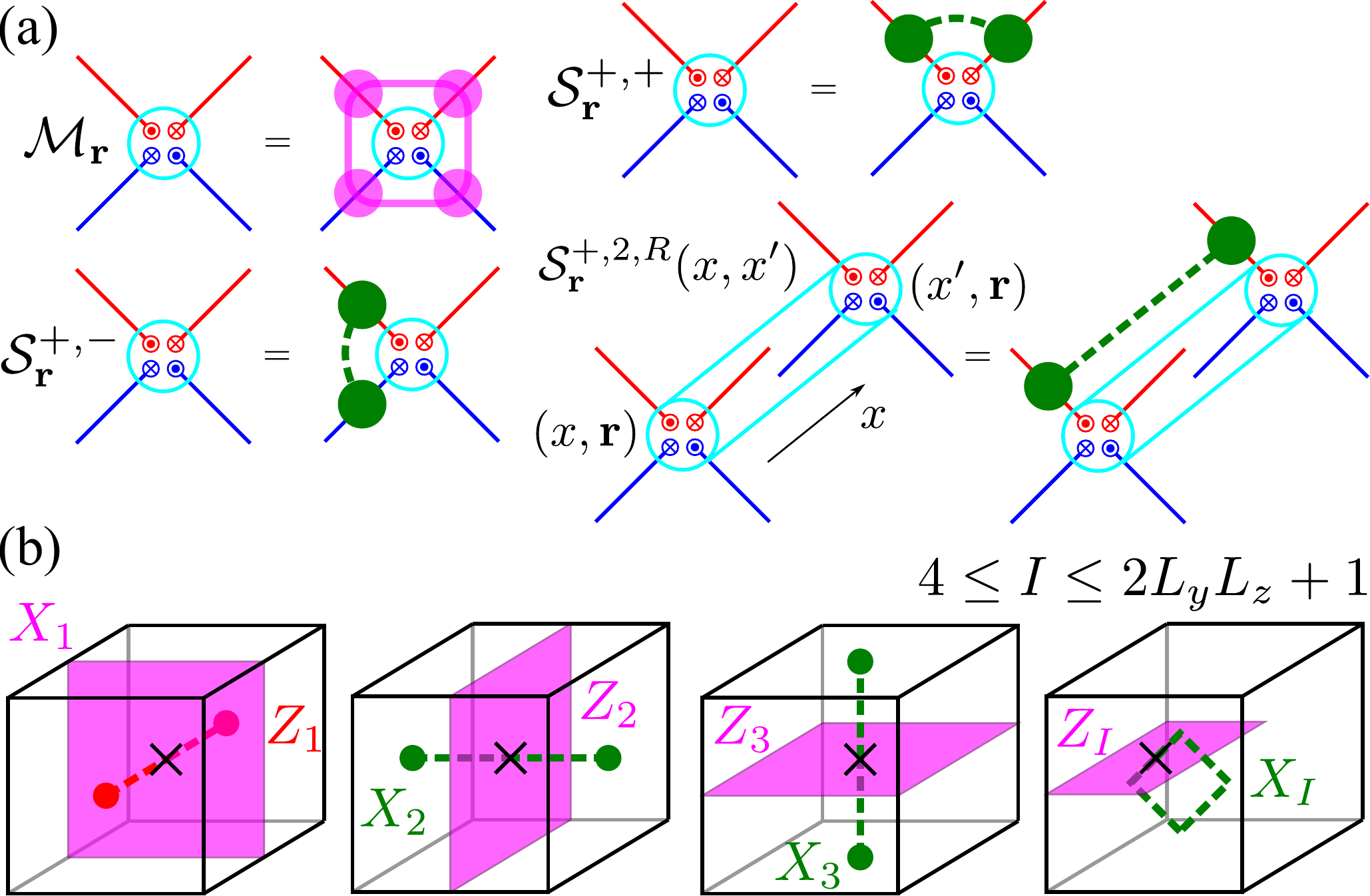}
\caption{(a) Actions of various local operators on a ground state. 
The purple (green) circles denote the kinks created by the $h=1/8$ ($h=1/2$) operators of the $U(1)_4$ CFT, respectively. 
(b) Schematic pictures of the operators $\{ X_I, Z_I \}$ that span the degenerate ground states.}
\label{fig:Operators}
\end{figure}

\emph{Ground state and quasiparticles.---}This extensive degeneracy is in sharp contrast to finite degeneracy expected for a pure 3D topological order and originates from less constrained energentics of quasiparticles in our model. 
As in the case of 2D coupled-wire models \cite{Teo14}, acting local operators on a ground state creates kinks on some links, at which the expectation values of the link fields deviate from those in the ground state; these kinks are interpreted as quasiparticles. 
The actions of such local operators are schematically shown in Fig.~\ref{fig:Operators}~(a).
There are two types of operators creating quasiparticles within the $yz$ plane: 
One is given by $\calS^{\rho,\rho'}_{\bfr}(x) \equiv e^{i(\tphi^\rho_{R,\bfr} -\tphi^{\rho'}_{L,\bfr})}$, up to multiplications of the $U(1)_4$ currents, which creates a pair of kinks on two of the four links centering on $\bfr$. 
Successive application of $\calS^{\rho,\rho'}_\bfr$ along a string separates two kinks far apart and may be viewed as creating a pair of point-like quasiparticles at the ends of the string. 
Since $e^{i\tphi^\rho_{P,\bfr}}$ has $h=1/2$, these point-like quasiparticles are fermions. 
The other is given by $\calM_\bfr(x) \equiv e^{(i/2)(\tphi^S_{R,\bfr} -\tphi^S_{L,\bfr} +\tphi^A_{R,\bfr} -\tphi^A_{L,\bfr})}$, up to multiplications of $\calS^{\rho,\rho'}_\bfr$, which creates kinks on all four links around $\bfr$. 
It consists of two primary fields of $U(1)_4$ with $h=1/8$.
Successively applying $\calM_\bfr$ over a membrane creates a loop-like excitation at the boundary of the membrane.
To have quasiparticles deconfined in the full 3D space, we need operators transferring quasiparticles along the $x$ direction. 
Such operators are given by $\calS^{\rho,q,P}_\bfr(x,x') \equiv e^{(iq/2) [\tphi^\rho_{P,\bfr}(x') -\tphi^\rho_{P,\bfr}(x)]}$ with $q \in \mathbb{Z}$, which transfer a quasiparticle from $x$ to $x'$ along the wire $\bfr$.
These operators, together with the fermionic strings in the $yz$ plane, make fermionic quasiparticles fully deconfined in the 3D space. 
It turns out, however, that we can also create a pair of the $h=1/8$ quasiparticles deconfined within individual wires but not between wires. 
This is actually the source of the extensive degeneracy that blows up with increasing the number of wires. 

In order to see this, we construct a set of operators that map the GS manifold to itself and span the $4 \cdot 2^{2L_y L_z}$-dimensional Hilbert space \footnotemark[1].
One operator is the closed membrane operator $X_1 \equiv \prod_\bfr \calM_\bfr$ that covers the whole $yz$ plane wrapping the torus. 
The other operators are formed by closed fermion string operators. 
There are two operators winding the torus along the $y$ or $z$ axis: $X_2 \equiv \prod_{\bfr \parallel \hat{\mathbf{y}}} \calS^{S,S}_\bfr$ and $X_3 \equiv \prod_{\bfr \parallel \hat{\mathbf{z}}} S^{S,A}_\bfr$. 
There is also an operator encircling a plaquette $p$ of the square lattice: $\calS_p \equiv \prod_{\bfr \in p} \calS^{\rho,\rho'}_\bfr$.  
Although there are $2L_y L_z$ square plaquettes, there are only $2L_y L_z-2$ independent $\calS_p$'s. 
We label these plaquette operators by $\{ X_4, X_5, \cdots, X_{2L_yL_z+1} \}$. 
We can then find a set of $2L_y L_z +1$ operators $\{ Z_I \}$ that consist of string operators $\calS^{\rho,q,P}_\bfr(0,L_x)$ and satisfies $X_1 Z_1 =iZ_1 X_1$ and $X_I Z_I =-Z_I X_I$ for $I>1$. 
Such $Z_I$ contains a string of the $h=1/8$ primaries winding the torus along the $x$ axis and intersecting only \emph{once} with $X_I$. 
These operators can be chosen to satisfy $[X_I, X_J] = [Z_I, Z_J] = [X_I, Z_J]=0$ for $I \neq J$ (for the precise definitions of $X_I,Z_I$, see \footnotemark[1]). 
The nontrivial algebra of $X_I$ and $Z_I$ is related to the statistical angle $\pi/4$ of the fundamental quasiparticle of $U(1)_4$. 
Hence, a pair of operators $(X_1,Z_1)$ span a four-dimensional space, and the remaining pairs $(X_I,Z_I)$ span two-dimensional spaces each, yielding a $4 \cdot 2^{2L_y L_z}$-dimensional Hilbert space of the degenerate ground states. 

\begin{figure}
\includegraphics[clip,width=0.48\textwidth]{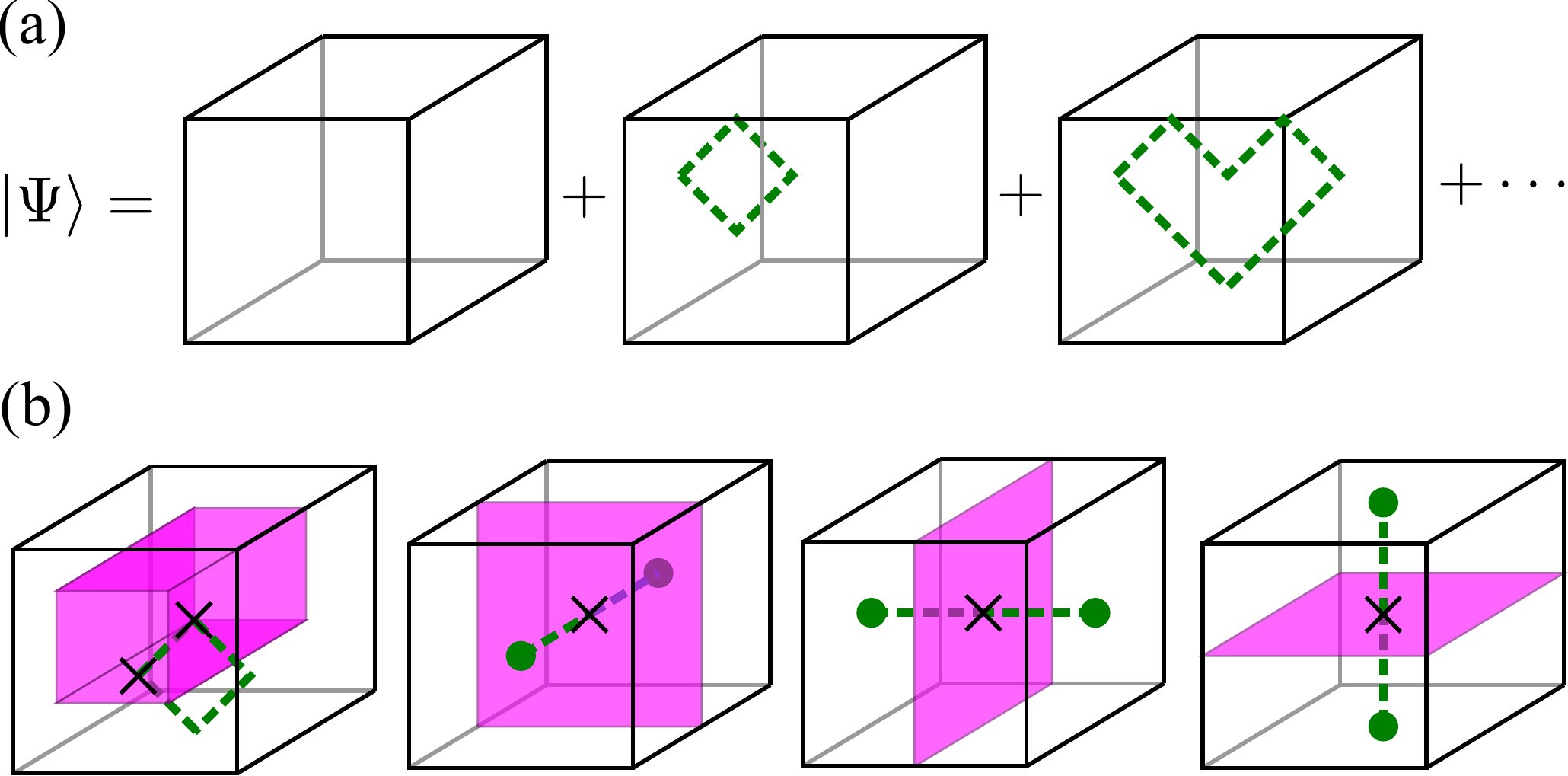}
\caption{(a) Schematic picture of $\left| \Psi \right>$, which is the condensate of closed fermion strings. 
(b) After the condensation, membrane operators composed of the $h=1/8$ operators must be closed as shown in the left-most panel. 
It results in the $2^3$-fold degenerate GS manifold spanned by the membrane and string operators shown in the right three panels.}
\label{fig:StringCondensation}
\end{figure}

\emph{Splitting degeneracy.---}The GS degeneracy is the very consequence of the existence of deconfined quasiparticles.
However, the extensive degeneracy obscures the braiding statistics of quasiparticles due to the lack of a well-defined adiabatic process.
We thus want to lift the extensive degeneracy to have ground states with finitely bound degeneracy. 
This should be possible since 1D deconfined excitations within each wire are not robust against local perturbations. 
For our purpose, we want a further kinetic constraint that prohibits individual motion of the $h=1/8$ quasipaticles and realizes loop-like quasiparticles deconfied in the $xy$ and $zx$ planes in addition to the $yz$ plane. 
Recalling the physics of spin chains, deconfined excitations are spinons (domain walls) and created by a string of the spin-$\frac{1}{2}$ primary field, which is now further fractionalized in our model. 
Interchain coupling of the form $\bfn_{\bfr,1} \cdot \bfn_{\bfr,2}$ or $\epsilon_{\bfr,1} \epsilon_{\bfr,2}$ binds these spinons and let the bound spinons coherently move on the chains. 
Adding such interactions to the Hamiltonian and performing degenerate perturbation theory, we obtain nonvanishing contributions at forth order, which are operators creating a closed fermion string; they are precisely $\calS_p$ defined above. 
We can then consider the effective Hamiltonian $H_\textrm{eff} = u \sum_p \int dx \, (\calS_p +\textrm{H.c.})$, where the sum is taken over all square plaquettes. 
In analogy with the string-net model \cite{Levin05}, a part of the Hilbert space is spanned by all possible configurations of closed fermion strings. 
The operators $\calS_p$ are local and commuting with each other and create a ``resonance'' between these string configurations. 
The Hamiltonian $H_\textrm{eff}$ thus induces a condensation of the fermion strings and its ground state $\left| \Psi \right>$ is given by the equal-weight superposition of all string configurations [see Fig.~\ref{fig:StringCondensation}~(a)]. 
The fermion strings created by $X_2$ or $X_3$ wind the torus and cannot be removed by the local resonance. 
They contribute the GS degeneracy of $2^2$. 
Projecting $X_1,Z_1$ onto $\left| \Psi \right>$, $X_1^2=1$ and $Z_1$ vanishes but $Z_1^2$ survives. 
From $X_1 Z_1^2 = -Z_1^2 X_1$, they also contribute a factor of $2$ to the degeneracy. 
Thus, we find in total $2^3$-fold degenerate ground states.
This is consistent with the following physical picture: 
The condensation of closed fermion strings allows only closed membrane operators composed of the string operators transferring the $h=1/8$ quasiparticles along the $x$ axis. 
Hence, we have three membrane operators of the $h=1/8$ quasiparticles in the $xy$, $yz$, and $zx$ planes, which have nontrivial algebra with the fermion string operators along the $z$, $x$, and $y$ axes, respectively, yielding the $2^3$-fold degeneracy on the torus [see Fig.~\ref{fig:StringCondensation}~(b)]. 
Therefore, with proper perturbations to lift the extensive degeneracy, our model will have nontrivial braiding between point-like and loop-like quasiparticles and describe a $Z_2$ gauge theory with a fermionic charge.

\emph{Generalization.---}We now consider the case of general embedding $\algg \supset \algg_1 \times \algg_2$. 
Let $\{ \Phi \}$, $\{ \Phi_1 \}$, and $\{ \Phi_2 \}$ be sets of primary fields of the current algebras $\algg$, $\algg_1$, and $\algg_2$, respectively. 
How the primary field $\Phi$ is written in terms of $\Phi_1$ and $\Phi_2$ is dictated by the branching rule (BR) \cite{dFMS}, $\Phi \mapsto \bigoplus_{\Phi_1, \Phi_2} b^\Phi_{\Phi_1,\Phi_2} (\Phi_1 \otimes \Phi_2)$, where the branching index $b^\Phi_{\Phi_1,\Phi_2}$ is some integer. 
This determines the allowed patterns of ``fractionalization'' of quasiparticles created by local operators, which are written as left-right products of $\Phi$. 
We here assume that $\Phi_1$ and $\Phi_2$ are always deconfined along the $x$ axis within each wire, as shown in the previous example, unless additional interactions to $H_1$ are considered.
We make the following observations.
(i) If the BR for given $\Phi$ has $b^\Phi_{\bfone_1,\Phi_2} \neq 0$ and $b^\Phi_{\Phi_1,\bfone_2} \neq 0$ for some nontrivial $\Phi_1$ and $\Phi_2$ ($\bfone$ denotes the trivial primary field), then local operators associated with $\Phi$ can create 3D point-like quasiparticles. 
We anticipate that this happens only for the primary fields with integer or half-integer conformal weight, as the particle statistics in 3D space is either bosonic or fermionic. 
(ii) If the BR has $b^\Phi_{\bfone_1, \Phi_2} \neq 0$ and $b^\Phi_{\Phi_1, \bfone_2} = 0$ (or vice versa), then there are point-like quasiparticles confined in a 2D subspace since wires connected via the $\algg_2$ ($\algg_1$) currents can only form a 2D network. 
(iii) If the BR has $b^\Phi_{\Phi_1,\Phi_2} \neq 0$ for nontrivial $\Phi_1$ and $\Phi_2$, there are loop-like quasiparicles in the $yz$ plane, but their energetics along the $x$ axis is not fully constrained. 

An example of embeddings that will admit fermionic point-like quasiparticles is $SO(p+q)_1 \supset SO(p)_1 \times SO(q)_1$. 
Indeed, the previous example is the special case of this type for $p=q=2$. 
For $p=q=3$, this embedding is equivalent to $SU(4)_1 \supset SU(2)_2 \times SU(2)_2$. 
The primary fields $\mathbf{6}$ of $SU(4)_1$ and $\mathbf{3}$ of $SU(2)_2$ have the conformal weight $h=1/2$ and thus are fermions. 
Since they follow the BR $\mathbf{6} \mapsto (\mathbf{1}_1 \otimes \mathbf{3}_2) \oplus (\mathbf{3}_1 \otimes \mathbf{1}_2)$ \footnotemark[1], we expect from (i) that local operators associated with $\mathbf{6}$ create fermionic quasiparticles. 
The other primary fields $\mathbf{4}$ and $\overline{\mathbf{4}}$ follow the BRs $\mathbf{4} \mapsto\mathbf{2}_1 \otimes \mathbf{2}_2$ and $\overline{\mathbf{4}} \mapsto \mathbf{2}_1 \otimes \mathbf{2}_2$ and will create loop-like quasiparticles. 
As models with bosonic quasiparticles, we may consider the embeddings $SU(8)_1 \supset SU(2)_4 \times SU(4)_2$ and $SU(9)_1 \supset SU(3)_3 \times SU(3)_3$. 
In the former case, the primary fields $\mathbf{70}$ of $SU(8)_1$, $\mathbf{5}$ of $SU(2)_4$, and $\mathbf{20}^a$ (the self-conjugate representation) of $SU(4)_2$ all have the weight $h=1$ and thus are bosons. 
From the BR $\mathbf{70} \mapsto (\mathbf{1}_1 \otimes \mathbf{20}^a_2) \oplus (\mathbf{5}_1 \otimes \mathbf{1}_2) \oplus (\mathbf{3}_1 \otimes \mathbf{15}_2)$, local operators associated with $\mathbf{70}$ create bosonic quasiparticles. 
Similarly, the primary fields $\mathbf{84}$ and $\overline{\mathbf{84}}$ of $SU(9)_1$ and $\mathbf{10}$ and $\overline{\mathbf{10}}$ of $SU(3)_3$ all have the weight $h=1$. 
From the BR $\mathbf{84} \mapsto (\mathbf{1}_1 \otimes \mathbf{10}_2) \oplus (\mathbf{10}_1 \otimes \mathbf{1}_2) \oplus (\mathbf{8}_1 \otimes \mathbf{8}_2) \oplus (\overline{\mathbf{10}}_1 \otimes \overline{\mathbf{10}}_2)$, local operators associated with $\mathbf{84}$ or $\overline{\mathbf{84}}$ will also create bosonic quasiparticles (the BR for $\overline{\mathbf{84}}$ is obtained by conjugation).
In the models corresponding to these embeddings, we expect that the GS manifold still has extensively degeneracy due to the presence of 1D deconfined quasiparticles in each wire, but a condensation of closed particle strings may lift such degeneracy. 
However, since these theories are non-Abelian, the energetics of quasiparticles will become more complicated than the solvable example considered before. 
There is also the case where there are no point-like quasiparticles deconfined in the full 3D space, such as $SU(6)_1 \supset SU(2)_3 \times SU(3)_2$. 
Complete understanding of models related to general embeddings is left for the future study. 

\emph{Discussion.---}We proposed a systematic way to construct 3D topological orders from coupled quantum wires using conformal embedding. 
We have considered only bulk properties of the models here, but we anticipate interesting surface physics in these models. 
In particular, if some appropriate CFT and embedding are chosen for each wire, one may find 3D symmetry-protected topological (SPT) phases with no nontrivial bulk quasiparticles. 
Such SPT phases are expected to have surface topological orders when the surface is gapped with preserving the symmetry \cite{Senthil15}. 
The existing coupled-layer constructions \cite{CMJian14, ChongWang13, Fidkowski13, BenZion16} will give us a hint on building microscopic Hamiltonians along with our proposal. 
We also expect from their similarity that there may be some connection between our models and the Walker-Wang models \cite{Walker12, vonKeyserlingk13, ZitaoWang17}, since our model has a degenerate GS manifold which is spanned by configurations of closed strings whose condensation leads to a 3D topological order. 
We believe that our construction gives a promising route to obtain microscopic Hamiltonians for various 3D topological phases.

\emph{Acknowledgments.---}Y.F. and A.F. are partially supported by JSPS KAKENHI Grant Nos. JP17H07362 and JP15K05141, respectively.

\bibliography{CoupledWire3DTopo}

\begin{thebibliography}{72}%
\makeatletter
\providecommand \@ifxundefined [1]{%
 \@ifx{#1\undefined}
}%
\providecommand \@ifnum [1]{%
 \ifnum #1\expandafter \@firstoftwo
 \else \expandafter \@secondoftwo
 \fi
}%
\providecommand \@ifx [1]{%
 \ifx #1\expandafter \@firstoftwo
 \else \expandafter \@secondoftwo
 \fi
}%
\providecommand \natexlab [1]{#1}%
\providecommand \enquote  [1]{``#1''}%
\providecommand \bibnamefont  [1]{#1}%
\providecommand \bibfnamefont [1]{#1}%
\providecommand \citenamefont [1]{#1}%
\providecommand \href@noop [0]{\@secondoftwo}%
\providecommand \href [0]{\begingroup \@sanitize@url \@href}%
\providecommand \@href[1]{\@@startlink{#1}\@@href}%
\providecommand \@@href[1]{\endgroup#1\@@endlink}%
\providecommand \@sanitize@url [0]{\catcode `\\12\catcode `\$12\catcode
  `\&12\catcode `\#12\catcode `\^12\catcode `\_12\catcode `\%12\relax}%
\providecommand \@@startlink[1]{}%
\providecommand \@@endlink[0]{}%
\providecommand \url  [0]{\begingroup\@sanitize@url \@url }%
\providecommand \@url [1]{\endgroup\@href {#1}{\urlprefix }}%
\providecommand \urlprefix  [0]{URL }%
\providecommand \Eprint [0]{\href }%
\providecommand \doibase [0]{http://dx.doi.org/}%
\providecommand \selectlanguage [0]{\@gobble}%
\providecommand \bibinfo  [0]{\@secondoftwo}%
\providecommand \bibfield  [0]{\@secondoftwo}%
\providecommand \translation [1]{[#1]}%
\providecommand \BibitemOpen [0]{}%
\providecommand \bibitemStop [0]{}%
\providecommand \bibitemNoStop [0]{.\EOS\space}%
\providecommand \EOS [0]{\spacefactor3000\relax}%
\providecommand \BibitemShut  [1]{\csname bibitem#1\endcsname}%
\let\auto@bib@innerbib\@empty
\bibitem [{\citenamefont {Tsui}\ \emph {et~al.}(1982)\citenamefont {Tsui},
  \citenamefont {Stormer},\ and\ \citenamefont {Gossard}}]{Tsui82}%
  \BibitemOpen
  \bibfield  {author} {\bibinfo {author} {\bibfnamefont {D.~C.}\ \bibnamefont
  {Tsui}}, \bibinfo {author} {\bibfnamefont {H.~L.}\ \bibnamefont {Stormer}}, \
  and\ \bibinfo {author} {\bibfnamefont {A.~C.}\ \bibnamefont {Gossard}},\
  }\href {\doibase 10.1103/PhysRevLett.48.1559} {\bibfield  {journal} {\bibinfo
   {journal} {Phys. Rev. Lett.}\ }\textbf {\bibinfo {volume} {48}},\ \bibinfo
  {pages} {1559} (\bibinfo {year} {1982})}\BibitemShut {NoStop}%
\bibitem [{\citenamefont {Laughlin}(1983)}]{Laughlin83}%
  \BibitemOpen
  \bibfield  {author} {\bibinfo {author} {\bibfnamefont {R.~B.}\ \bibnamefont
  {Laughlin}},\ }\href {\doibase 10.1103/PhysRevLett.50.1395} {\bibfield
  {journal} {\bibinfo  {journal} {Phys. Rev. Lett.}\ }\textbf {\bibinfo
  {volume} {50}},\ \bibinfo {pages} {1395} (\bibinfo {year}
  {1983})}\BibitemShut {NoStop}%
\bibitem [{\citenamefont {Kitaev}(2003)}]{Kitaev03}%
  \BibitemOpen
  \bibfield  {author} {\bibinfo {author} {\bibfnamefont {A.}~\bibnamefont
  {Kitaev}},\ }\href {\doibase 10.1016/S0003-4916(02)00018-0} {\bibfield
  {journal} {\bibinfo  {journal} {Ann. Phys.}\ }\textbf {\bibinfo {volume}
  {303}},\ \bibinfo {pages} {2 } (\bibinfo {year} {2003})}\BibitemShut
  {NoStop}%
\bibitem [{\citenamefont {Nayak}\ \emph {et~al.}(2008)\citenamefont {Nayak},
  \citenamefont {Simon}, \citenamefont {Stern}, \citenamefont {Freedman},\ and\
  \citenamefont {Das~Sarma}}]{Nayak08}%
  \BibitemOpen
  \bibfield  {author} {\bibinfo {author} {\bibfnamefont {C.}~\bibnamefont
  {Nayak}}, \bibinfo {author} {\bibfnamefont {S.~H.}\ \bibnamefont {Simon}},
  \bibinfo {author} {\bibfnamefont {A.}~\bibnamefont {Stern}}, \bibinfo
  {author} {\bibfnamefont {M.}~\bibnamefont {Freedman}}, \ and\ \bibinfo
  {author} {\bibfnamefont {S.}~\bibnamefont {Das~Sarma}},\ }\href {\doibase
  10.1103/RevModPhys.80.1083} {\bibfield  {journal} {\bibinfo  {journal} {Rev.
  Mod. Phys.}\ }\textbf {\bibinfo {volume} {80}},\ \bibinfo {pages} {1083}
  (\bibinfo {year} {2008})}\BibitemShut {NoStop}%
\bibitem [{\citenamefont {Wang}\ and\ \citenamefont
  {Levin}(2014)}]{ChenjieWang14}%
  \BibitemOpen
  \bibfield  {author} {\bibinfo {author} {\bibfnamefont {C.}~\bibnamefont
  {Wang}}\ and\ \bibinfo {author} {\bibfnamefont {M.}~\bibnamefont {Levin}},\
  }\href {\doibase 10.1103/PhysRevLett.113.080403} {\bibfield  {journal}
  {\bibinfo  {journal} {Phys. Rev. Lett.}\ }\textbf {\bibinfo {volume} {113}},\
  \bibinfo {pages} {080403} (\bibinfo {year} {2014})}\BibitemShut {NoStop}%
\bibitem [{\citenamefont {Jiang}\ \emph {et~al.}(2014)\citenamefont {Jiang},
  \citenamefont {Mesaros},\ and\ \citenamefont {Ran}}]{SJiang14}%
  \BibitemOpen
  \bibfield  {author} {\bibinfo {author} {\bibfnamefont {S.}~\bibnamefont
  {Jiang}}, \bibinfo {author} {\bibfnamefont {A.}~\bibnamefont {Mesaros}}, \
  and\ \bibinfo {author} {\bibfnamefont {Y.}~\bibnamefont {Ran}},\ }\href
  {\doibase 10.1103/PhysRevX.4.031048} {\bibfield  {journal} {\bibinfo
  {journal} {Phys. Rev. X}\ }\textbf {\bibinfo {volume} {4}},\ \bibinfo {pages}
  {031048} (\bibinfo {year} {2014})}\BibitemShut {NoStop}%
\bibitem [{\citenamefont {Jian}\ and\ \citenamefont {Qi}(2014)}]{CMJian14}%
  \BibitemOpen
  \bibfield  {author} {\bibinfo {author} {\bibfnamefont {C.-M.}\ \bibnamefont
  {Jian}}\ and\ \bibinfo {author} {\bibfnamefont {X.-L.}\ \bibnamefont {Qi}},\
  }\href {\doibase 10.1103/PhysRevX.4.041043} {\bibfield  {journal} {\bibinfo
  {journal} {Phys. Rev. X}\ }\textbf {\bibinfo {volume} {4}},\ \bibinfo {pages}
  {041043} (\bibinfo {year} {2014})}\BibitemShut {NoStop}%
\bibitem [{\citenamefont {Wang}\ and\ \citenamefont {Wen}(2015)}]{JCWang15}%
  \BibitemOpen
  \bibfield  {author} {\bibinfo {author} {\bibfnamefont {J.~C.}\ \bibnamefont
  {Wang}}\ and\ \bibinfo {author} {\bibfnamefont {X.-G.}\ \bibnamefont {Wen}},\
  }\href {\doibase 10.1103/PhysRevB.91.035134} {\bibfield  {journal} {\bibinfo
  {journal} {Phys. Rev. B}\ }\textbf {\bibinfo {volume} {91}},\ \bibinfo
  {pages} {035134} (\bibinfo {year} {2015})}\BibitemShut {NoStop}%
\bibitem [{\citenamefont {Lin}\ and\ \citenamefont {Levin}(2015)}]{CHLin15}%
  \BibitemOpen
  \bibfield  {author} {\bibinfo {author} {\bibfnamefont {C.-H.}\ \bibnamefont
  {Lin}}\ and\ \bibinfo {author} {\bibfnamefont {M.}~\bibnamefont {Levin}},\
  }\href {\doibase 10.1103/PhysRevB.92.035115} {\bibfield  {journal} {\bibinfo
  {journal} {Phys. Rev. B}\ }\textbf {\bibinfo {volume} {92}},\ \bibinfo
  {pages} {035115} (\bibinfo {year} {2015})}\BibitemShut {NoStop}%
\bibitem [{\citenamefont {Putrov}\ \emph {et~al.}(2017)\citenamefont {Putrov},
  \citenamefont {Wang},\ and\ \citenamefont {Yau}}]{Putrov17}%
  \BibitemOpen
  \bibfield  {author} {\bibinfo {author} {\bibfnamefont {P.}~\bibnamefont
  {Putrov}}, \bibinfo {author} {\bibfnamefont {J.}~\bibnamefont {Wang}}, \ and\
  \bibinfo {author} {\bibfnamefont {S.-T.}\ \bibnamefont {Yau}},\ }\href
  {\doibase 10.1016/j.aop.2017.06.019} {\bibfield  {journal} {\bibinfo
  {journal} {Ann. Phys.}\ }\textbf {\bibinfo {volume} {384}},\ \bibinfo {pages}
  {254 } (\bibinfo {year} {2017})}\BibitemShut {NoStop}%
\bibitem [{\citenamefont {Chan}\ \emph {et~al.}(2018)\citenamefont {Chan},
  \citenamefont {Ye},\ and\ \citenamefont {Ryu}}]{AtMaChan18}%
  \BibitemOpen
  \bibfield  {author} {\bibinfo {author} {\bibfnamefont {A.~P.~O.}\
  \bibnamefont {Chan}}, \bibinfo {author} {\bibfnamefont {P.}~\bibnamefont
  {Ye}}, \ and\ \bibinfo {author} {\bibfnamefont {S.}~\bibnamefont {Ryu}},\
  }\href {\doibase 10.1103/PhysRevLett.121.061601} {\bibfield  {journal}
  {\bibinfo  {journal} {Phys. Rev. Lett.}\ }\textbf {\bibinfo {volume} {121}},\
  \bibinfo {pages} {061601} (\bibinfo {year} {2018})}\BibitemShut {NoStop}%
\bibitem [{\citenamefont {Wen}(2016)}]{Wen16}%
  \BibitemOpen
  \bibfield  {author} {\bibinfo {author} {\bibfnamefont {X.-G.}\ \bibnamefont
  {Wen}},\ }\href {\doibase 10.1093/nsr/nwv077} {\bibfield  {journal} {\bibinfo
   {journal} {Nat. Sci. Rev.}\ }\textbf {\bibinfo {volume} {3}},\ \bibinfo
  {pages} {68} (\bibinfo {year} {2016})}\BibitemShut {NoStop}%
\bibitem [{\citenamefont {Gu}\ \emph {et~al.}(2015)\citenamefont {Gu},
  \citenamefont {Wang},\ and\ \citenamefont {Wen}}]{ZCGu15}%
  \BibitemOpen
  \bibfield  {author} {\bibinfo {author} {\bibfnamefont {Z.-C.}\ \bibnamefont
  {Gu}}, \bibinfo {author} {\bibfnamefont {Z.}~\bibnamefont {Wang}}, \ and\
  \bibinfo {author} {\bibfnamefont {X.-G.}\ \bibnamefont {Wen}},\ }\href
  {\doibase 10.1103/PhysRevB.91.125149} {\bibfield  {journal} {\bibinfo
  {journal} {Phys. Rev. B}\ }\textbf {\bibinfo {volume} {91}},\ \bibinfo
  {pages} {125149} (\bibinfo {year} {2015})}\BibitemShut {NoStop}%
\bibitem [{\citenamefont {Lan}\ \emph {et~al.}(2016)\citenamefont {Lan},
  \citenamefont {Kong},\ and\ \citenamefont {Wen}}]{TLan16}%
  \BibitemOpen
  \bibfield  {author} {\bibinfo {author} {\bibfnamefont {T.}~\bibnamefont
  {Lan}}, \bibinfo {author} {\bibfnamefont {L.}~\bibnamefont {Kong}}, \ and\
  \bibinfo {author} {\bibfnamefont {X.-G.}\ \bibnamefont {Wen}},\ }\href
  {\doibase 10.1103/PhysRevB.94.155113} {\bibfield  {journal} {\bibinfo
  {journal} {Phys. Rev. B}\ }\textbf {\bibinfo {volume} {94}},\ \bibinfo
  {pages} {155113} (\bibinfo {year} {2016})}\BibitemShut {NoStop}%
\bibitem [{\citenamefont {Lan}\ \emph {et~al.}(2018)\citenamefont {Lan},
  \citenamefont {Kong},\ and\ \citenamefont {Wen}}]{TLan18}%
  \BibitemOpen
  \bibfield  {author} {\bibinfo {author} {\bibfnamefont {T.}~\bibnamefont
  {Lan}}, \bibinfo {author} {\bibfnamefont {L.}~\bibnamefont {Kong}}, \ and\
  \bibinfo {author} {\bibfnamefont {X.-G.}\ \bibnamefont {Wen}},\ }\href
  {\doibase 10.1103/PhysRevX.8.021074} {\bibfield  {journal} {\bibinfo
  {journal} {Phys. Rev. X}\ }\textbf {\bibinfo {volume} {8}},\ \bibinfo {pages}
  {021074} (\bibinfo {year} {2018})}\BibitemShut {NoStop}%
\bibitem [{\citenamefont {Lan}\ and\ \citenamefont {Wen}(2019)}]{TLan19}%
  \BibitemOpen
  \bibfield  {author} {\bibinfo {author} {\bibfnamefont {T.}~\bibnamefont
  {Lan}}\ and\ \bibinfo {author} {\bibfnamefont {X.-G.}\ \bibnamefont {Wen}},\
  }\href {\doibase 10.1103/PhysRevX.9.021005} {\bibfield  {journal} {\bibinfo
  {journal} {Phys. Rev. X}\ }\textbf {\bibinfo {volume} {9}},\ \bibinfo {pages}
  {021005} (\bibinfo {year} {2019})}\BibitemShut {NoStop}%
\bibitem [{\citenamefont {Levin}\ and\ \citenamefont {Wen}(2005)}]{Levin05}%
  \BibitemOpen
  \bibfield  {author} {\bibinfo {author} {\bibfnamefont {M.~A.}\ \bibnamefont
  {Levin}}\ and\ \bibinfo {author} {\bibfnamefont {X.-G.}\ \bibnamefont
  {Wen}},\ }\href {\doibase 10.1103/PhysRevB.71.045110} {\bibfield  {journal}
  {\bibinfo  {journal} {Phys. Rev. B}\ }\textbf {\bibinfo {volume} {71}},\
  \bibinfo {pages} {045110} (\bibinfo {year} {2005})}\BibitemShut {NoStop}%
\bibitem [{\citenamefont {Dijkgraaf}\ and\ \citenamefont
  {Witten}(1990)}]{Dijkgraaf90}%
  \BibitemOpen
  \bibfield  {author} {\bibinfo {author} {\bibfnamefont {R.}~\bibnamefont
  {Dijkgraaf}}\ and\ \bibinfo {author} {\bibfnamefont {E.}~\bibnamefont
  {Witten}},\ }\href {\doibase 10.1007/BF02096988} {\bibfield  {journal}
  {\bibinfo  {journal} {Commun. Math. Phys.}\ }\textbf {\bibinfo {volume}
  {129}},\ \bibinfo {pages} {393} (\bibinfo {year} {1990})}\BibitemShut
  {NoStop}%
\bibitem [{\citenamefont {Wan}\ \emph {et~al.}(2015)\citenamefont {Wan},
  \citenamefont {Wang},\ and\ \citenamefont {He}}]{YidunWan15}%
  \BibitemOpen
  \bibfield  {author} {\bibinfo {author} {\bibfnamefont {Y.}~\bibnamefont
  {Wan}}, \bibinfo {author} {\bibfnamefont {J.~C.}\ \bibnamefont {Wang}}, \
  and\ \bibinfo {author} {\bibfnamefont {H.}~\bibnamefont {He}},\ }\href
  {\doibase 10.1103/PhysRevB.92.045101} {\bibfield  {journal} {\bibinfo
  {journal} {Phys. Rev. B}\ }\textbf {\bibinfo {volume} {92}},\ \bibinfo
  {pages} {045101} (\bibinfo {year} {2015})}\BibitemShut {NoStop}%
\bibitem [{\citenamefont {Walker}\ and\ \citenamefont {Wang}(2012)}]{Walker12}%
  \BibitemOpen
  \bibfield  {author} {\bibinfo {author} {\bibfnamefont {K.}~\bibnamefont
  {Walker}}\ and\ \bibinfo {author} {\bibfnamefont {Z.}~\bibnamefont {Wang}},\
  }\href {\doibase 10.1007/s11467-011-0194-z} {\bibfield  {journal} {\bibinfo
  {journal} {Front. Phys.}\ }\textbf {\bibinfo {volume} {7}},\ \bibinfo {pages}
  {150} (\bibinfo {year} {2012})}\BibitemShut {NoStop}%
\bibitem [{\citenamefont {von Keyserlingk}\ \emph {et~al.}(2013)\citenamefont
  {von Keyserlingk}, \citenamefont {Burnell},\ and\ \citenamefont
  {Simon}}]{vonKeyserlingk13}%
  \BibitemOpen
  \bibfield  {author} {\bibinfo {author} {\bibfnamefont {C.~W.}\ \bibnamefont
  {von Keyserlingk}}, \bibinfo {author} {\bibfnamefont {F.~J.}\ \bibnamefont
  {Burnell}}, \ and\ \bibinfo {author} {\bibfnamefont {S.~H.}\ \bibnamefont
  {Simon}},\ }\href {\doibase 10.1103/PhysRevB.87.045107} {\bibfield  {journal}
  {\bibinfo  {journal} {Phys. Rev. B}\ }\textbf {\bibinfo {volume} {87}},\
  \bibinfo {pages} {045107} (\bibinfo {year} {2013})}\BibitemShut {NoStop}%
\bibitem [{\citenamefont {Wang}\ and\ \citenamefont
  {Chen}(2017)}]{ZitaoWang17}%
  \BibitemOpen
  \bibfield  {author} {\bibinfo {author} {\bibfnamefont {Z.}~\bibnamefont
  {Wang}}\ and\ \bibinfo {author} {\bibfnamefont {X.}~\bibnamefont {Chen}},\
  }\href {\doibase 10.1103/PhysRevB.95.115142} {\bibfield  {journal} {\bibinfo
  {journal} {Phys. Rev. B}\ }\textbf {\bibinfo {volume} {95}},\ \bibinfo
  {pages} {115142} (\bibinfo {year} {2017})}\BibitemShut {NoStop}%
\bibitem [{\citenamefont {Williamson}\ and\ \citenamefont
  {Wang}(2017)}]{Williamson17}%
  \BibitemOpen
  \bibfield  {author} {\bibinfo {author} {\bibfnamefont {D.~J.}\ \bibnamefont
  {Williamson}}\ and\ \bibinfo {author} {\bibfnamefont {Z.}~\bibnamefont
  {Wang}},\ }\href {\doibase 10.1016/j.aop.2016.12.018} {\bibfield  {journal}
  {\bibinfo  {journal} {Ann. Phys.}\ }\textbf {\bibinfo {volume} {377}},\
  \bibinfo {pages} {311 } (\bibinfo {year} {2017})}\BibitemShut {NoStop}%
\bibitem [{\citenamefont {Kane}\ \emph {et~al.}(2002)\citenamefont {Kane},
  \citenamefont {Mukhopadhyay},\ and\ \citenamefont {Lubensky}}]{Kane02}%
  \BibitemOpen
  \bibfield  {author} {\bibinfo {author} {\bibfnamefont {C.~L.}\ \bibnamefont
  {Kane}}, \bibinfo {author} {\bibfnamefont {R.}~\bibnamefont {Mukhopadhyay}},
  \ and\ \bibinfo {author} {\bibfnamefont {T.~C.}\ \bibnamefont {Lubensky}},\
  }\href {\doibase 10.1103/PhysRevLett.88.036401} {\bibfield  {journal}
  {\bibinfo  {journal} {Phys. Rev. Lett.}\ }\textbf {\bibinfo {volume} {88}},\
  \bibinfo {pages} {036401} (\bibinfo {year} {2002})}\BibitemShut {NoStop}%
\bibitem [{\citenamefont {Teo}\ and\ \citenamefont {Kane}(2014)}]{Teo14}%
  \BibitemOpen
  \bibfield  {author} {\bibinfo {author} {\bibfnamefont {J.~C.~Y.}\
  \bibnamefont {Teo}}\ and\ \bibinfo {author} {\bibfnamefont {C.~L.}\
  \bibnamefont {Kane}},\ }\href {\doibase 10.1103/PhysRevB.89.085101}
  {\bibfield  {journal} {\bibinfo  {journal} {Phys. Rev. B}\ }\textbf {\bibinfo
  {volume} {89}},\ \bibinfo {pages} {085101} (\bibinfo {year}
  {2014})}\BibitemShut {NoStop}%
\bibitem [{\citenamefont {Francesco}\ \emph {et~al.}(1997)\citenamefont
  {Francesco}, \citenamefont {Mathieu},\ and\ \citenamefont
  {S\'en\'echal}}]{dFMS}%
  \BibitemOpen
  \bibfield  {author} {\bibinfo {author} {\bibfnamefont {P.~D.}\ \bibnamefont
  {Francesco}}, \bibinfo {author} {\bibfnamefont {P.}~\bibnamefont {Mathieu}},
  \ and\ \bibinfo {author} {\bibfnamefont {D.}~\bibnamefont {S\'en\'echal}},\
  }\href@noop {} {\emph {\bibinfo {title} {Conformal Field Theory}}}\ (\bibinfo
   {publisher} {Splinger-Verlag},\ \bibinfo {address} {New York},\ \bibinfo
  {year} {1997})\BibitemShut {NoStop}%
\bibitem [{\citenamefont {Lu}\ and\ \citenamefont {Vishwanath}(2012)}]{YMLu12}%
  \BibitemOpen
  \bibfield  {author} {\bibinfo {author} {\bibfnamefont {Y.-M.}\ \bibnamefont
  {Lu}}\ and\ \bibinfo {author} {\bibfnamefont {A.}~\bibnamefont
  {Vishwanath}},\ }\href {\doibase 10.1103/PhysRevB.86.125119} {\bibfield
  {journal} {\bibinfo  {journal} {Phys. Rev. B}\ }\textbf {\bibinfo {volume}
  {86}},\ \bibinfo {pages} {125119} (\bibinfo {year} {2012})}\BibitemShut
  {NoStop}%
\bibitem [{\citenamefont {Mong}\ \emph {et~al.}(2014)\citenamefont {Mong},
  \citenamefont {Clarke}, \citenamefont {Alicea}, \citenamefont {Lindner},
  \citenamefont {Fendley}, \citenamefont {Nayak}, \citenamefont {Oreg},
  \citenamefont {Stern}, \citenamefont {Berg}, \citenamefont {Shtengel},\ and\
  \citenamefont {Fisher}}]{Mong14}%
  \BibitemOpen
  \bibfield  {author} {\bibinfo {author} {\bibfnamefont {R.~S.~K.}\
  \bibnamefont {Mong}}, \bibinfo {author} {\bibfnamefont {D.~J.}\ \bibnamefont
  {Clarke}}, \bibinfo {author} {\bibfnamefont {J.}~\bibnamefont {Alicea}},
  \bibinfo {author} {\bibfnamefont {N.~H.}\ \bibnamefont {Lindner}}, \bibinfo
  {author} {\bibfnamefont {P.}~\bibnamefont {Fendley}}, \bibinfo {author}
  {\bibfnamefont {C.}~\bibnamefont {Nayak}}, \bibinfo {author} {\bibfnamefont
  {Y.}~\bibnamefont {Oreg}}, \bibinfo {author} {\bibfnamefont {A.}~\bibnamefont
  {Stern}}, \bibinfo {author} {\bibfnamefont {E.}~\bibnamefont {Berg}},
  \bibinfo {author} {\bibfnamefont {K.}~\bibnamefont {Shtengel}}, \ and\
  \bibinfo {author} {\bibfnamefont {M.~P.~A.}\ \bibnamefont {Fisher}},\ }\href
  {\doibase 10.1103/PhysRevX.4.011036} {\bibfield  {journal} {\bibinfo
  {journal} {Phys. Rev. X}\ }\textbf {\bibinfo {volume} {4}},\ \bibinfo {pages}
  {011036} (\bibinfo {year} {2014})}\BibitemShut {NoStop}%
\bibitem [{\citenamefont {Oreg}\ \emph {et~al.}(2014)\citenamefont {Oreg},
  \citenamefont {Sela},\ and\ \citenamefont {Stern}}]{Oreg14}%
  \BibitemOpen
  \bibfield  {author} {\bibinfo {author} {\bibfnamefont {Y.}~\bibnamefont
  {Oreg}}, \bibinfo {author} {\bibfnamefont {E.}~\bibnamefont {Sela}}, \ and\
  \bibinfo {author} {\bibfnamefont {A.}~\bibnamefont {Stern}},\ }\href
  {\doibase 10.1103/PhysRevB.89.115402} {\bibfield  {journal} {\bibinfo
  {journal} {Phys. Rev. B}\ }\textbf {\bibinfo {volume} {89}},\ \bibinfo
  {pages} {115402} (\bibinfo {year} {2014})}\BibitemShut {NoStop}%
\bibitem [{\citenamefont {Neupert}\ \emph {et~al.}(2014)\citenamefont
  {Neupert}, \citenamefont {Chamon}, \citenamefont {Mudry},\ and\ \citenamefont
  {Thomale}}]{Neupert14}%
  \BibitemOpen
  \bibfield  {author} {\bibinfo {author} {\bibfnamefont {T.}~\bibnamefont
  {Neupert}}, \bibinfo {author} {\bibfnamefont {C.}~\bibnamefont {Chamon}},
  \bibinfo {author} {\bibfnamefont {C.}~\bibnamefont {Mudry}}, \ and\ \bibinfo
  {author} {\bibfnamefont {R.}~\bibnamefont {Thomale}},\ }\href {\doibase
  10.1103/PhysRevB.90.205101} {\bibfield  {journal} {\bibinfo  {journal} {Phys.
  Rev. B}\ }\textbf {\bibinfo {volume} {90}},\ \bibinfo {pages} {205101}
  (\bibinfo {year} {2014})}\BibitemShut {NoStop}%
\bibitem [{\citenamefont {Sagi}\ and\ \citenamefont {Oreg}(2014)}]{Sagi14}%
  \BibitemOpen
  \bibfield  {author} {\bibinfo {author} {\bibfnamefont {E.}~\bibnamefont
  {Sagi}}\ and\ \bibinfo {author} {\bibfnamefont {Y.}~\bibnamefont {Oreg}},\
  }\href {\doibase 10.1103/PhysRevB.90.201102} {\bibfield  {journal} {\bibinfo
  {journal} {Phys. Rev. B}\ }\textbf {\bibinfo {volume} {90}},\ \bibinfo
  {pages} {201102} (\bibinfo {year} {2014})}\BibitemShut {NoStop}%
\bibitem [{\citenamefont {Klinovaja}\ and\ \citenamefont
  {Tserkovnyak}(2014)}]{Klinovaja14}%
  \BibitemOpen
  \bibfield  {author} {\bibinfo {author} {\bibfnamefont {J.}~\bibnamefont
  {Klinovaja}}\ and\ \bibinfo {author} {\bibfnamefont {Y.}~\bibnamefont
  {Tserkovnyak}},\ }\href {\doibase 10.1103/PhysRevB.90.115426} {\bibfield
  {journal} {\bibinfo  {journal} {Phys. Rev. B}\ }\textbf {\bibinfo {volume}
  {90}},\ \bibinfo {pages} {115426} (\bibinfo {year} {2014})}\BibitemShut
  {NoStop}%
\bibitem [{\citenamefont {Vaezi}(2014)}]{Vaezi14}%
  \BibitemOpen
  \bibfield  {author} {\bibinfo {author} {\bibfnamefont {A.}~\bibnamefont
  {Vaezi}},\ }\href {\doibase 10.1103/PhysRevX.4.031009} {\bibfield  {journal}
  {\bibinfo  {journal} {Phys. Rev. X}\ }\textbf {\bibinfo {volume} {4}},\
  \bibinfo {pages} {031009} (\bibinfo {year} {2014})}\BibitemShut {NoStop}%
\bibitem [{\citenamefont {Huang}\ \emph {et~al.}(2016)\citenamefont {Huang},
  \citenamefont {Chen}, \citenamefont {Gomes}, \citenamefont {Neupert},
  \citenamefont {Chamon},\ and\ \citenamefont {Mudry}}]{PHHuang16}%
  \BibitemOpen
  \bibfield  {author} {\bibinfo {author} {\bibfnamefont {P.-H.}\ \bibnamefont
  {Huang}}, \bibinfo {author} {\bibfnamefont {J.-H.}\ \bibnamefont {Chen}},
  \bibinfo {author} {\bibfnamefont {P.~R.~S.}\ \bibnamefont {Gomes}}, \bibinfo
  {author} {\bibfnamefont {T.}~\bibnamefont {Neupert}}, \bibinfo {author}
  {\bibfnamefont {C.}~\bibnamefont {Chamon}}, \ and\ \bibinfo {author}
  {\bibfnamefont {C.}~\bibnamefont {Mudry}},\ }\href {\doibase
  10.1103/PhysRevB.93.205123} {\bibfield  {journal} {\bibinfo  {journal} {Phys.
  Rev. B}\ }\textbf {\bibinfo {volume} {93}},\ \bibinfo {pages} {205123}
  (\bibinfo {year} {2016})}\BibitemShut {NoStop}%
\bibitem [{\citenamefont {Fuji}\ and\ \citenamefont
  {Lecheminant}(2017)}]{Fuji17}%
  \BibitemOpen
  \bibfield  {author} {\bibinfo {author} {\bibfnamefont {Y.}~\bibnamefont
  {Fuji}}\ and\ \bibinfo {author} {\bibfnamefont {P.}~\bibnamefont
  {Lecheminant}},\ }\href {\doibase 10.1103/PhysRevB.95.125130} {\bibfield
  {journal} {\bibinfo  {journal} {Phys. Rev. B}\ }\textbf {\bibinfo {volume}
  {95}},\ \bibinfo {pages} {125130} (\bibinfo {year} {2017})}\BibitemShut
  {NoStop}%
\bibitem [{\citenamefont {Kane}\ \emph {et~al.}(2017)\citenamefont {Kane},
  \citenamefont {Stern},\ and\ \citenamefont {Halperin}}]{Kane17}%
  \BibitemOpen
  \bibfield  {author} {\bibinfo {author} {\bibfnamefont {C.~L.}\ \bibnamefont
  {Kane}}, \bibinfo {author} {\bibfnamefont {A.}~\bibnamefont {Stern}}, \ and\
  \bibinfo {author} {\bibfnamefont {B.~I.}\ \bibnamefont {Halperin}},\ }\href
  {\doibase 10.1103/PhysRevX.7.031009} {\bibfield  {journal} {\bibinfo
  {journal} {Phys. Rev. X}\ }\textbf {\bibinfo {volume} {7}},\ \bibinfo {pages}
  {031009} (\bibinfo {year} {2017})}\BibitemShut {NoStop}%
\bibitem [{\citenamefont {Kane}\ and\ \citenamefont {Stern}(2018)}]{Kane18}%
  \BibitemOpen
  \bibfield  {author} {\bibinfo {author} {\bibfnamefont {C.~L.}\ \bibnamefont
  {Kane}}\ and\ \bibinfo {author} {\bibfnamefont {A.}~\bibnamefont {Stern}},\
  }\href {\doibase 10.1103/PhysRevB.98.085302} {\bibfield  {journal} {\bibinfo
  {journal} {Phys. Rev. B}\ }\textbf {\bibinfo {volume} {98}},\ \bibinfo
  {pages} {085302} (\bibinfo {year} {2018})}\BibitemShut {NoStop}%
\bibitem [{\citenamefont {Mross}\ \emph {et~al.}(2015)\citenamefont {Mross},
  \citenamefont {Essin},\ and\ \citenamefont {Alicea}}]{Mross15}%
  \BibitemOpen
  \bibfield  {author} {\bibinfo {author} {\bibfnamefont {D.~F.}\ \bibnamefont
  {Mross}}, \bibinfo {author} {\bibfnamefont {A.}~\bibnamefont {Essin}}, \ and\
  \bibinfo {author} {\bibfnamefont {J.}~\bibnamefont {Alicea}},\ }\href
  {\doibase 10.1103/PhysRevX.5.011011} {\bibfield  {journal} {\bibinfo
  {journal} {Phys. Rev. X}\ }\textbf {\bibinfo {volume} {5}},\ \bibinfo {pages}
  {011011} (\bibinfo {year} {2015})}\BibitemShut {NoStop}%
\bibitem [{\citenamefont {Sahoo}\ \emph {et~al.}(2016)\citenamefont {Sahoo},
  \citenamefont {Zhang},\ and\ \citenamefont {Teo}}]{Sahoo16}%
  \BibitemOpen
  \bibfield  {author} {\bibinfo {author} {\bibfnamefont {S.}~\bibnamefont
  {Sahoo}}, \bibinfo {author} {\bibfnamefont {Z.}~\bibnamefont {Zhang}}, \ and\
  \bibinfo {author} {\bibfnamefont {J.~C.~Y.}\ \bibnamefont {Teo}},\ }\href
  {\doibase 10.1103/PhysRevB.94.165142} {\bibfield  {journal} {\bibinfo
  {journal} {Phys. Rev. B}\ }\textbf {\bibinfo {volume} {94}},\ \bibinfo
  {pages} {165142} (\bibinfo {year} {2016})}\BibitemShut {NoStop}%
\bibitem [{\citenamefont {Lu}\ \emph {et~al.}(2017)\citenamefont {Lu},
  \citenamefont {Shi},\ and\ \citenamefont {Lu}}]{FLu17}%
  \BibitemOpen
  \bibfield  {author} {\bibinfo {author} {\bibfnamefont {F.}~\bibnamefont
  {Lu}}, \bibinfo {author} {\bibfnamefont {B.}~\bibnamefont {Shi}}, \ and\
  \bibinfo {author} {\bibfnamefont {Y.-M.}\ \bibnamefont {Lu}},\ }\href
  {\doibase 10.1088/1367-2630/aa769b} {\bibfield  {journal} {\bibinfo
  {journal} {New J. Phys.}\ }\textbf {\bibinfo {volume} {19}},\ \bibinfo
  {pages} {073002} (\bibinfo {year} {2017})}\BibitemShut {NoStop}%
\bibitem [{\citenamefont {Hong}\ and\ \citenamefont {Fu}()}]{SHong17}%
  \BibitemOpen
  \bibfield  {author} {\bibinfo {author} {\bibfnamefont {S.}~\bibnamefont
  {Hong}}\ and\ \bibinfo {author} {\bibfnamefont {L.}~\bibnamefont {Fu}},\
  }\href {http://arxiv.org/abs/1707.02594} {\enquote {\bibinfo {title}
  {Topological order and symmetry anomaly on the surface of topological
  crystalline insulators},}\ }\Eprint {http://arxiv.org/abs/arXiv:1707.02594}
  {arXiv:1707.02594} \BibitemShut {NoStop}%
\bibitem [{\citenamefont {Volpez}\ \emph {et~al.}(2017)\citenamefont {Volpez},
  \citenamefont {Loss},\ and\ \citenamefont {Klinovaja}}]{Volpez17}%
  \BibitemOpen
  \bibfield  {author} {\bibinfo {author} {\bibfnamefont {Y.}~\bibnamefont
  {Volpez}}, \bibinfo {author} {\bibfnamefont {D.}~\bibnamefont {Loss}}, \ and\
  \bibinfo {author} {\bibfnamefont {J.}~\bibnamefont {Klinovaja}},\ }\href
  {\doibase 10.1103/PhysRevB.96.085422} {\bibfield  {journal} {\bibinfo
  {journal} {Phys. Rev. B}\ }\textbf {\bibinfo {volume} {96}},\ \bibinfo
  {pages} {085422} (\bibinfo {year} {2017})}\BibitemShut {NoStop}%
\bibitem [{\citenamefont {Cheng}(2018)}]{MCheng18}%
  \BibitemOpen
  \bibfield  {author} {\bibinfo {author} {\bibfnamefont {M.}~\bibnamefont
  {Cheng}},\ }\href {\doibase 10.1103/PhysRevLett.120.036801} {\bibfield
  {journal} {\bibinfo  {journal} {Phys. Rev. Lett.}\ }\textbf {\bibinfo
  {volume} {120}},\ \bibinfo {pages} {036801} (\bibinfo {year}
  {2018})}\BibitemShut {NoStop}%
\bibitem [{\citenamefont {Han}\ and\ \citenamefont {Teo}(2019)}]{BHan18}%
  \BibitemOpen
  \bibfield  {author} {\bibinfo {author} {\bibfnamefont {B.}~\bibnamefont
  {Han}}\ and\ \bibinfo {author} {\bibfnamefont {J.~C.~Y.}\ \bibnamefont
  {Teo}},\ }\href {\doibase 10.1103/PhysRevB.99.235102} {\bibfield  {journal}
  {\bibinfo  {journal} {Phys. Rev. B}\ }\textbf {\bibinfo {volume} {99}},\
  \bibinfo {pages} {235102} (\bibinfo {year} {2019})}\BibitemShut {NoStop}%
\bibitem [{\citenamefont {Sagi}\ and\ \citenamefont {Oreg}(2015)}]{Sagi15}%
  \BibitemOpen
  \bibfield  {author} {\bibinfo {author} {\bibfnamefont {E.}~\bibnamefont
  {Sagi}}\ and\ \bibinfo {author} {\bibfnamefont {Y.}~\bibnamefont {Oreg}},\
  }\href {\doibase 10.1103/PhysRevB.92.195137} {\bibfield  {journal} {\bibinfo
  {journal} {Phys. Rev. B}\ }\textbf {\bibinfo {volume} {92}},\ \bibinfo
  {pages} {195137} (\bibinfo {year} {2015})}\BibitemShut {NoStop}%
\bibitem [{\citenamefont {Meng}(2015)}]{Meng15}%
  \BibitemOpen
  \bibfield  {author} {\bibinfo {author} {\bibfnamefont {T.}~\bibnamefont
  {Meng}},\ }\href {\doibase 10.1103/PhysRevB.92.115152} {\bibfield  {journal}
  {\bibinfo  {journal} {Phys. Rev. B}\ }\textbf {\bibinfo {volume} {92}},\
  \bibinfo {pages} {115152} (\bibinfo {year} {2015})}\BibitemShut {NoStop}%
\bibitem [{\citenamefont {Iadecola}\ \emph {et~al.}(2016)\citenamefont
  {Iadecola}, \citenamefont {Neupert}, \citenamefont {Chamon},\ and\
  \citenamefont {Mudry}}]{Iadecola16}%
  \BibitemOpen
  \bibfield  {author} {\bibinfo {author} {\bibfnamefont {T.}~\bibnamefont
  {Iadecola}}, \bibinfo {author} {\bibfnamefont {T.}~\bibnamefont {Neupert}},
  \bibinfo {author} {\bibfnamefont {C.}~\bibnamefont {Chamon}}, \ and\ \bibinfo
  {author} {\bibfnamefont {C.}~\bibnamefont {Mudry}},\ }\href {\doibase
  10.1103/PhysRevB.93.195136} {\bibfield  {journal} {\bibinfo  {journal} {Phys.
  Rev. B}\ }\textbf {\bibinfo {volume} {93}},\ \bibinfo {pages} {195136}
  (\bibinfo {year} {2016})}\BibitemShut {NoStop}%
\bibitem [{\citenamefont {Iadecola}\ \emph {et~al.}()\citenamefont {Iadecola},
  \citenamefont {Neupert}, \citenamefont {Chamon},\ and\ \citenamefont
  {Mudry}}]{Iadecola17}%
  \BibitemOpen
  \bibfield  {author} {\bibinfo {author} {\bibfnamefont {T.}~\bibnamefont
  {Iadecola}}, \bibinfo {author} {\bibfnamefont {T.}~\bibnamefont {Neupert}},
  \bibinfo {author} {\bibfnamefont {C.}~\bibnamefont {Chamon}}, \ and\ \bibinfo
  {author} {\bibfnamefont {C.}~\bibnamefont {Mudry}},\ }\href
  {http://arxiv.org/abs/1703.03418} {\enquote {\bibinfo {title} {{Non-Abelian}
  topological phases in three spatial dimensions from coupled wires},}\
  }\Eprint {http://arxiv.org/abs/arXiv:1703.03418} {arXiv:1703.03418}
  \BibitemShut {NoStop}%
\bibitem [{\citenamefont {Park}\ \emph {et~al.}(2018)\citenamefont {Park},
  \citenamefont {Raza}, \citenamefont {Gilbert},\ and\ \citenamefont
  {Teo}}]{MJPark18}%
  \BibitemOpen
  \bibfield  {author} {\bibinfo {author} {\bibfnamefont {M.~J.}\ \bibnamefont
  {Park}}, \bibinfo {author} {\bibfnamefont {S.}~\bibnamefont {Raza}}, \bibinfo
  {author} {\bibfnamefont {M.~J.}\ \bibnamefont {Gilbert}}, \ and\ \bibinfo
  {author} {\bibfnamefont {J.~C.~Y.}\ \bibnamefont {Teo}},\ }\href {\doibase
  10.1103/PhysRevB.98.184514} {\bibfield  {journal} {\bibinfo  {journal} {Phys.
  Rev. B}\ }\textbf {\bibinfo {volume} {98}},\ \bibinfo {pages} {184514}
  (\bibinfo {year} {2018})}\BibitemShut {NoStop}%
\bibitem [{\citenamefont {Raza}\ \emph {et~al.}(2019)\citenamefont {Raza},
  \citenamefont {Sirota},\ and\ \citenamefont {Teo}}]{Raza19}%
  \BibitemOpen
  \bibfield  {author} {\bibinfo {author} {\bibfnamefont {S.}~\bibnamefont
  {Raza}}, \bibinfo {author} {\bibfnamefont {A.}~\bibnamefont {Sirota}}, \ and\
  \bibinfo {author} {\bibfnamefont {J.~C.~Y.}\ \bibnamefont {Teo}},\ }\href
  {\doibase 10.1103/PhysRevX.9.011039} {\bibfield  {journal} {\bibinfo
  {journal} {Phys. Rev. X}\ }\textbf {\bibinfo {volume} {9}},\ \bibinfo {pages}
  {011039} (\bibinfo {year} {2019})}\BibitemShut {NoStop}%
\bibitem [{\citenamefont {Wang}\ and\ \citenamefont
  {Senthil}(2013)}]{ChongWang13}%
  \BibitemOpen
  \bibfield  {author} {\bibinfo {author} {\bibfnamefont {C.}~\bibnamefont
  {Wang}}\ and\ \bibinfo {author} {\bibfnamefont {T.}~\bibnamefont {Senthil}},\
  }\href {\doibase 10.1103/PhysRevB.87.235122} {\bibfield  {journal} {\bibinfo
  {journal} {Phys. Rev. B}\ }\textbf {\bibinfo {volume} {87}},\ \bibinfo
  {pages} {235122} (\bibinfo {year} {2013})}\BibitemShut {NoStop}%
\bibitem [{\citenamefont {Fidkowski}\ \emph {et~al.}(2013)\citenamefont
  {Fidkowski}, \citenamefont {Chen},\ and\ \citenamefont
  {Vishwanath}}]{Fidkowski13}%
  \BibitemOpen
  \bibfield  {author} {\bibinfo {author} {\bibfnamefont {L.}~\bibnamefont
  {Fidkowski}}, \bibinfo {author} {\bibfnamefont {X.}~\bibnamefont {Chen}}, \
  and\ \bibinfo {author} {\bibfnamefont {A.}~\bibnamefont {Vishwanath}},\
  }\href {\doibase 10.1103/PhysRevX.3.041016} {\bibfield  {journal} {\bibinfo
  {journal} {Phys. Rev. X}\ }\textbf {\bibinfo {volume} {3}},\ \bibinfo {pages}
  {041016} (\bibinfo {year} {2013})}\BibitemShut {NoStop}%
\bibitem [{\citenamefont {Ben-Zion}\ \emph {et~al.}(2016)\citenamefont
  {Ben-Zion}, \citenamefont {Das},\ and\ \citenamefont {McGreevy}}]{BenZion16}%
  \BibitemOpen
  \bibfield  {author} {\bibinfo {author} {\bibfnamefont {D.}~\bibnamefont
  {Ben-Zion}}, \bibinfo {author} {\bibfnamefont {D.}~\bibnamefont {Das}}, \
  and\ \bibinfo {author} {\bibfnamefont {J.}~\bibnamefont {McGreevy}},\ }\href
  {\doibase 10.1103/PhysRevB.93.155147} {\bibfield  {journal} {\bibinfo
  {journal} {Phys. Rev. B}\ }\textbf {\bibinfo {volume} {93}},\ \bibinfo
  {pages} {155147} (\bibinfo {year} {2016})}\BibitemShut {NoStop}%
\bibitem [{\citenamefont {Bais}\ and\ \citenamefont
  {Slingerland}(2009)}]{Bais09}%
  \BibitemOpen
  \bibfield  {author} {\bibinfo {author} {\bibfnamefont {F.~A.}\ \bibnamefont
  {Bais}}\ and\ \bibinfo {author} {\bibfnamefont {J.~K.}\ \bibnamefont
  {Slingerland}},\ }\href {\doibase 10.1103/PhysRevB.79.045316} {\bibfield
  {journal} {\bibinfo  {journal} {Phys. Rev. B}\ }\textbf {\bibinfo {volume}
  {79}},\ \bibinfo {pages} {045316} (\bibinfo {year} {2009})}\BibitemShut
  {NoStop}%
\bibitem [{\citenamefont {Schoutens}\ and\ \citenamefont
  {Wen}(2016)}]{Schoutens16}%
  \BibitemOpen
  \bibfield  {author} {\bibinfo {author} {\bibfnamefont {K.}~\bibnamefont
  {Schoutens}}\ and\ \bibinfo {author} {\bibfnamefont {X.-G.}\ \bibnamefont
  {Wen}},\ }\href {\doibase 10.1103/PhysRevB.93.045109} {\bibfield  {journal}
  {\bibinfo  {journal} {Phys. Rev. B}\ }\textbf {\bibinfo {volume} {93}},\
  \bibinfo {pages} {045109} (\bibinfo {year} {2016})}\BibitemShut {NoStop}%
\bibitem [{\citenamefont {James}\ \emph {et~al.}(2018)\citenamefont {James},
  \citenamefont {Konik}, \citenamefont {Lecheminant}, \citenamefont
  {Robinson},\ and\ \citenamefont {Tsvelik}}]{James18}%
  \BibitemOpen
  \bibfield  {author} {\bibinfo {author} {\bibfnamefont {A.~J.~A.}\
  \bibnamefont {James}}, \bibinfo {author} {\bibfnamefont {R.~M.}\ \bibnamefont
  {Konik}}, \bibinfo {author} {\bibfnamefont {P.}~\bibnamefont {Lecheminant}},
  \bibinfo {author} {\bibfnamefont {N.~J.}\ \bibnamefont {Robinson}}, \ and\
  \bibinfo {author} {\bibfnamefont {A.~M.}\ \bibnamefont {Tsvelik}},\ }\href
  {\doibase 10.1088/1361-6633/aa91ea} {\bibfield  {journal} {\bibinfo
  {journal} {Rep. Prog. Phys.}\ }\textbf {\bibinfo {volume} {81}},\ \bibinfo
  {pages} {046002} (\bibinfo {year} {2018})}\BibitemShut {NoStop}%
\bibitem [{\citenamefont {Gogolin}\ \emph {et~al.}(1998)\citenamefont
  {Gogolin}, \citenamefont {Nersesyan},\ and\ \citenamefont {Tsvelik}}]{GNT}%
  \BibitemOpen
  \bibfield  {author} {\bibinfo {author} {\bibfnamefont {A.~O.}\ \bibnamefont
  {Gogolin}}, \bibinfo {author} {\bibfnamefont {A.~A.}\ \bibnamefont
  {Nersesyan}}, \ and\ \bibinfo {author} {\bibfnamefont {A.~M.}\ \bibnamefont
  {Tsvelik}},\ }\href@noop {} {\emph {\bibinfo {title} {Bosonization and
  Strongly Correlated Systems}}}\ (\bibinfo  {publisher} {Cambridge University
  Press},\ \bibinfo {address} {Cambridge},\ \bibinfo {year} {1998})\BibitemShut
  {NoStop}%
\bibitem [{\citenamefont {Fradkin}(2013)}]{Fradkin}%
  \BibitemOpen
  \bibfield  {author} {\bibinfo {author} {\bibfnamefont {E.}~\bibnamefont
  {Fradkin}},\ }\href@noop {} {\emph {\bibinfo {title} {Field Theories of
  Condensed Matter Physics}}},\ \bibinfo {edition} {2nd}\ ed.\ (\bibinfo
  {publisher} {Cambridge University Press},\ \bibinfo {year}
  {2013})\BibitemShut {NoStop}%
\bibitem [{\citenamefont {Schulz}(1986)}]{Schulz86}%
  \BibitemOpen
  \bibfield  {author} {\bibinfo {author} {\bibfnamefont {H.~J.}\ \bibnamefont
  {Schulz}},\ }\href {\doibase 10.1103/PhysRevB.34.6372} {\bibfield  {journal}
  {\bibinfo  {journal} {Phys. Rev. B}\ }\textbf {\bibinfo {volume} {34}},\
  \bibinfo {pages} {6372} (\bibinfo {year} {1986})}\BibitemShut {NoStop}%
\bibitem [{\citenamefont {Strong}\ and\ \citenamefont
  {Millis}(1992)}]{Strong92}%
  \BibitemOpen
  \bibfield  {author} {\bibinfo {author} {\bibfnamefont {S.~P.}\ \bibnamefont
  {Strong}}\ and\ \bibinfo {author} {\bibfnamefont {A.~J.}\ \bibnamefont
  {Millis}},\ }\href {\doibase 10.1103/PhysRevLett.69.2419} {\bibfield
  {journal} {\bibinfo  {journal} {Phys. Rev. Lett.}\ }\textbf {\bibinfo
  {volume} {69}},\ \bibinfo {pages} {2419} (\bibinfo {year}
  {1992})}\BibitemShut {NoStop}%
\bibitem [{\citenamefont {Shelton}\ \emph {et~al.}(1996)\citenamefont
  {Shelton}, \citenamefont {Nersesyan},\ and\ \citenamefont
  {Tsvelik}}]{Shelton96}%
  \BibitemOpen
  \bibfield  {author} {\bibinfo {author} {\bibfnamefont {D.~G.}\ \bibnamefont
  {Shelton}}, \bibinfo {author} {\bibfnamefont {A.~A.}\ \bibnamefont
  {Nersesyan}}, \ and\ \bibinfo {author} {\bibfnamefont {A.~M.}\ \bibnamefont
  {Tsvelik}},\ }\href {\doibase 10.1103/PhysRevB.53.8521} {\bibfield  {journal}
  {\bibinfo  {journal} {Phys. Rev. B}\ }\textbf {\bibinfo {volume} {53}},\
  \bibinfo {pages} {8521} (\bibinfo {year} {1996})}\BibitemShut {NoStop}%
\bibitem [{\citenamefont {Gorohovsky}\ \emph {et~al.}(2015)\citenamefont
  {Gorohovsky}, \citenamefont {Pereira},\ and\ \citenamefont
  {Sela}}]{Gorohovsky15}%
  \BibitemOpen
  \bibfield  {author} {\bibinfo {author} {\bibfnamefont {G.}~\bibnamefont
  {Gorohovsky}}, \bibinfo {author} {\bibfnamefont {R.~G.}\ \bibnamefont
  {Pereira}}, \ and\ \bibinfo {author} {\bibfnamefont {E.}~\bibnamefont
  {Sela}},\ }\href {\doibase 10.1103/PhysRevB.91.245139} {\bibfield  {journal}
  {\bibinfo  {journal} {Phys. Rev. B}\ }\textbf {\bibinfo {volume} {91}},\
  \bibinfo {pages} {245139} (\bibinfo {year} {2015})}\BibitemShut {NoStop}%
\bibitem [{\citenamefont {Lecheminant}\ and\ \citenamefont
  {Tsvelik}(2017)}]{Lecheminant17}%
  \BibitemOpen
  \bibfield  {author} {\bibinfo {author} {\bibfnamefont {P.}~\bibnamefont
  {Lecheminant}}\ and\ \bibinfo {author} {\bibfnamefont {A.~M.}\ \bibnamefont
  {Tsvelik}},\ }\href {\doibase 10.1103/PhysRevB.95.140406} {\bibfield
  {journal} {\bibinfo  {journal} {Phys. Rev. B}\ }\textbf {\bibinfo {volume}
  {95}},\ \bibinfo {pages} {140406} (\bibinfo {year} {2017})}\BibitemShut
  {NoStop}%
\bibitem [{\citenamefont {Chen}\ \emph {et~al.}(2017)\citenamefont {Chen},
  \citenamefont {Mudry}, \citenamefont {Chamon},\ and\ \citenamefont
  {Tsvelik}}]{JHChen17}%
  \BibitemOpen
  \bibfield  {author} {\bibinfo {author} {\bibfnamefont {J.-H.}\ \bibnamefont
  {Chen}}, \bibinfo {author} {\bibfnamefont {C.}~\bibnamefont {Mudry}},
  \bibinfo {author} {\bibfnamefont {C.}~\bibnamefont {Chamon}}, \ and\ \bibinfo
  {author} {\bibfnamefont {A.~M.}\ \bibnamefont {Tsvelik}},\ }\href {\doibase
  10.1103/PhysRevB.96.224420} {\bibfield  {journal} {\bibinfo  {journal} {Phys.
  Rev. B}\ }\textbf {\bibinfo {volume} {96}},\ \bibinfo {pages} {224420}
  (\bibinfo {year} {2017})}\BibitemShut {NoStop}%
\bibitem [{\citenamefont {Haldane}(1995)}]{Haldane95}%
  \BibitemOpen
  \bibfield  {author} {\bibinfo {author} {\bibfnamefont {F.~D.~M.}\
  \bibnamefont {Haldane}},\ }\href {\doibase 10.1103/PhysRevLett.74.2090}
  {\bibfield  {journal} {\bibinfo  {journal} {Phys. Rev. Lett.}\ }\textbf
  {\bibinfo {volume} {74}},\ \bibinfo {pages} {2090} (\bibinfo {year}
  {1995})}\BibitemShut {NoStop}%
\bibitem [{\citenamefont {Ganeshan}\ and\ \citenamefont
  {Levin}(2016)}]{Ganeshan16}%
  \BibitemOpen
  \bibfield  {author} {\bibinfo {author} {\bibfnamefont {S.}~\bibnamefont
  {Ganeshan}}\ and\ \bibinfo {author} {\bibfnamefont {M.}~\bibnamefont
  {Levin}},\ }\href {\doibase 10.1103/PhysRevB.93.075118} {\bibfield  {journal}
  {\bibinfo  {journal} {Phys. Rev. B}\ }\textbf {\bibinfo {volume} {93}},\
  \bibinfo {pages} {075118} (\bibinfo {year} {2016})}\BibitemShut {NoStop}%
\bibitem [{Note1()}]{Note1}%
  \BibitemOpen
  \bibinfo {note} {See Supplemental Material, which includes Refs.~\cite
  {Newman, Ganeshan17, Walton89, Altschuler90}, for a detailed discussion on
  the ground state and full forms of the branching rules.}\BibitemShut {Stop}%
\bibitem [{\citenamefont {Senthil}(2015)}]{Senthil15}%
  \BibitemOpen
  \bibfield  {author} {\bibinfo {author} {\bibfnamefont {T.}~\bibnamefont
  {Senthil}},\ }\href {\doibase 10.1146/annurev-conmatphys-031214-014740}
  {\bibfield  {journal} {\bibinfo  {journal} {Ann. Rev. Condens. Matter Phys.}\
  }\textbf {\bibinfo {volume} {6}},\ \bibinfo {pages} {299} (\bibinfo {year}
  {2015})}\BibitemShut {NoStop}%
\bibitem [{\citenamefont {Newman}(1972)}]{Newman}%
  \BibitemOpen
  \bibfield  {author} {\bibinfo {author} {\bibfnamefont {M.}~\bibnamefont
  {Newman}},\ }\href@noop {} {\emph {\bibinfo {title} {Integral Matrices}}}\
  (\bibinfo  {publisher} {Academic Press},\ \bibinfo {address} {New York},\
  \bibinfo {year} {1972})\BibitemShut {NoStop}%
\bibitem [{\citenamefont {Ganeshan}\ \emph {et~al.}(2017)\citenamefont
  {Ganeshan}, \citenamefont {Gorshkov}, \citenamefont {Gurarie},\ and\
  \citenamefont {Galitski}}]{Ganeshan17}%
  \BibitemOpen
  \bibfield  {author} {\bibinfo {author} {\bibfnamefont {S.}~\bibnamefont
  {Ganeshan}}, \bibinfo {author} {\bibfnamefont {A.~V.}\ \bibnamefont
  {Gorshkov}}, \bibinfo {author} {\bibfnamefont {V.}~\bibnamefont {Gurarie}}, \
  and\ \bibinfo {author} {\bibfnamefont {V.~M.}\ \bibnamefont {Galitski}},\
  }\href {\doibase 10.1103/PhysRevB.95.045309} {\bibfield  {journal} {\bibinfo
  {journal} {Phys. Rev. B}\ }\textbf {\bibinfo {volume} {95}},\ \bibinfo
  {pages} {045309} (\bibinfo {year} {2017})}\BibitemShut {NoStop}%
\bibitem [{\citenamefont {Walton}(1989)}]{Walton89}%
  \BibitemOpen
  \bibfield  {author} {\bibinfo {author} {\bibfnamefont {M.~A.}\ \bibnamefont
  {Walton}},\ }\href {\doibase 10.1016/0550-3213(89)90237-X} {\bibfield
  {journal} {\bibinfo  {journal} {Nucl. Phys. B}\ }\textbf {\bibinfo {volume}
  {322}},\ \bibinfo {pages} {775 } (\bibinfo {year} {1989})}\BibitemShut
  {NoStop}%
\bibitem [{\citenamefont {Altschuler}\ \emph {et~al.}(1990)\citenamefont
  {Altschuler}, \citenamefont {Bauer},\ and\ \citenamefont
  {Itzykson}}]{Altschuler90}%
  \BibitemOpen
  \bibfield  {author} {\bibinfo {author} {\bibfnamefont {D.}~\bibnamefont
  {Altschuler}}, \bibinfo {author} {\bibfnamefont {M.}~\bibnamefont {Bauer}}, \
  and\ \bibinfo {author} {\bibfnamefont {C.}~\bibnamefont {Itzykson}},\ }\href
  {\doibase 10.1007/BF02096653} {\bibfield  {journal} {\bibinfo  {journal}
  {Commun. Math. Phys.}\ }\textbf {\bibinfo {volume} {132}},\ \bibinfo {pages}
  {349} (\bibinfo {year} {1990})}\BibitemShut {NoStop}%
\end{thebibliography}%

\begin{widetext}
\newpage
\section*{Supplemental materials for ``Coupled-wire models with three-dimensional fractional excitations''}

\subsection{Solution of the $SU(2)_1 \times SU(2)_1 \supset U(1)_4 \times U(1)_4$ model}

We here present the solution of the 3D coupled-wire model based on the conformal embedding $SU(2)_1 \times SU(2)_1 \supset U(1)_4 \times U(1)_4$. 
Let us first summarize the structure of the model.
We consider the 3D coupled-wire model given by the Hamiltonian, 
\begin{align}
H &= H_0 +H_1.
\end{align}
Here, the first term $H_0$ describes the Hamiltonian for decoupled wires, 
\begin{align}
H_0 &= \sum_\bfR (H_{0,\bfR} +H_{0,\bfR+\bfdelta_y+\bfdelta_z}), \\
H_{0,\bfr} &= \frac{v}{4\pi} \int dx \sum_{l=1,2} \bigl[ (\partial_x \phi^l_{R,\bfr})^2 +(\partial_x \phi^l_{L,\bfr})^2 \bigr],
\end{align}
where $\bfR=(y,z) \in (\mathbb{Z},\mathbb{Z})$ specifies the position of a unit cell in the $yz$ plane and each unit cell contains two wires at the positions $\bfR$ and $\bfR+\bfdelta_y+\bfdelta_z$ with $\bfdelta_y=(\frac{1}{2},0)$ and $\bfdelta_z=(0,\frac{1}{2})$. 
The chiral bosonic fields are defined through the nonchiral ones by $\phi^l_{R/L,\bfr}(x) = \varphi^l_\bfr(x) \pm \theta^l_{\bfr}(x)$. 
The latter fields obey the commutation relations $[\theta^l_\bfr(x), \varphi^{l'}_{\bfr'}(x')] = i\pi \delta_{\bfr, \bfr'} \delta_{l,l'} \Theta (x-x')$. 
The second term $H_1$ describes the interactions between wires, 
\begin{align}
H_1 &= 2\gamma \sum_\bfR \int dx \ \Bigl[ \cos \bigl( 2\ttheta_{\bfR,\bfdelta_y+\bfdelta_z}(x) \bigr) +\cos \bigl( 2\ttheta_{\bfR,-\bfdelta_y+\bfdelta_z}(x) \bigr) +\cos \bigl( 2\ttheta_{\bfR,-\bfdelta_y-\bfdelta_z}(x) \bigr) +\cos \bigl( 2\ttheta_{\bfR,\bfdelta_y-\bfdelta_z}(x) \bigr) \Bigr], 
\end{align}
where we have defined the bosonic fields on the four links encircling $\bfR$ by 
\begin{align}
\begin{split}
\ttheta_{\bfR,\bfdelta_y+\bfdelta_z}(x) &\equiv -\tphi^S_{L,\bfR}(x) +\tphi^S_{R,\bfR+\bfdelta_y+\bfdelta_z}(x), \\
\ttheta_{\bfR,-\bfdelta_y+\bfdelta_z}(x) &\equiv \tphi^S_{R,\bfR}(x) -\tphi^S_{L,\bfR-\bfdelta_y+\bfdelta_z}(x), \\
\ttheta_{\bfR,-\bfdelta_y-\bfdelta_z}(x) &\equiv -\tphi^A_{L,\bfR}(x) +\tphi^A_{R,\bfR-\bfdelta_y-\bfdelta_z}(x), \\
\ttheta_{\bfR,\bfdelta_y-\bfdelta_z}(x) &\equiv \tphi^A_{R,\bfR}(x) -\tphi^A_{L,\bfR+\bfdelta_y-\bfdelta_z}(x),
\end{split}
\end{align}
and $\tphi^S_{R/L,\bfr}(x)$ and $\tphi^A_{R/L,\bfr}(x)$ are the symmetric and antisymmetric combinations of the bosonic fields $\phi^l_{R/L,\bfr}(x)$,
\begin{align}
\begin{split}
\tphi^S_{R/L,\bfr}(x) &\equiv \frac{1}{\sqrt{2}} \bigl( \phi^1_{R/L,\bfr}(x) +\phi^2_{R/L,\bfr}(x) \bigr), \\
\tphi^A_{R/L,\bfr}(x) &\equiv \frac{1}{\sqrt{2}} \bigl( \phi^1_{R/L,\bfr}(x) -\phi^2_{R/L,\bfr}(x) \bigr).
\end{split}
\end{align}

\subsubsection{Setup of the problem}

We now consider the model put on a torus $T^3$ with linear sizes $L_x \times L_y \times L_z$, so that the sum over $\bfR=(y,z)$ is restricted to the ranges $y \in [0, L_y-1)$ and $z \in [0, L_z-1)$ and the periodic boundary condition is imposed along each axis.
Since the Hamiltonian $H_0+H_1$ is just a sine-Gordon model with $4L_y L_z$ cosine terms, we can solve the model in the limit $|\gamma| \to \infty$ by expanding the cosine terms around the minima of $\ttheta_{\bfR,\bfdelta}(x)$ and keeping only quadratic terms. 
To do so, we adopt the formal recipe proposed by Ganeshan and Levin in Ref.~\cite{Ganeshan16}.

Since in this formalism the arguments of cosine terms are assumed to be real-valued, the Hamiltonian $H_0 +H_1$ does not solely maintain the compactification conditions of the bosonic fields, 
\begin{align} \label{eq:Compactification}
\begin{split}
\varphi^l_\bfr &\sim \varphi^l_\bfr +\sqrt{2} \pi \mathbb{Z}, \\
\theta^l_\bfr &\sim \theta^l_\bfr +\sqrt{2} \pi \mathbb{Z}.
\end{split}
\end{align}
We thus add the following terms \cite{Ganeshan16}, 
\begin{align}
H_U = -U \sum_\bfR \sum_{l=1,2} \bigl[ \cos (\sqrt{2} \pi P^l_\bfR) +\cos (\sqrt{2} \pi P^l_{\bfR+\bfdelta_y+\bfdelta_z}) +\cos (\sqrt{2} \pi Q^l_\bfR) +\cos (\sqrt{2} \pi Q^l_{\bfR+\bfdelta_y+\bfdelta_z}) \bigr], 
\end{align}
to the Hamiltonian $H_0+H_1$, where $P^l_\bfr$ and $Q^l_\bfr$ satisfy the commutation relations, 
\begin{align}
\begin{split}
[P^l_\bfr, \theta^{l'}_{\bfr'}(x')] &= [Q^l_\bfr, \varphi^{l'}_{\bfr'}(x')] = i\delta_{\bfr,\bfr'} \delta_{l,l'}, \\
[P^l_{\bfr}, \varphi^{l'}_{\bfr'}(x')] &= [Q^l_\bfr, \theta^{l'}_{\bfr'}(x')] = [P^l_\bfr, P^{l'}_{\bfr'}] = [Q^l_\bfr, Q^{l'}_{\bfr'}] = [P^l_\bfr, Q^{l'}_{\bfr'}] =0.
\end{split}
\end{align}
By letting $U \to \infty$, $H_U$ dynamically enforces the compactificatiion conditions in Eqs.~\eqref{eq:Compactification} and also the discreteness of $P^l_\bfr$ and $Q^l_\bfr$ such that $P^l_\bfr, Q^l_\bfr \in \sqrt{2} \mathbb{Z}$. 
The operators $P^l_\bfr$ and $Q^l_\bfr$ represent zero-mode parts in the mode expansions of the bosonic fields $\varphi^l_\bfr(x)$ and $\theta^l_\bfr(x)$, respectively, and are formally written as 
\begin{align}
\begin{split}
P^l_\bfr &= \frac{1}{\pi} \int^{L_x}_0 dx \ \partial_x \varphi^l_\bfr (x), \\
Q^l_\bfr &= \frac{1}{\pi} \int^{L_x}_0 dx \ \partial_x \theta^l_\bfr (x).
\end{split}
\end{align}
Once these operators are exponentiated in the form $e^{i\sqrt{2} \pi \sum_l (f^l_P P^l_\bfr +f^l_Q Q^l_\bfr)}$ with appropriate coefficients $f^l_P$ and $f^l_Q$, they have a nice physical meaning: 
They are string operators that transfer certain quasiparticles on the wire $\bfr$ along the cycle in $x$ on the torus. 
For example, 
\begin{align}
e^{(i\sqrt{2} \pi/4) (P^1_\bfr +P^2_\bfr +Q^1_\bfr +Q^2_\bfr)} = e^{(i/2) \int^{L_x}_0 dx \ \partial_x \tphi^S_{R,\bfr}(x)} = \calS^{S,1,R}_\bfr(0,L_x)
\end{align}
can be seen as the string operator transferring the $h=1/8$ quasiparticle of $U(1)_4$ in the symmetric sector. 

Labeling the $12L_y L_z$ field variables in the cosine terms by $C_I$ as 
\begin{align} \label{eq:DefC1}
\begin{split}
\{ C_1, \cdots, C_{4L_yL_z} \} &= \{ C^\theta_{(0,0)}, \cdots, C^\theta_{(L_y-1,0)}, C^\theta_{(0,1)}, \cdots, C^\theta_{(L_y-1,1)}, \cdots, C^\theta_{(0,L_z-1)}, \cdots, C^\theta_{(L_y-1,L_z-1)} \}, \\
\{ C_{4L_yL_z+1}, \cdots, C_{8L_yL_z} \} &= \{ C^P_{(0,0)}, \cdots, C^P_{(L_y-1,0)}, \cdots, C^P_{(0,1)}, \cdots, C^P_{(L_y-1,1)}, \cdots, C^P_{(0,L_z-1)}, \cdots, C^P_{(L_y-1,L_z-1)} \}, \\
\{ C_{8L_yL_z+1}, \cdots, C_{12L_yL_z} \} &= \{ C^Q_{(0,0)}, \cdots, C^Q_{(L_y-1,0)}, \cdots, C^Q_{(0,1)}, \cdots, C^Q_{(L_y-1,1)}, \cdots, C^Q_{(0,L_z-1)}, \cdots, C^Q_{(L_y-1,L_z-1)} \}, 
\end{split}
\end{align}
with 
\begin{align} \label{eq:DefC2}
\begin{split}
C^\theta_\bfR &\equiv \{ C^\theta_{\bfR,1}, C^\theta_{\bfR,2}, C^\theta_{\bfR,3}, C^\theta_{\bfR,4} \} = \{ 2\ttheta_{\bfR,\bfdelta_y+\bfdelta_z}, 2\ttheta_{\bfR,-\bfdelta_y+\bfdelta_z}, 2\ttheta_{\bfR,-\bfdelta_y-\bfdelta_z}, 2\ttheta_{\bfR,\bfdelta_y-\bfdelta_z} \}, \\
C^P_\bfR &\equiv \{ C^P_{\bfR,1}, C^P_{\bfR,2}, C^P_{\bfR,3}, C^P_{\bfR,4} \} = \{ \sqrt{2} \pi P^1_\bfR, \sqrt{2} \pi P^2_\bfR, \sqrt{2} \pi P^1_{\bfR+\bfdelta_y+\bfdelta_z}, \sqrt{2} \pi P^2_{\bfR+\bfdelta_y+\bfdelta_z} \}, \\
C^Q_\bfR &\equiv \{ C^Q_{\bfR,1}, C^Q_{\bfR,2}, C^Q_{\bfR,3}, C^Q_{\bfR,4} \} = \{ \sqrt{2} \pi Q^1_\bfR, \sqrt{2} \pi Q^2_\bfR, \sqrt{2} \pi Q^1_{\bfR+\bfdelta_y+\bfdelta_z}, \sqrt{2} \pi Q^2_{\bfR+\bfdelta_y+\bfdelta_z} \}, 
\end{split}
\end{align}
we consider the Hamiltonian, 
\begin{align}
H' = \gamma \sum_{I=1}^{4L_y L_z} \int dx \ \cos \bigl( C_I(x) \bigr) -U \sum_{I=4L_yL_z+1}^{12L_yL_z} \cos (C_I), 
\end{align}
in the limit of $\gamma \to -\infty$ and $U \to \infty$. 
The limit $\gamma \to \infty$ can also be studied by replacing $C_I \to C_I +\pi$ for $1 \leq I \leq 4L_y L_z$, but it does not alter the conclusions below. 
In this study, we are just interested in the ground state of the Hamiltonian $H'$. 
Then the problem to solve boils down to find an integer unimodular matrix $\calV_{IJ}$ that brings the skew-symmetric (antisymmetric) integer matrix $\calZ_{IJ}$ formed by the commutators \cite{Ganeshan16}, 
\begin{align} \label{eq:ZCommMat}
\calZ_{IJ} = \frac{1}{2\pi i} [C_I,C_J],
\end{align}
into the skew normal form \cite{Newman}, 
\begin{align}
\calZ' = \calV \calZ \calV^T, 
\end{align}
with 
\begin{align} \label{eq:ZSkewNormal}
\calZ' = \left( \begin{array}{ccccccc|c} 
0 & -d_1 & 0 & 0 & \cdots & 0 & 0 & \mathbf{0}_M \\
d_1 & 0 & 0 & 0 & \cdots & 0 & 0 & \mathbf{0}_M \\
0 & 0 & 0 & -d_2 & & 0 & 0 & \mathbf{0}_M \\
0 & 0 & d_2 & 0 & & 0 & 0 & \mathbf{0}_M \\
\vdots & \vdots & & & \ddots & \vdots & \vdots & \vdots \\
0 & 0 & 0 & 0 & \cdots & 0 & -d_r & \mathbf{0}_M \\
0 & 0 & 0 & 0 & \cdots & d_r & 0 & \mathbf{0}_M \\ \hline
\mathbf{0}_M^T & \mathbf{0}_M^T & \mathbf{0}_M^T & \mathbf{0}_M^T & \cdots & \mathbf{0}_M^T & \mathbf{0}_M^T & \mathbf{O}_M
\end{array} \right).
\end{align}
Here, $2r$ corresponds to the rank of $\calZ$, $M=2r-12L_yL_z$, $\mathbf{0}_M$ and $\mathbf{O}_M$ are the $M$-dimensional vector and matrix whose entries are all zero, respectively, and $d_1,\cdots,d_r$ are positive integers aligned such that $d_I$ divides all $d_J$'s with $I< J \leq r$. 
There always exists such a transformation $\calV$ although it is not unique. 
Once a transformation $\calV$ is found, the ground-state degeneracy is given by $\prod_{I=1}^r d_I$. 
[If one only wants to compute the ground-state degeneracy, it suffices to find the Smith normal form of $\calZ$ that takes the form of $\textrm{diag}(d_1,d_1,d_2,d_2,\cdots,d_r,d_r,\mathbf{0}_M)$.] 
Furthermore, we can obtain a set of operators, 
\begin{align}
\begin{split}
X'_I &= e^{(i/d_I)\sum_J \calV_{2I-1,J} C_J}, \\
Z'_I &= e^{(i/d_I) \sum_J \calV_{2I,J} C_J}, 
\end{split}
\end{align}
that obey $X'_I Z'_I = e^{2\pi i/d_I} Z'_I X'_I$ and span the $d_I$-dimensional subspace of the ground-state manifold \cite{Ganeshan17}. 
As we will show below, we find that $r=4L_y L_z$ and 
\begin{align} \label{eq:Degeneracy}
d_I=
\begin{cases}
1 & (1 \leq I \leq 2L_y L_z-1), \\
2 & (2L_y L_z \leq I \leq 4L_y L_z-1), \\
4 & (I=4L_y L_z),
\end{cases}
\end{align}
and thus the ground state degeneracy is $4 \cdot 2^{2L_y L_z}$. 

Before proceeding, we here present the explicit form of $\calZ$ for the completeness. 
Let us first write $\calZ$ in the block form: 
\begin{align}
\calZ = \begin{pmatrix} \calZ^{\theta,\theta} & \calZ^{\theta,P} & \calZ^{\theta,Q} \\ -(\calZ^{\theta,P})^T & \mathbf{O}_{4L_yL_z} & \mathbf{O}_{4L_yL_z} \\ -(\calZ^{\theta,Q})^T & \mathbf{O}_{4L_yL_z} & \mathbf{O}_{4L_yL_z} \end{pmatrix}, 
\end{align}
where each block is a $4L_y L_z \times 4L_y L_z$ matrix. 
By letting $I=(\bfR,\ell)$ and $J=(\bfR',\ell')$ with $\ell,\ell'=1,\cdots,4$ from Eqs.~\eqref{eq:DefC1} and \eqref{eq:DefC2}, we have 
\begin{align}
\calZ^{\theta,\theta}_{(\bfR,\ell),(\bfR',\ell')} 
&= 2\delta_{\ell,1} \delta_{\ell',2} (\delta_{\bfR,\bfR'} -\delta_{\bfR+\bfe_y,\bfR'}) 
+2\delta_{\ell,2} \delta_{\ell',1} (-\delta_{\bfR,\bfR'} +\delta_{\bfR-\bfe_y,\bfR'}) \nonumber \\
&\ \ \ +2\delta_{\ell,3} \delta_{\ell',4} (\delta_{\bfR,\bfR'} -\delta_{\bfR-\bfe_y,\bfR'}) 
+2\delta_{\ell,4} \delta_{\ell',3} (-\delta_{\bfR,\bfR'} +\delta_{\bfR+\bfe_y,\bfR'}), \\
\calZ^{\theta,P}_{(\bfR,\ell),(\bfR',\ell')} 
&= \delta_{\ell,1} \delta_{\bfR,\bfR'} (-\delta_{\ell',1}-\delta_{\ell',2} -\delta_{\ell',3} -\delta_{\ell',4}) 
+\delta_{\ell,2} [\delta_{\bfR,\bfR'} (-\delta_{\ell',1} -\delta_{\ell',2}) +\delta_{\bfR-\bfe_y,\bfR'} (-\delta_{\ell',3} -\delta_{\ell',4})] \nonumber \\
&\ \ \ +\delta_{\ell,3} [\delta_{\bfR,\bfR'} (-\delta_{\ell',1} +\delta_{\ell',2}) +\delta_{\bfR-\bfe_y-\bfe_z,\bfR'} (-\delta_{\ell',3} +\delta_{\ell',4})] \nonumber \\
&\ \ \ +\delta_{\ell,4} [\delta_{\bfR,\bfR'}(-\delta_{\ell',1}+\delta_{\ell',2}) +\delta_{\bfR-\bfe_z,\bfR'}(-\delta_{\ell',3} +\delta_{\ell',4})], \\
\calZ^{\theta,Q}_{(\bfR,\ell),(\bfR',\ell')} 
&= \delta_{\ell,1} \delta_{\bfR,\bfR'} (\delta_{\ell',1}+\delta_{\ell',2} -\delta_{\ell',3} -\delta_{\ell',4}) 
+\delta_{\ell,2} [\delta_{\bfR,\bfR'} (-\delta_{\ell',1} -\delta_{\ell',2}) +\delta_{\bfR-\bfe_y,\bfR'} (\delta_{\ell',3} +\delta_{\ell',4})] \nonumber \\
&\ \ \ +\delta_{\ell,3} [\delta_{\bfR,\bfR'} (\delta_{\ell',1} -\delta_{\ell',2}) +\delta_{\bfR-\bfe_y-\bfe_z,\bfR'} (-\delta_{\ell',3} +\delta_{\ell',4})] \nonumber \\
&\ \ \ +\delta_{\ell,4} [\delta_{\bfR,\bfR'}(-\delta_{\ell',1}+\delta_{\ell',2}) +\delta_{\bfR-\bfe_z,\bfR'}(\delta_{\ell',3} -\delta_{\ell',4})], 
\end{align}
where $\bfe_y=(1,0)$ and $\bfe_z=(0,1)$. 

\subsubsection{Operators spanning the ground-state manifold}

While the transformation $\calV$ can be found in an algorithmic way by repeating elementary low and column operations \cite{Newman}, we here present a modified algorithm to obtain a physically sensible transformation for our purpose.

\begin{enumerate}
\item We first apply a unimodular transformation $C^{(1)}_I=\sum_{J} \calV^{(1)}_{IJ} C_J$ only acting on the link fields $C_I$ with $1 \leq I \leq 4L_y L_z$. 
Explicitly, we choose the transformation $\calV^{(1)}$ such that it brings the fields $C_I$ to the following forms:
\begin{align}
\label{eq:C1onLink1}
C^{(1)}_{2(y+zL_y)+1} &= \begin{cases} 
2\ttheta_{(0,z),-\bfdelta_y-\bfdelta_z} & \textrm{for $y=0$, $0 \leq z \leq L_z-1$}, \\
2\ttheta_{(y,z),-\bfdelta_y+\bfdelta_z} & \textrm{for $0 < y \leq L_y-1$, $0 \leq z \leq L_z-1$}, \end{cases} \\
\label{eq:C1onLink2}
C^{(1)}_{2(y+zL_y)+2} &= 2\ttheta_{(y,z),\bfdelta_y+\bfdelta_z} \ \ \ \textrm{for $(y,z) \neq (L_y-1,L_z-1)$}, \\
\label{eq:C1onPlaquetteA}
C^{(1)}_{2L_yL_z +2(y+zL_y)} &= 2\ttheta_{(y,z),\bfdelta_y+\bfdelta_z} +2\ttheta_{(y,z),-\bfdelta_y+\bfdelta_z} +2\ttheta_{(y,z+1),\bfdelta_y-\bfdelta_z} +2\ttheta_{(y,z+1),-\bfdelta_y-\bfdelta_z} \nonumber \\
&\hspace{200pt} \textrm{for $(y,z) \neq (L_y-1,L_z-1)$}, \\
\label{eq:C1onPlaquetteB}
C^{(1)}_{2L_yL_z +2(y+zL_y)+1} &= 2\ttheta_{(y,z),\bfdelta_y+\bfdelta_z} +2\ttheta_{(y,z),\bfdelta_y-\bfdelta_z} -2\ttheta_{(y+1,z),-\bfdelta_y-\bfdelta_z} -2\ttheta_{(y+1,z),-\bfdelta_y+\bfdelta_z} \nonumber \\
&\hspace{200pt} \textrm{for $(y,z) \neq (L_y-1,L_z-1)$}, \\
\label{eq:C1onCycleY}
C^{(1)}_{4L_yL_z-2} &= \sum_y \bigl( 2\ttheta_{(y,0),\bfdelta_y+\bfdelta_z} +2\ttheta_{(y,0),-\bfdelta_y+\bfdelta_z} \bigr), \\
\label{eq:C1onCycleZ}
C^{(1)}_{4L_yL_z-1} &= \sum_z \bigl( 2\ttheta_{(0,z),\bfdelta_y+\bfdelta_z} +2\ttheta_{(0,z),\bfdelta_y-\bfdelta_z} \bigr), \\
\label{eq:C1onYZPlane}
C^{(1)}_{4L_yL_z} &= \sum_\bfR \bigl( 2\ttheta_{\bfR,\bfdelta_y+\bfdelta_z} +2\ttheta_{\bfR,-\bfdelta_y+\bfdelta_z} +2\ttheta_{\bfR,-\bfdelta_y-\bfdelta_z} +2\ttheta_{\bfR,\bfdelta_y-\bfdelta_z} \bigr), \\
C^{(1)}_I &= C_I \ \ \ \textrm{for $4L_yL_z < I \leq 12 L_y L_z$}.
\end{align}
This is the most heuristic part of our algorithm; we numerically confirmed for many choices of $(L_y,L_z)$ that the corresponding transformation $\calV^{(1)}$ is unimodular: $|\det \calV^{(1)}|=1$. 
One can further check that the last $2L_y L_z+1$ fields, $\{ C^{(1)}_{2L_yL_z}, \cdots, C^{(1)}_{4L_yL_z} \}$, defined in Eqs.~\eqref{eq:C1onPlaquetteA}-\eqref{eq:C1onYZPlane} commute with each other.

\item We appropriately add integer multiples of $C^{(1)}_I \in \{ C^{(1)}_{4L_y L_z+1}, \cdots, C^{(1)}_{12 L_y L_z} \}$ to $C^{(1)}_I \in \{ C^{(1)}_1, \cdots, C^{(1)}_{2L_y L_z-1} \}$ and define the corresponding transformation by $C^{(2)}_I = \sum_J \calV^{(2)}_{IJ} C^{(1)}_J$ such that the first $4L_y L_z$ fields $\{ C^{(2)}_1, \cdots, C^{(2)}_{4L_yL_z} \}$ are all commuting with each other. 
Since the transformation $\calV^{(2)}$ only involves elementary row operations, it is unimodular. 
We note that $C^{(2)}_I = C^{(1)}_I$ for $2L_y L_z \leq I \leq 12 L_y L_z$.

\item After the above two steps, $\calZ_{IJ}$ in Eq.~\eqref{eq:ZCommMat} is transformed into 
\begin{align}
(\calV^{(2)} \calV^{(1)}) \calZ (\calV^{(2)} \calV^{(1)})^T = \begin{pmatrix} \mathbf{O}_{4L_yL_z} & -\mathbf{Y}^T \\ \mathbf{Y} & \mathbf{O}_{8L_yL_z} \end{pmatrix}, 
\end{align}
where $\mathbf{Y}$ is an $8L_yL_z \times 4L_yL_z$ integer matrix. 
We then apply elementary row operations to $\mathbf{Y}$ such that it becomes the Hermite normal form, 
\begin{align}
\mathbf{Y}' = \begin{pmatrix} \mathbf{D} \\ \mathbf{O}_{4L_yL_z} \end{pmatrix}.
\end{align}
One can find that $\mathbf{D}=\textrm{diag}(d_1, d_2, \cdots, d_{4L_yL_z})$ with $d_I$ given in Eq.~\eqref{eq:Degeneracy}. 
Writing the corresponding transformation as $\mathbf{Y}' = \mathbf{W} \mathbf{Y}$, we obtain the unimodular transformation $\calV^{(3)} = \mathbf{I}_{4L_yL_z} \oplus \mathbf{W}$ where $\mathbf{I}_m$ is the $m$-dimensional identity matrix. 
We thus find 
\begin{align} \label{eq:ZCommMat3}
(\calV^{(3)} \calV^{(2)} \calV^{(1)}) \calZ (\calV^{(3)} \calV^{(2)} \calV^{(1)})^T = \begin{pmatrix} \mathbf{O}_{4L_yL_z} & -\mathbf{D} & \mathbf{O}_{4L_yL_z} \\ \mathbf{D} & \mathbf{O}_{4L_yL_z} & \mathbf{O}_{4L_yL_z} \\ \mathbf{O}_{4L_yL_z} & \mathbf{O}_{4L_yL_z} & \mathbf{O}_{4L_yL_z} \end{pmatrix}.
\end{align}
We then define $C^{(3)}_I = \sum_J \calV^{(3)}_{IJ} C^{(2)}_J$.

\item We rearrange $C^{(3)}_I$ such that Eq.~\eqref{eq:ZCommMat3} is brought into $\calZ'$ defined in Eq.~\eqref{eq:ZSkewNormal}. 
Writing the corresponding transformation as $\calV^{(4)}$, we finally find $\calV = \calV^{(4)} \calV^{(3)} \calV^{(2)} \calV^{(1)}$.
\end{enumerate}

In the end, we obtain the transformed field variables $C'_I = \sum_J \calV_{IJ} C_J$. 
Now only $C'_{2I-1}$ with $1 \leq I \leq 4L_y L_z$ contain the link fields $\ttheta_{\bfR,\bfdelta}(x)$, while the other $C'_I$'s are composed solely of $P^l_\bfr$ and $Q^l_\bfr$. 
In particular, the forms of the link fields contained in $C'_{2I-1}$ with $1 \leq I \leq 4L_yL_z$ is not changed from $C^{(1)}_I$ defined in Eqs.~\eqref{eq:C1onLink1}-\eqref{eq:C1onYZPlane} even after the sequence of the transformations. 
The dimensions $d_I$ of the ground state subspaces are tied to how these link fields are written in terms of local operators of the theory. 
We below consider the operators $X'_I=e^{iC'_{2I-1}/d_I}$ and $Z'_I=e^{iC'_{2I}/d_I}$ that map the ground-state manifold to itself and span the $d_I$-dimensional subspace of the ground-state manifold. 
Both $X'_I$ and $Z'_I$ must be written as products of local operators.

The link fields in $\{ C'_1, C'_3, \cdots, C'_{4L_yL_z-3} \}$ and their arbitrary combinations can only form open strings in the $yz$ plane. 
The corresponding operators $\{ X'_1, X'_2, \cdots, X'_{2L_yL_z-1} \}$ must then be written in terms of left-right products of the $U(1)_4$ currents $(\bJ^\rho_\bfr)^+ \propto e^{i2 \tphi^\rho_{R,\bfr}}$ and $(J^\rho_\bfr)^- \propto e^{-i2 \phi^\rho_{L,\bfr}}$ on the same link, which thus trivially act on the ground state. 
Their partners $\{ Z'_1, Z'_2, \cdots, Z'_{2L_yL_z-1} \}$ transfer quasiparticles associated with the $h=1/8$ fields $e^{\pm (i/2) \tphi^\rho_{R/L,\bfr}}$ of the $U(1)_4$ CFT along the $x$ axis. 
Since $\{ X'_1, \cdots, X'_{2L_yL_z-1} \}$ consist of the $U(1)_4$ currents, their algebras can only be trivial: $X'_I Z'_I = Z'_I X'_I$. 
The fields $C'_{2I-1}$ with $2L_yL_z \leq I \leq 4L_yL_z$ are nothing but $C^{(1)}_I$ defined in Eqs.~\eqref{eq:C1onPlaquetteA}-\eqref{eq:C1onYZPlane}. 
The link fields in $\{ C'_{4L_yL_z-1}, C'_{4L_yL_z+1}, \cdots, C'_{8L_yL_z-3} \}$ form closed strings in the $yz$ plane. 
The corresponding operators $\{ X'_{2L_yL_z}, X'_{2L_yL_z+1}, \cdots, X'_{4L_yL_z-1} \}$ can be written in terms of left-right products of the $h=1/2$ fermionic primary fields of the $U(1)_4$ CFT, $e^{i \tphi^\rho_{R,\bfr} \pm i\tphi^{\rho'}_{L,\bfr}}$, since they do not create any excitation once they are multiplied along a closed string. 
Since the operators $\{ Z'_{2L_yL_z}, Z'_{2L_yL_z+1}, \cdots, Z'_{4L_yL_z-1} \}$ transfer the $h=1/8$ quasiparticle along a path parallel to the $x$ axis that intersects with the closed string of the link fields, they obey the algebra $X'_I Z'_I = -Z'_I X'_I$. 
Finally, $C'_{8L_yL_z-1}$ is composed of the link fields fully covering the $yz$ plane. 
The corresponding operator $X'_{4L_yL_z}$ is written in terms of products of the $h=1/8$ primary fields, $e^{(i/2) (\tphi^S_{R,\bfr}-\tphi^S_{L,\bfr}+\tphi^A_{R,\bfr}-\tphi^A_{L,\bfr})}$, multiplied over all the wires such that they do not have a boundary where excitations are created. 
The string operator $Z'_{4L_yL_z}$ transfers the $h=1/8$ quasiparticle along the $x$ axis and obeys the algebra $X'_{4L_yL_z} Z'_{4L_yL_z} = e^{i\pi/2} Z'_{4L_yL_z} X'_{4L_yL_z}$. 

In Tables~\ref{tab:Ops2x2a}-\ref{tab:Ops2x2c}, we illustrate a set of the operators obtained by the above algorithm for $(L_y,L_z)=(2,2)$.
\begin{table}
\caption{Graphical representations of the operators spanning the ground-state manifold for $(L_y,L_z)=(2,2)$. 
The green solid lines denote links where the link fields $\ttheta_{\bfr,\bfdelta}$ are acting. 
The associated symbols $J$, $\psi$, and $a$ correspond to $e^{i2\tphi^\rho_{P,\bfr}}$, $e^{i\tphi^\rho_{P,\bfr}}$, and $e^{(i/2)\tphi^\rho_{P,\bfr}}$, respectively, whose labels $\rho=S,A$ and $P=R,L$ should be assigned in accordance with the bosonic modes near the symbols, and $\bJ$, $\bar{\psi}$, and $\bar{a}$ are to their complex conjugates (they here do not mean left- or right-going modes). 
The purple ovals denote string operators along the wires. 
The associated symbols are similarly defined such as $J$ for $e^{i2\tphi^\rho_{P,\bfr}(L_x) -i2\tphi^\rho_{P,\bfr}(0)}$, $\bar{\psi}$ for $e^{-i\tphi^\rho_{P,\bfr}(L_x) +i\tphi^\rho_{P,\bfr}(0)}$, and so on.}
\label{tab:Ops2x2a}
\begin{tabular}{ll|cc||ll|cc}
\hline \hline
$I$ & $d_I$ & $\exp(iC'_{2I-1}/d_I)$ & $\exp(iC'_{2I}/d_I)$ & $I$ & $d_I$ & $\exp(iC'_{2I-1}/d_I)$ & $\exp(iC'_{2I}/d_I)$ \\ \hline
$1$ & $1$ & \includegraphics[clip,width=0.2\textwidth]{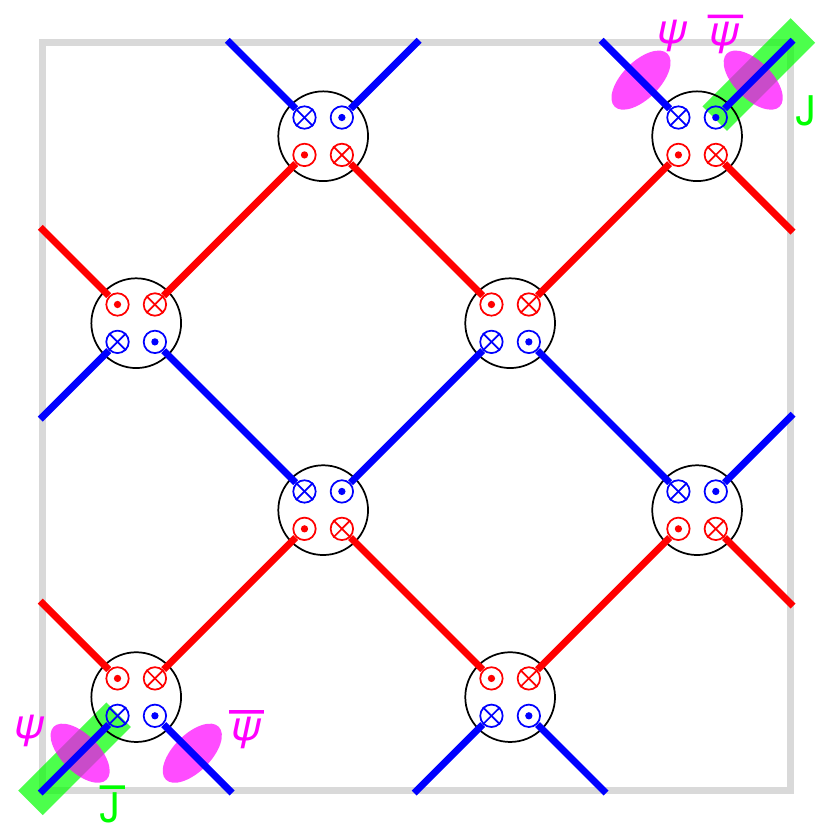} & \includegraphics[clip,width=0.2\textwidth]{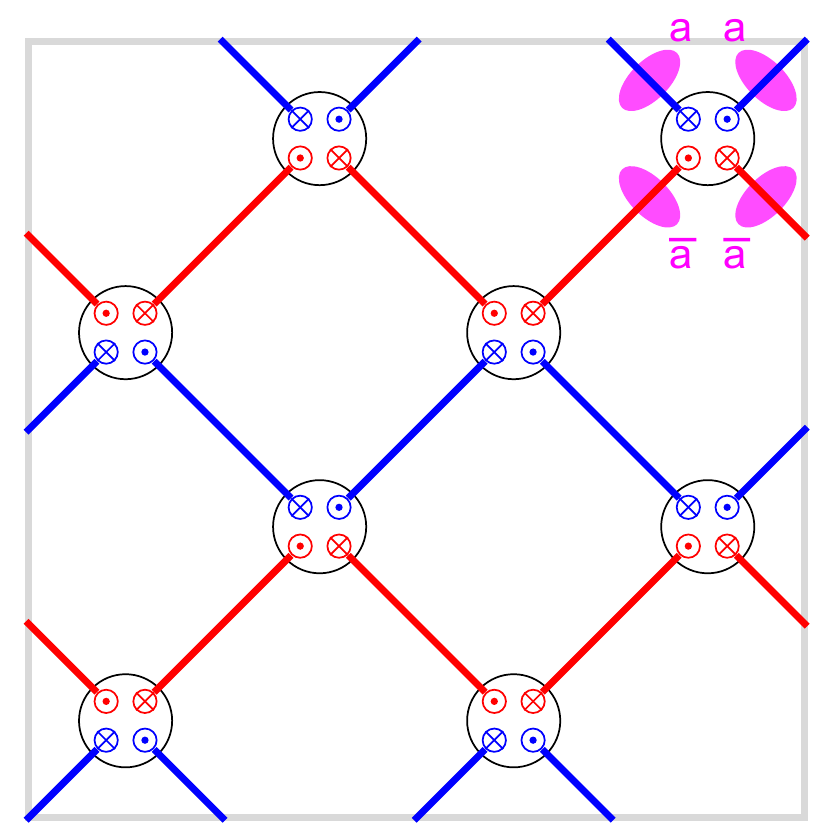} & 
$2$ & $1$ & \includegraphics[clip,width=0.2\textwidth]{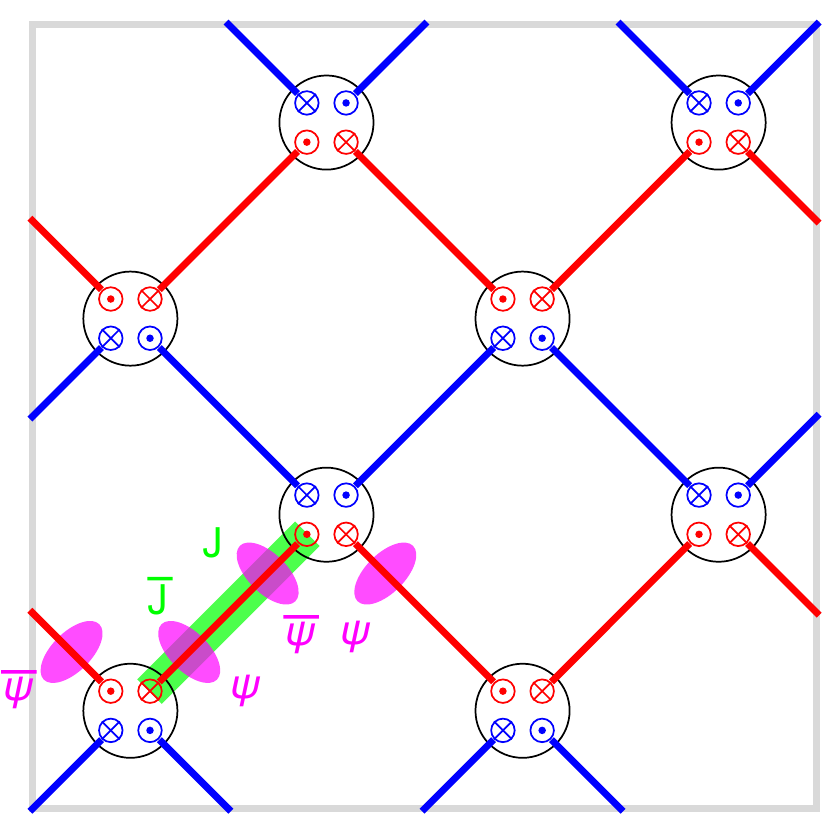} & \includegraphics[clip,width=0.2\textwidth]{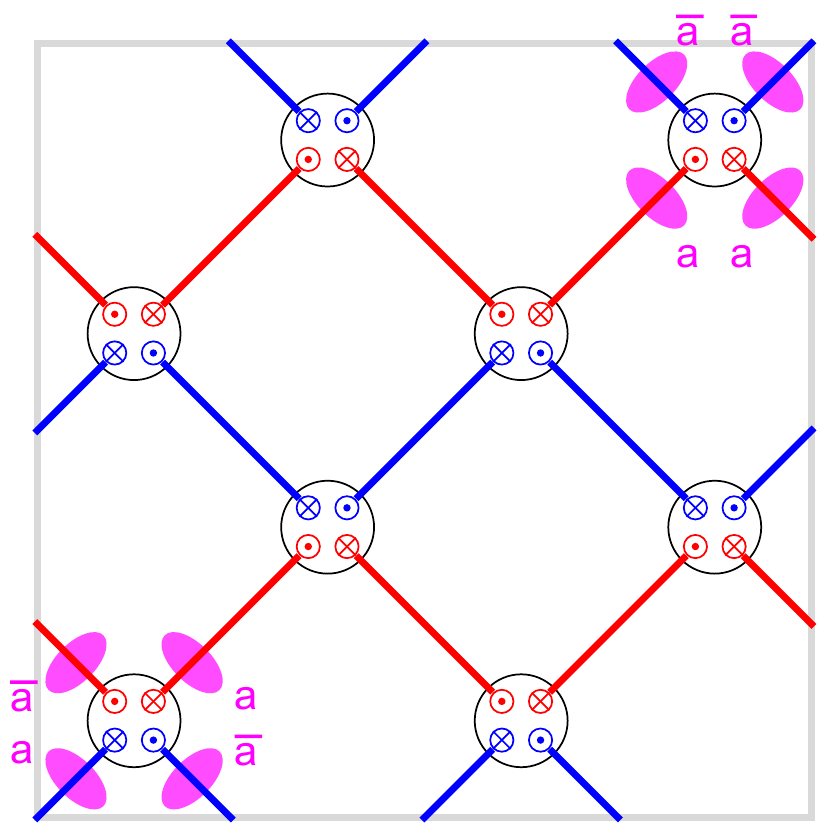} \\ \hline
$3$ & $1$ & \includegraphics[clip,width=0.2\textwidth]{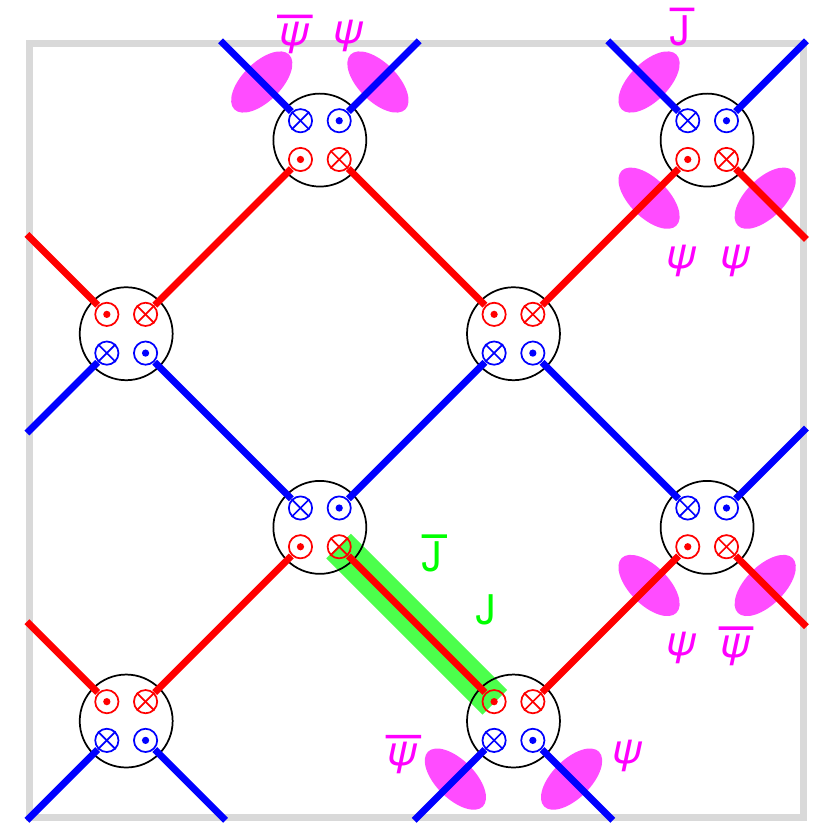} & \includegraphics[clip,width=0.2\textwidth]{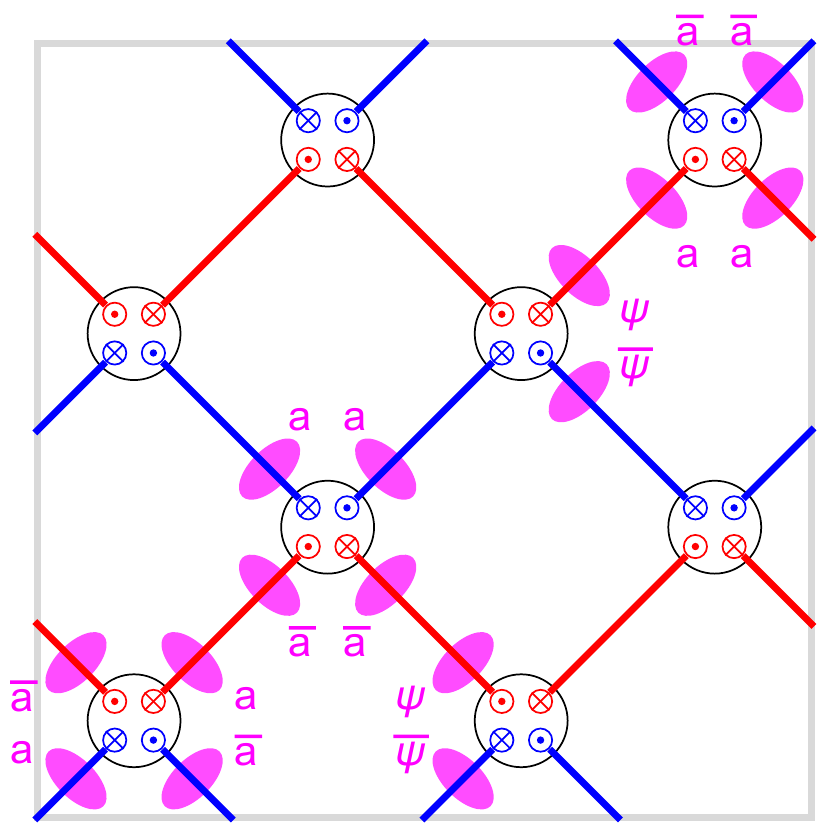} &
$4$ & $1$ & \includegraphics[clip,width=0.2\textwidth]{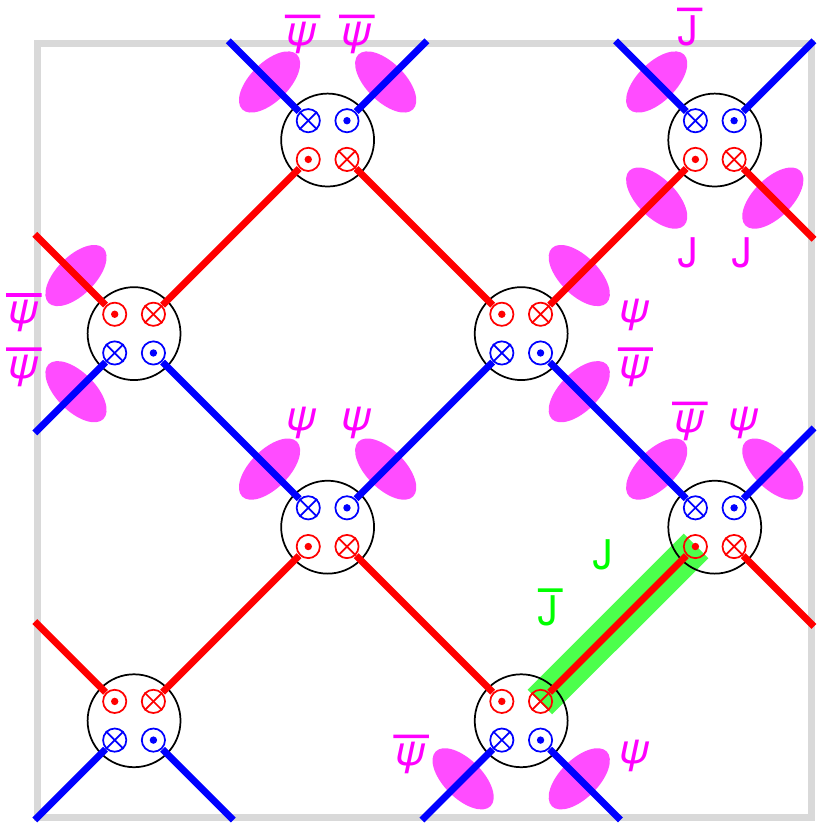} & \includegraphics[clip,width=0.2\textwidth]{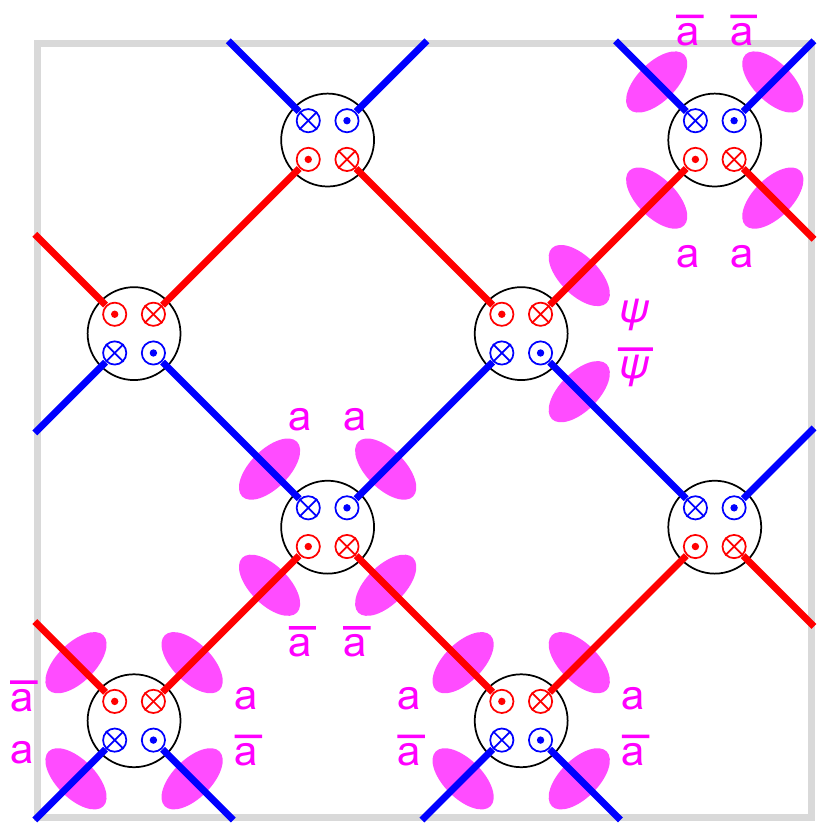} \\ \hline
$5$ & $1$ & \includegraphics[clip,width=0.2\textwidth]{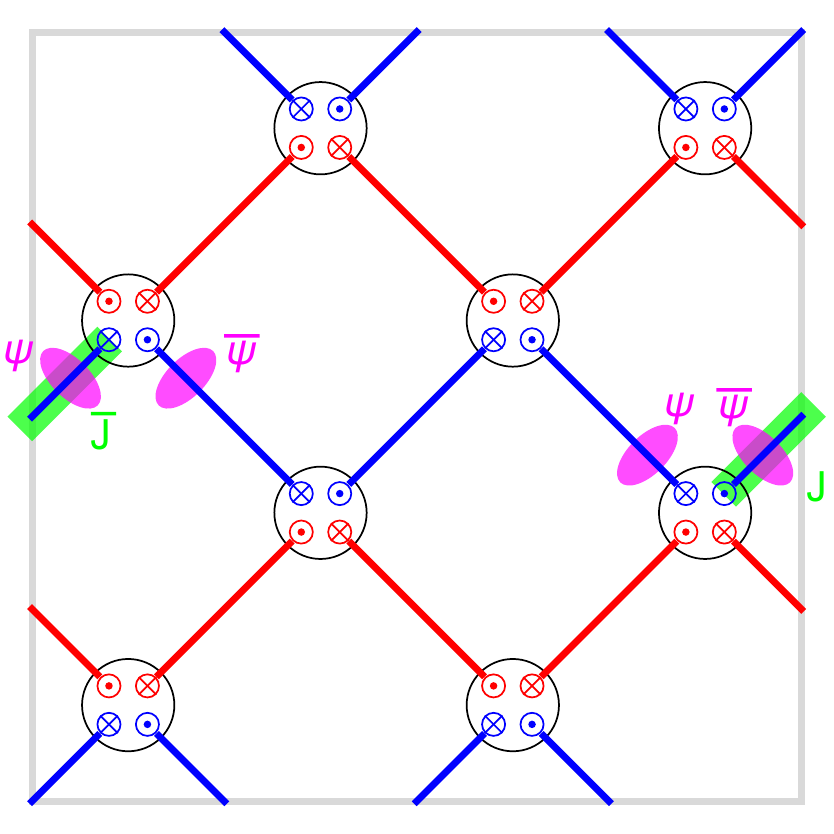} & \includegraphics[clip,width=0.2\textwidth]{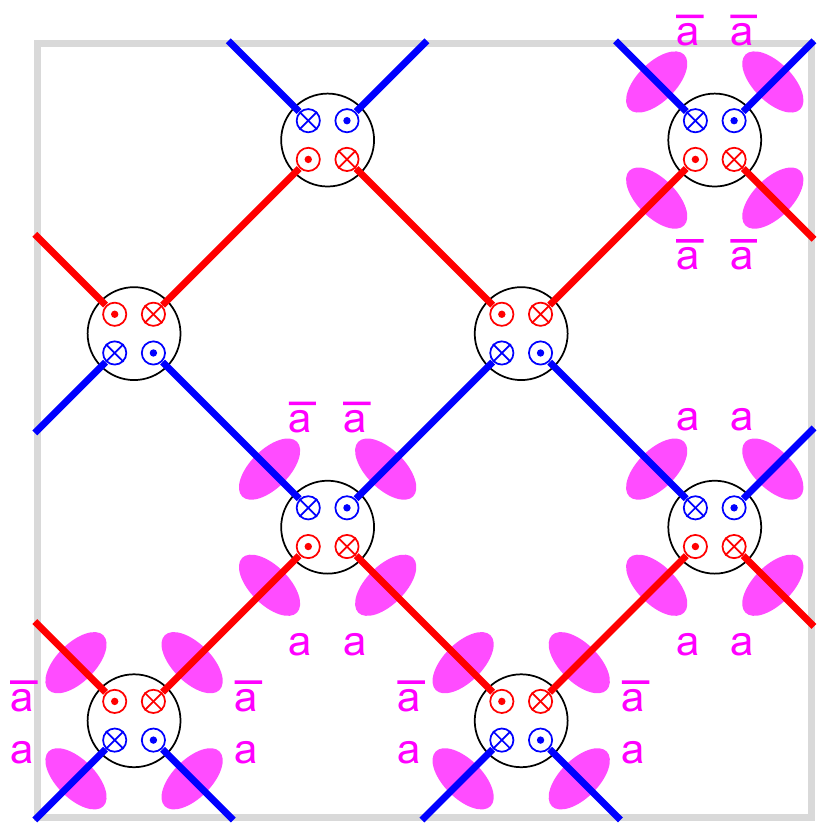} &
$6$ & $1$ & \includegraphics[clip,width=0.2\textwidth]{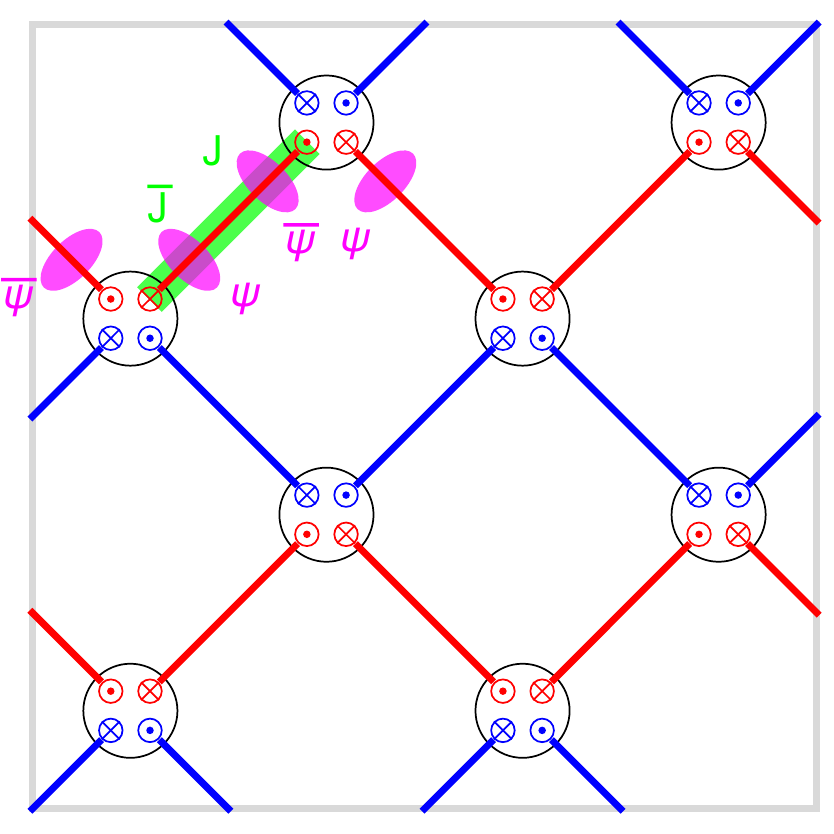} & \includegraphics[clip,width=0.2\textwidth]{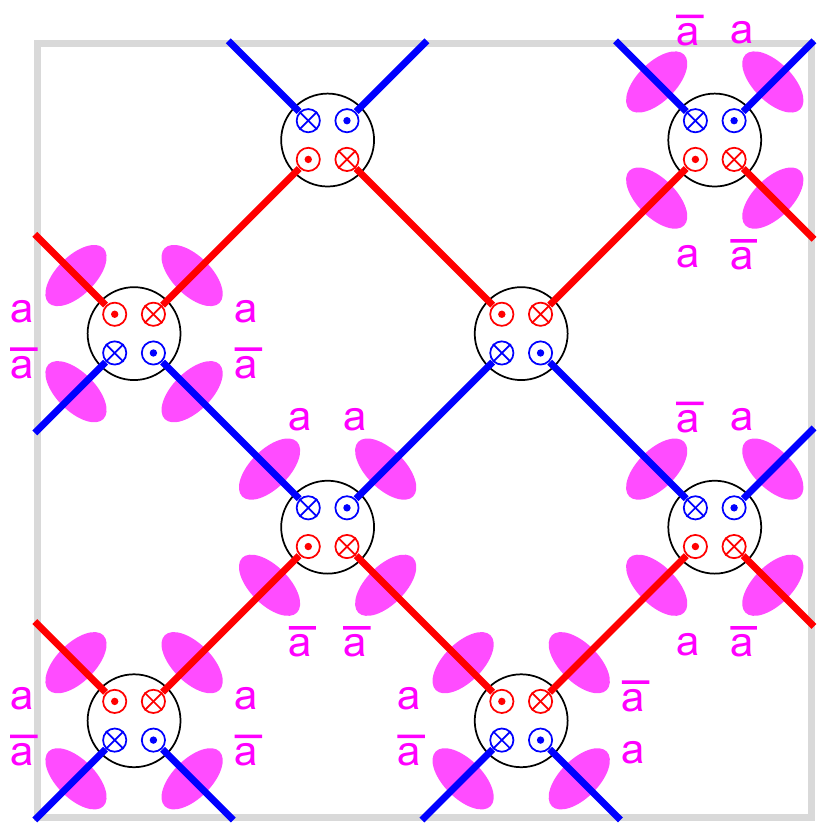} \\ \hline
$7$ & $1$ & \includegraphics[clip,width=0.2\textwidth]{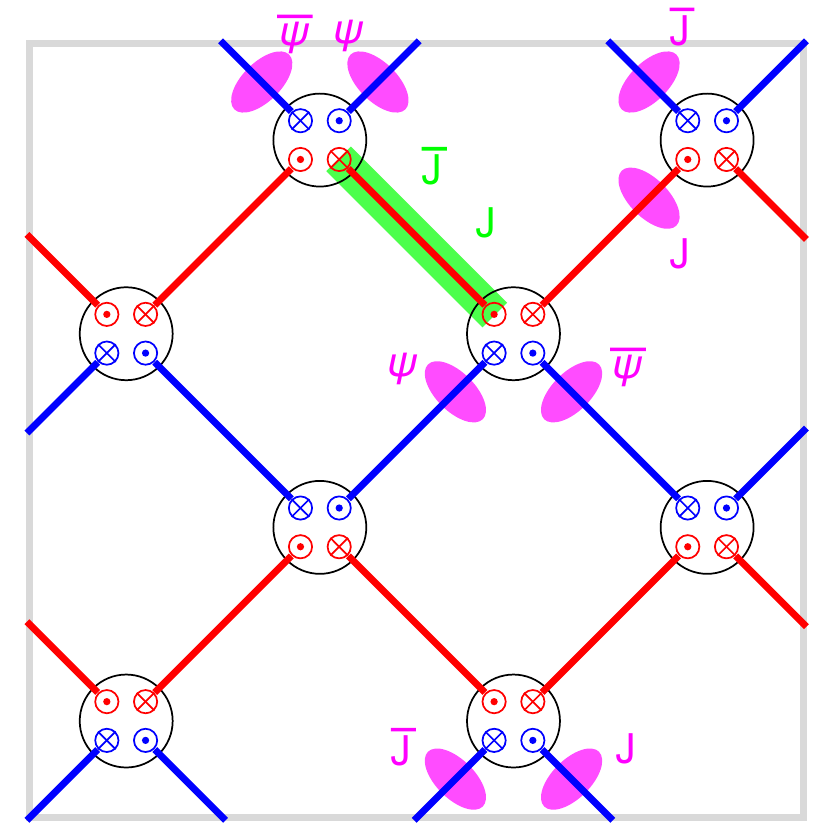} & \includegraphics[clip,width=0.2\textwidth]{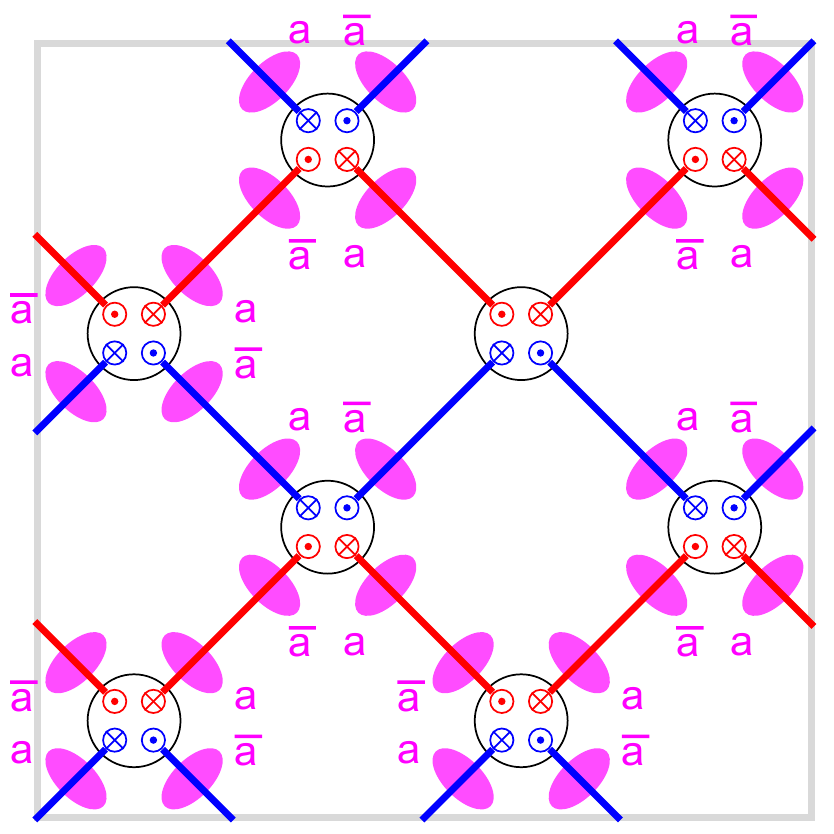} &
$8$ & $2$ & \includegraphics[clip,width=0.2\textwidth]{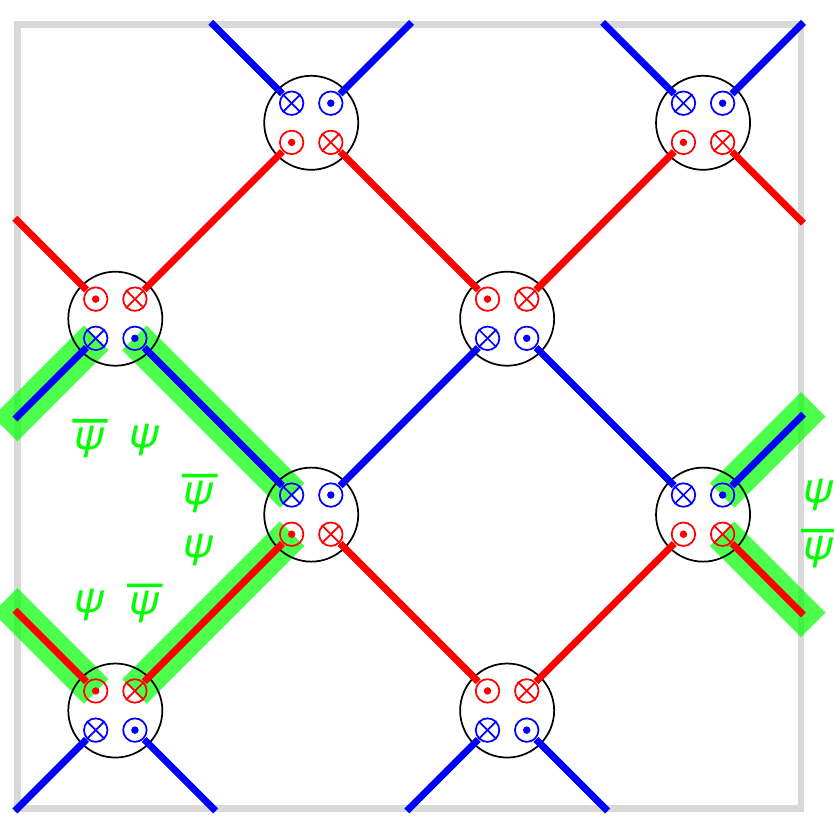} & \includegraphics[clip,width=0.2\textwidth]{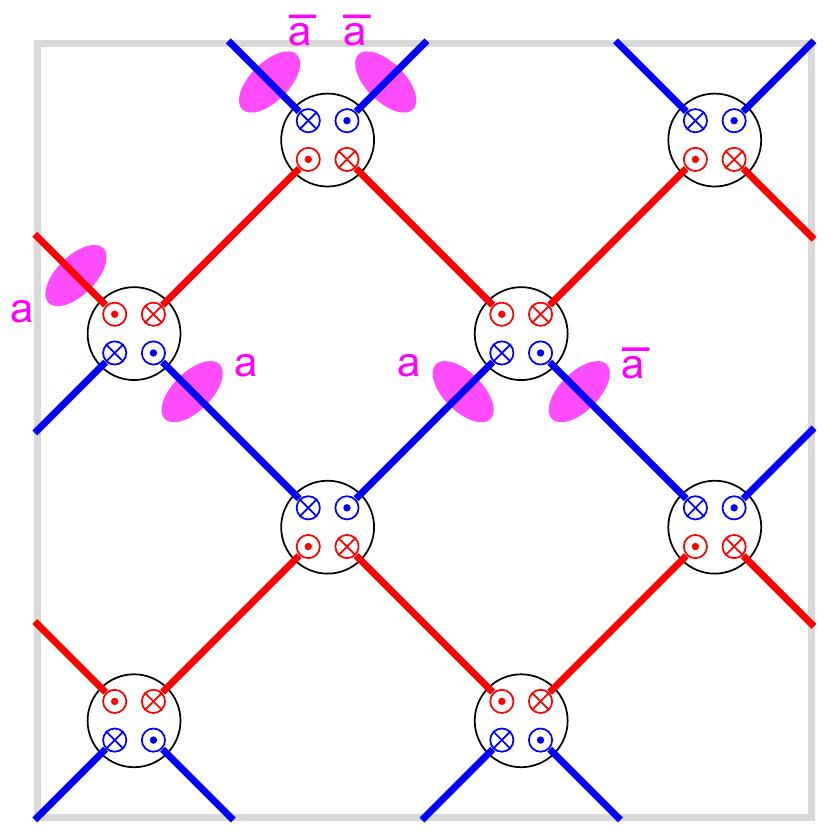} \\ \hline
$9$ & $1$ & \includegraphics[clip,width=0.2\textwidth]{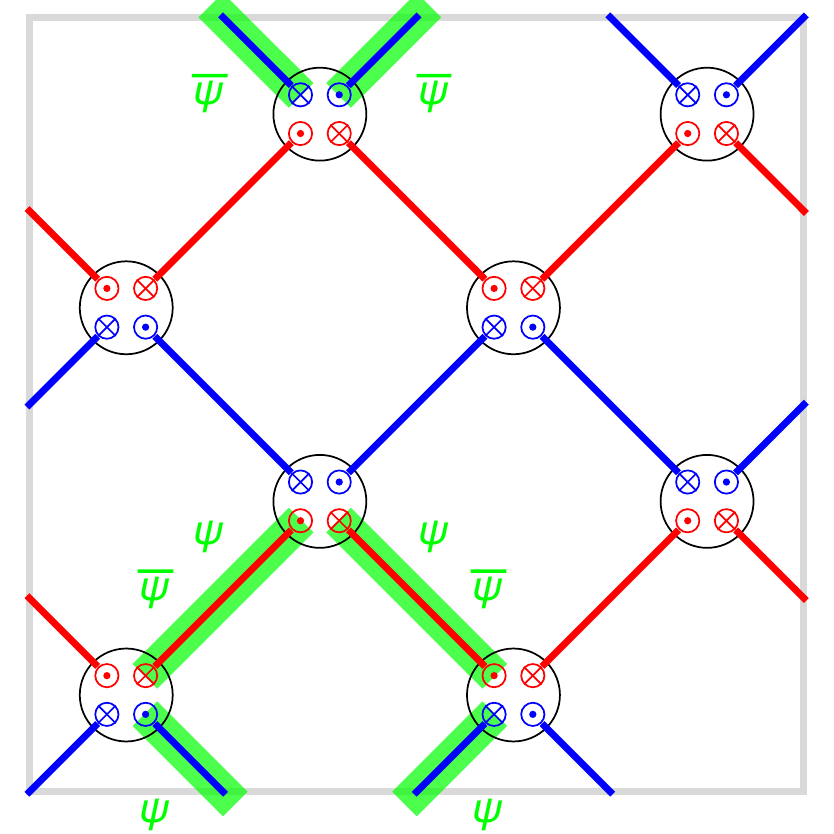} & \includegraphics[clip,width=0.2\textwidth]{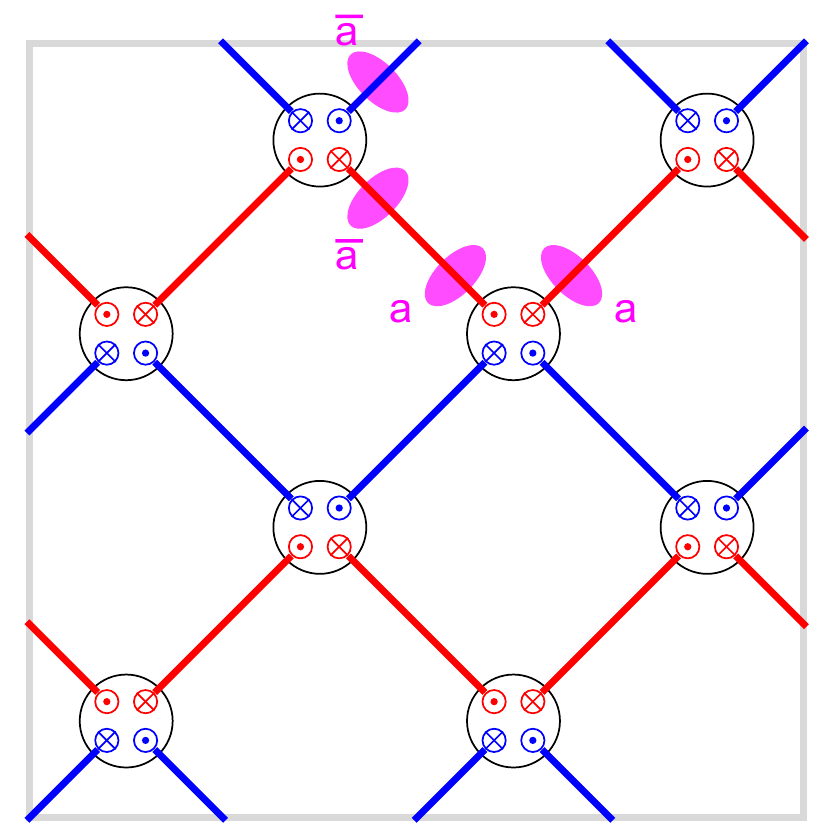} &
$10$ & $2$ & \includegraphics[clip,width=0.2\textwidth]{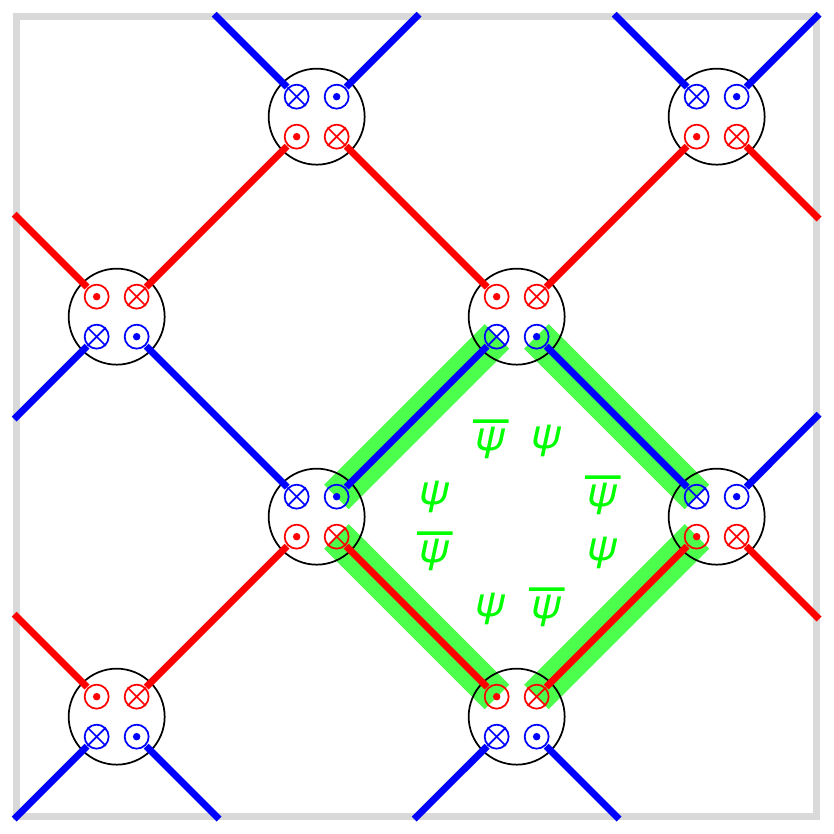} & \includegraphics[clip,width=0.2\textwidth]{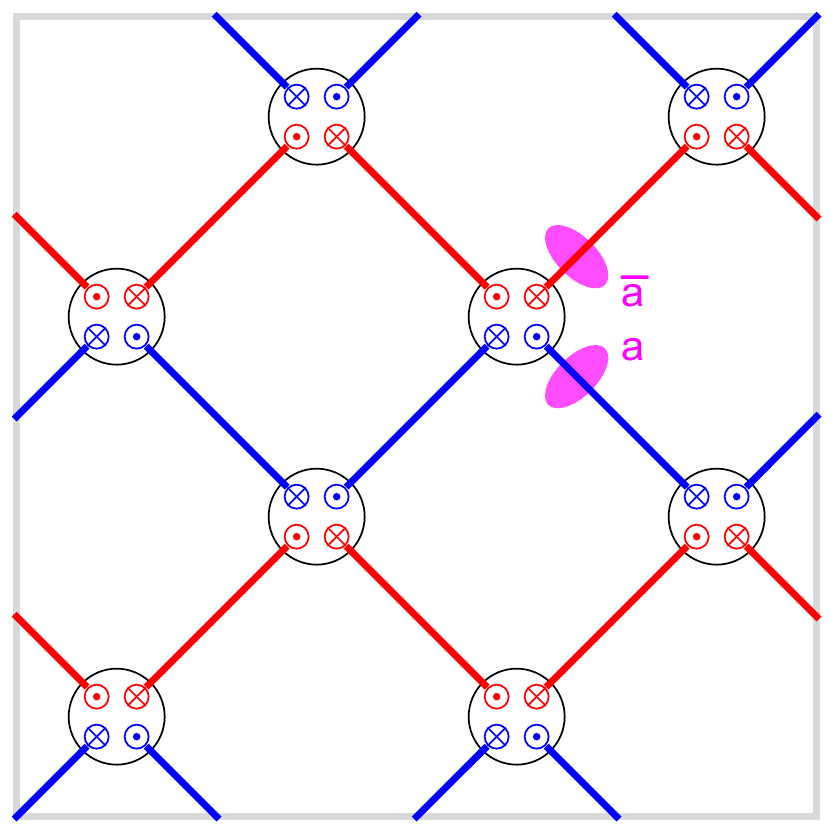} \\ \hline \hline
\end{tabular}
\end{table}
\begin{table}
\caption{(Continued) Graphical representations of the operators spanning the ground-state manifold for $(L_y,L_z)=(2,2)$.}
\label{tab:Ops2x2b}
\begin{tabular}{ll|cc||ll|cc}
\hline \hline
$I$ & $d_I$ & $\exp(iC'_{2I-1}/d_I)$ & $\exp(iC'_{2I}/d_I)$ & $I$ & $d_I$ & $\exp(iC'_{2I-1}/d_I)$ & $\exp(iC'_{2I}/d_I)$ \\ \hline
$11$ & $1$ & \includegraphics[clip,width=0.2\textwidth]{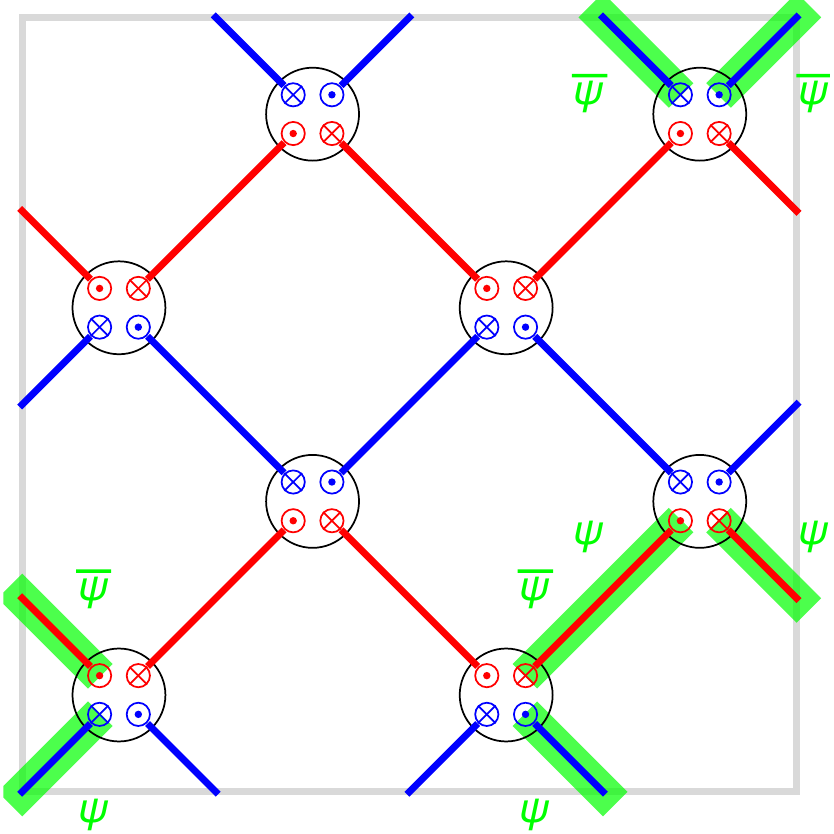} & \includegraphics[clip,width=0.2\textwidth]{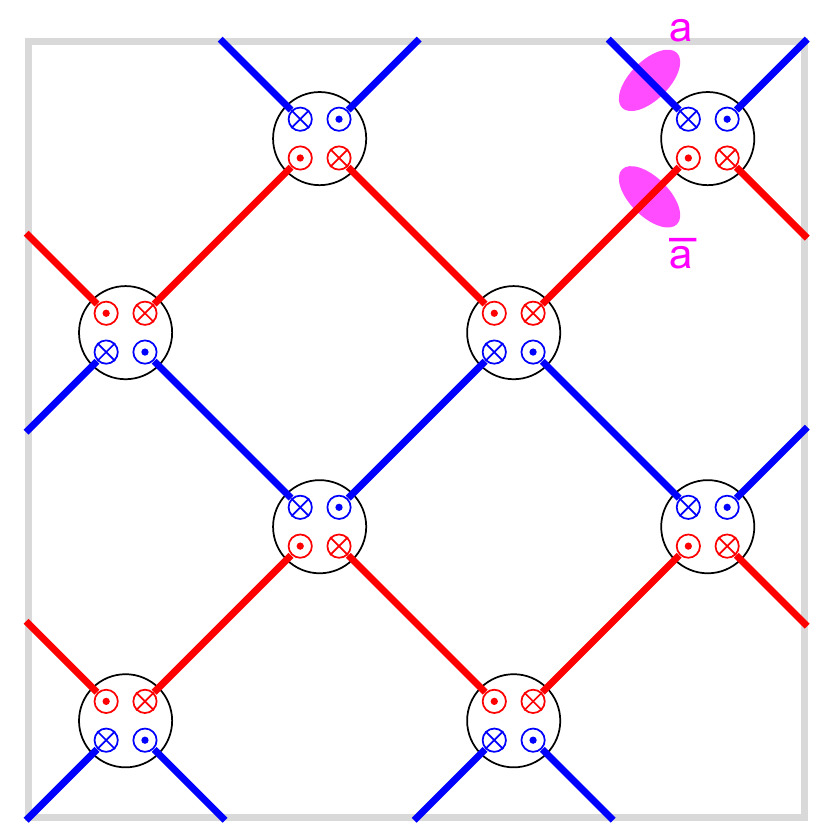} &
$12$ & $2$ & \includegraphics[clip,width=0.2\textwidth]{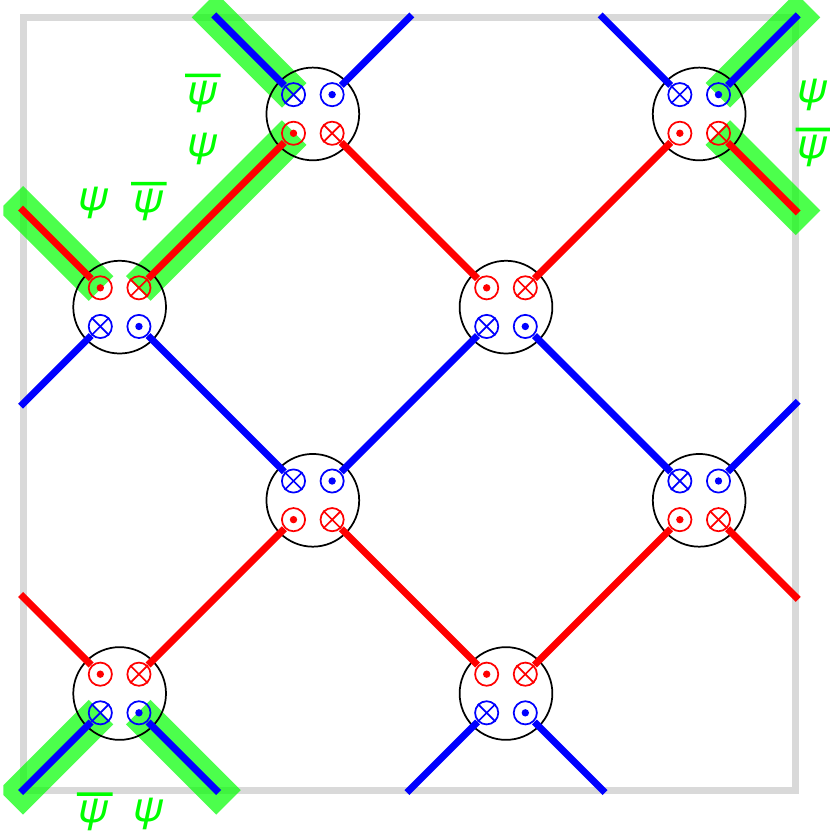} & \includegraphics[clip,width=0.2\textwidth]{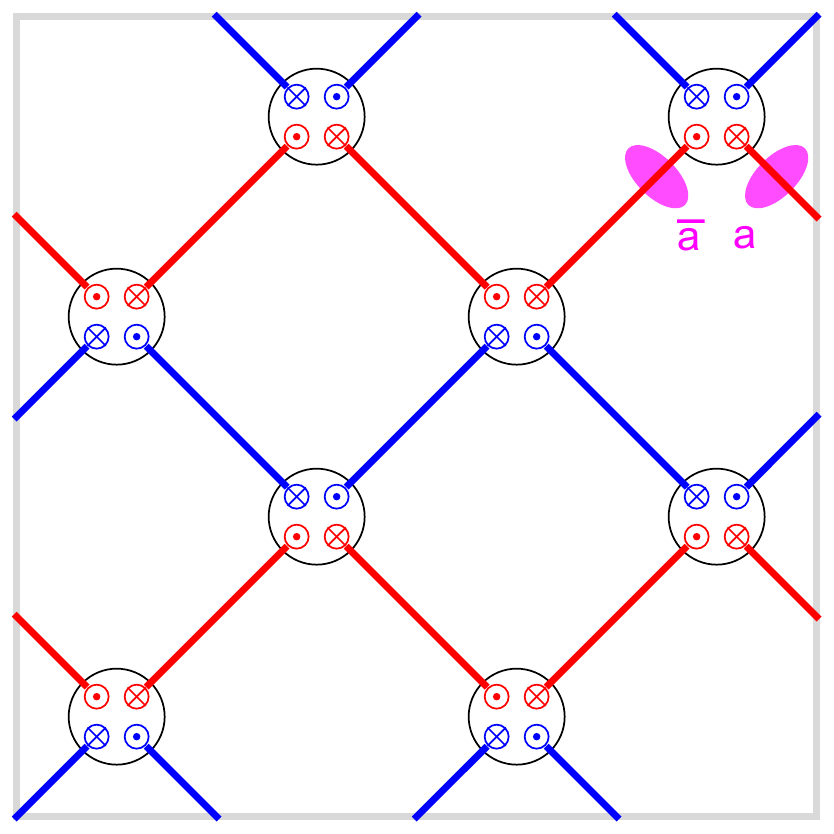} \\ \hline
$13$ & $2$ & \includegraphics[clip,width=0.2\textwidth]{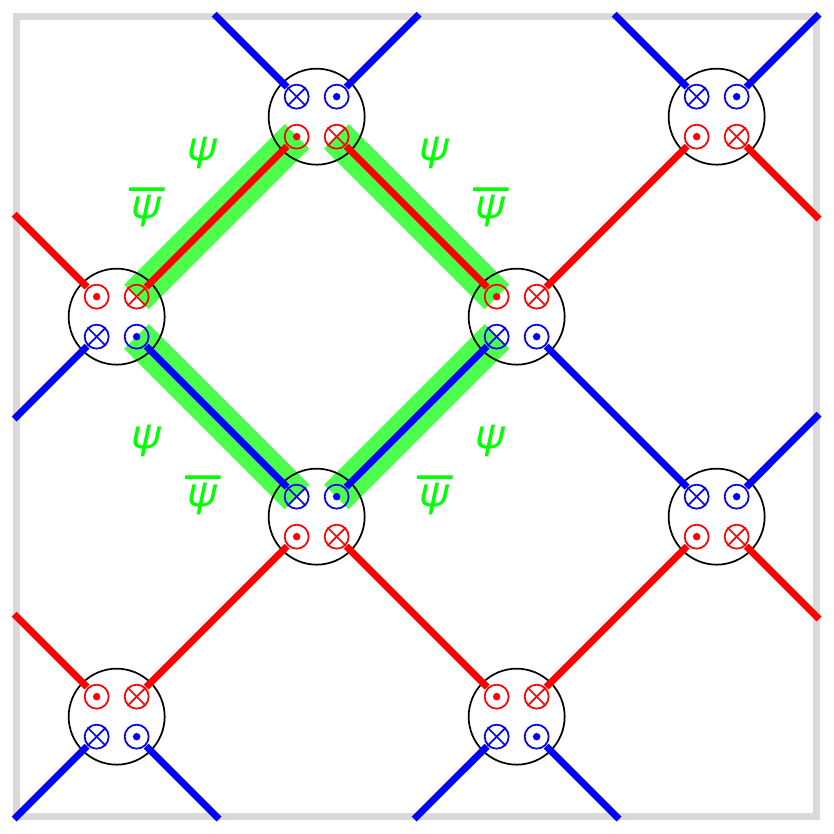} & \includegraphics[clip,width=0.2\textwidth]{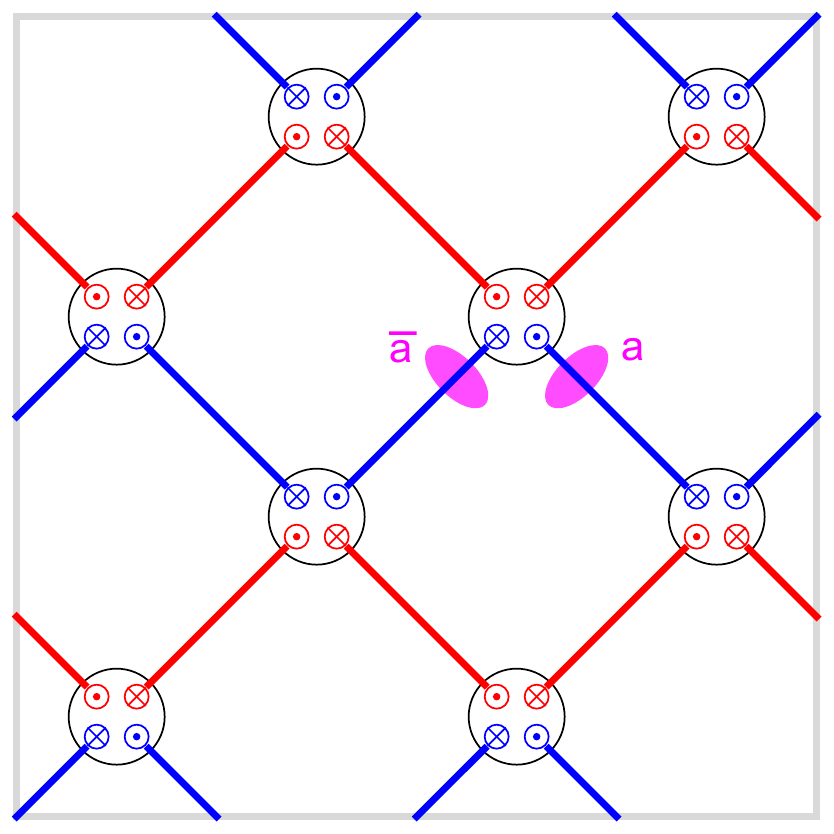} & 
$14$ & $2$ & \includegraphics[clip,width=0.2\textwidth]{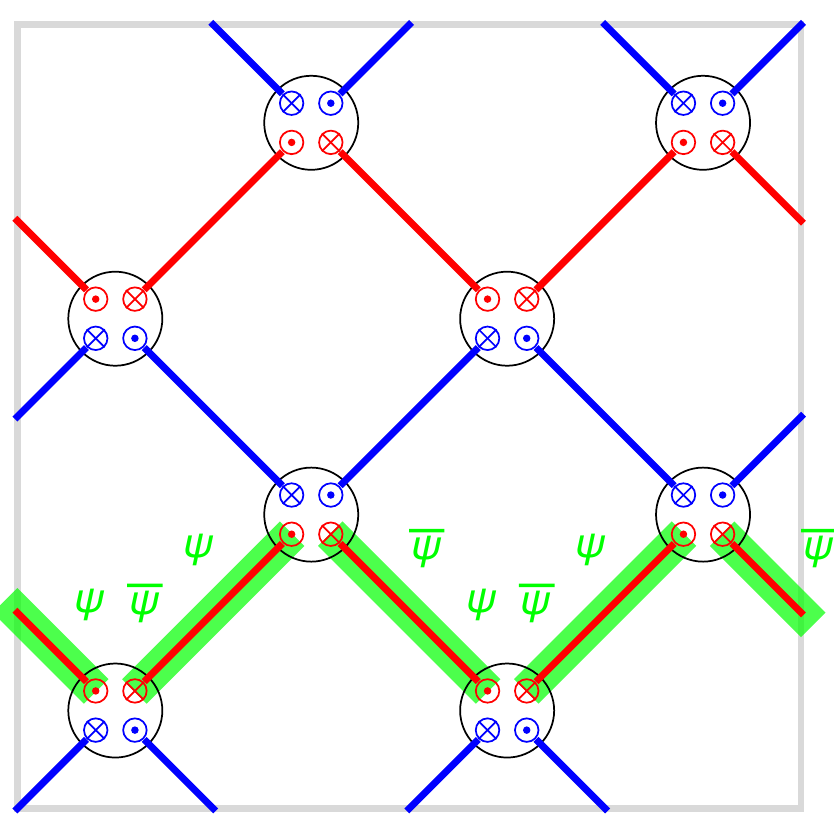} & \includegraphics[clip,width=0.2\textwidth]{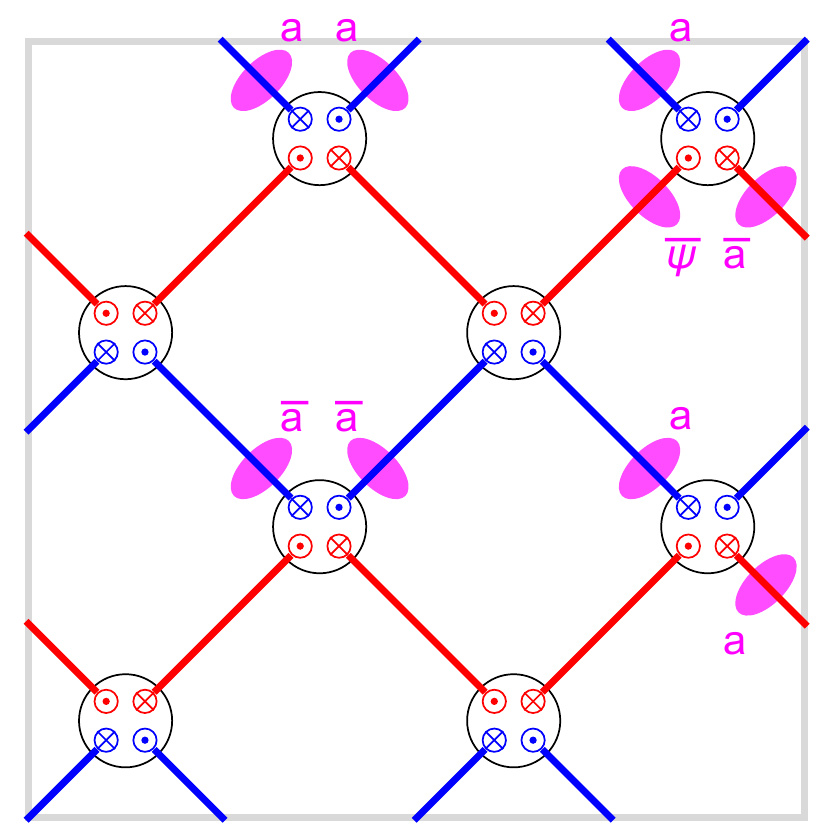} \\ \hline
$15$ & $2$ & \includegraphics[clip,width=0.2\textwidth]{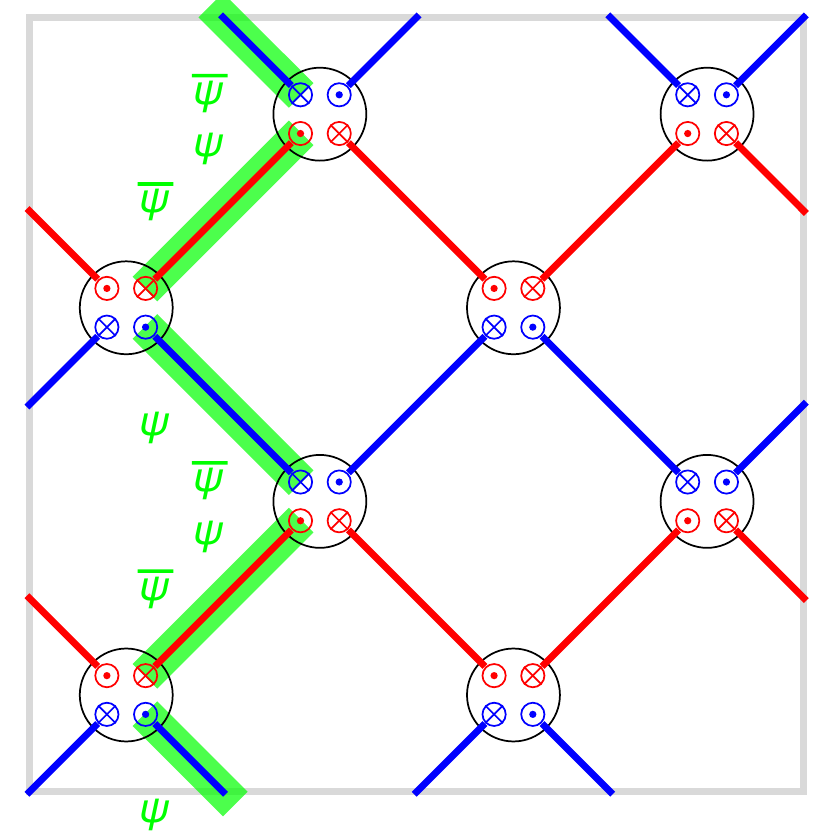} & \includegraphics[clip,width=0.2\textwidth]{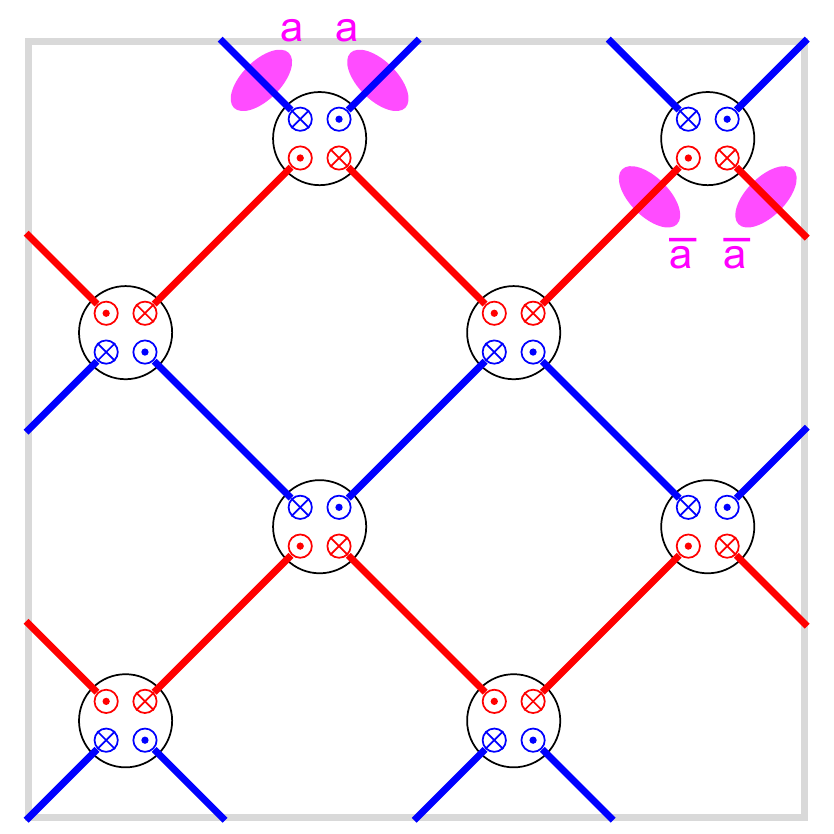} &
$16$ & $4$ & \includegraphics[clip,width=0.2\textwidth]{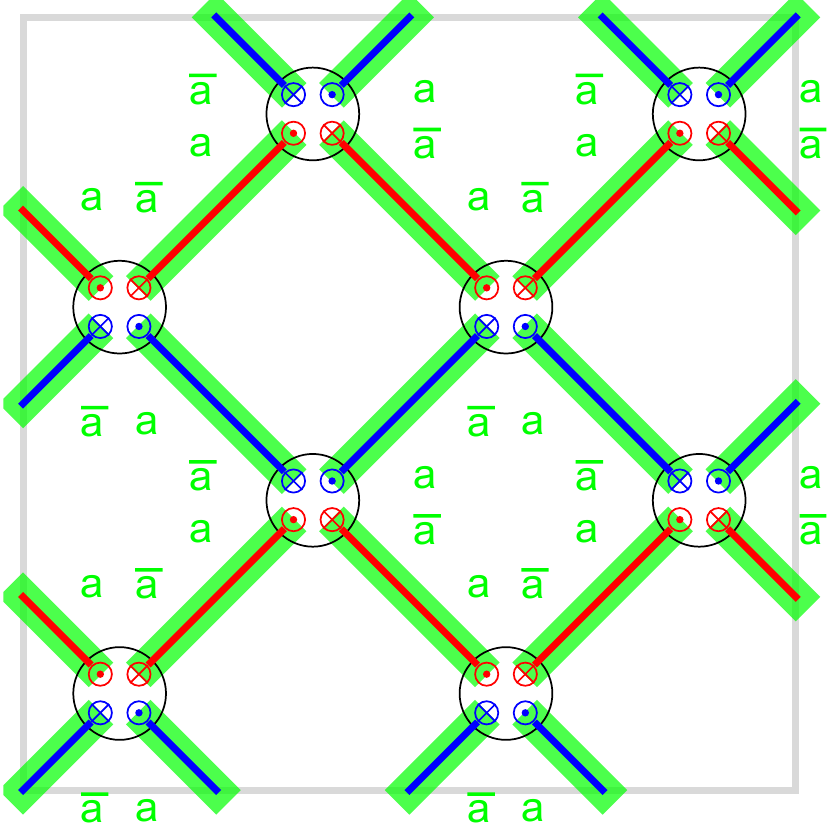} & \includegraphics[clip,width=0.2\textwidth]{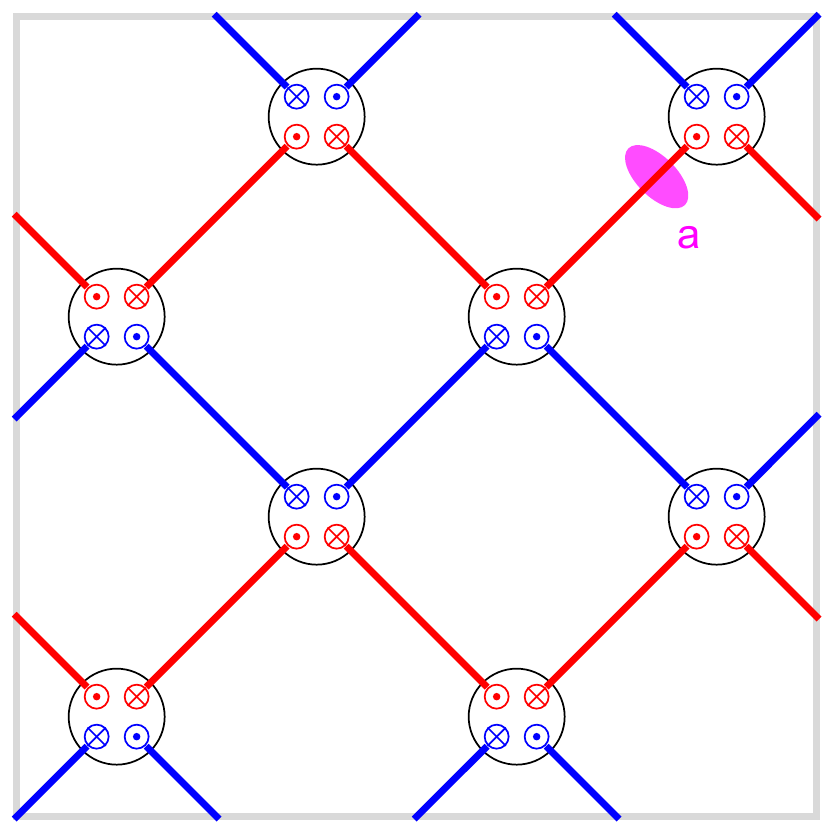} \\ \hline \hline
\end{tabular}
\end{table}
\begin{table}
\caption{Graphical representations of the operators trivially acting on the ground-state manifold for $(L_y,L_z)=(2,2)$.}
\label{tab:Ops2x2c}
\begin{tabular}{l|cccc}
\hline \hline
$I$ & $\exp(iC'_{4I-3})$ & $\exp(iC'_{4I-2})$ & $\exp(iC'_{4I-1})$ & $\exp(iC'_{4I})$ \\ \hline
$9$ & \includegraphics[clip,width=0.2\textwidth]{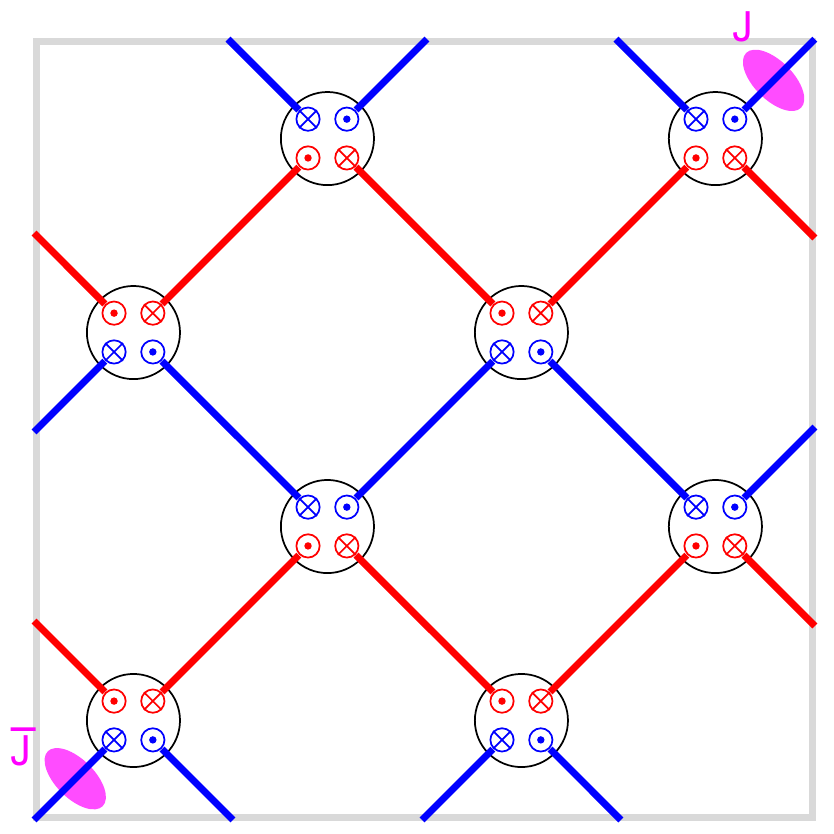} & \includegraphics[clip,width=0.2\textwidth]{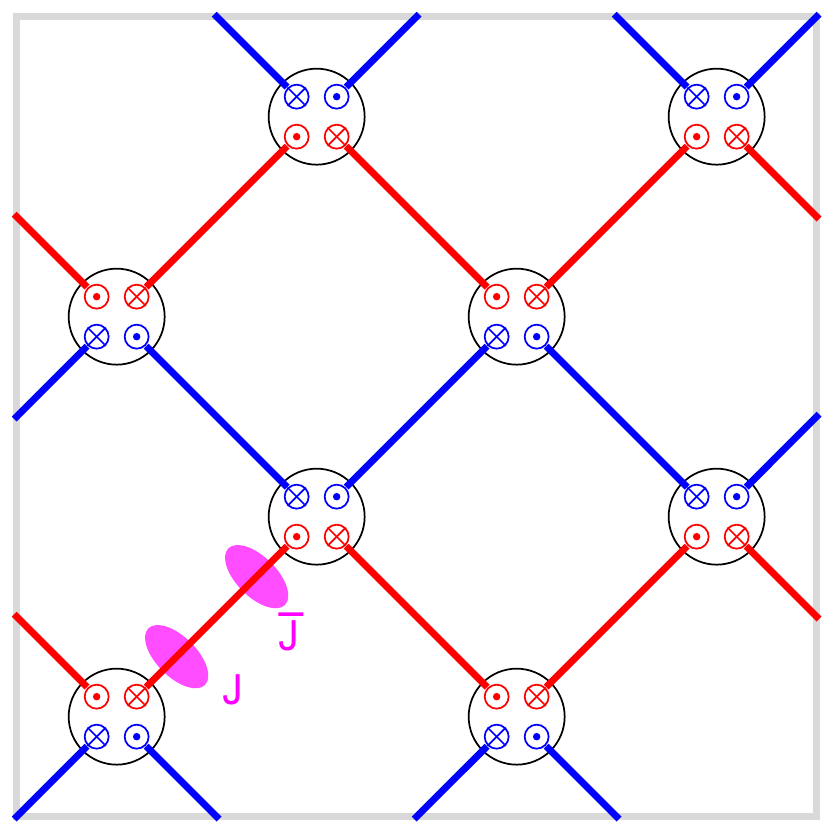} & 
\includegraphics[clip,width=0.2\textwidth]{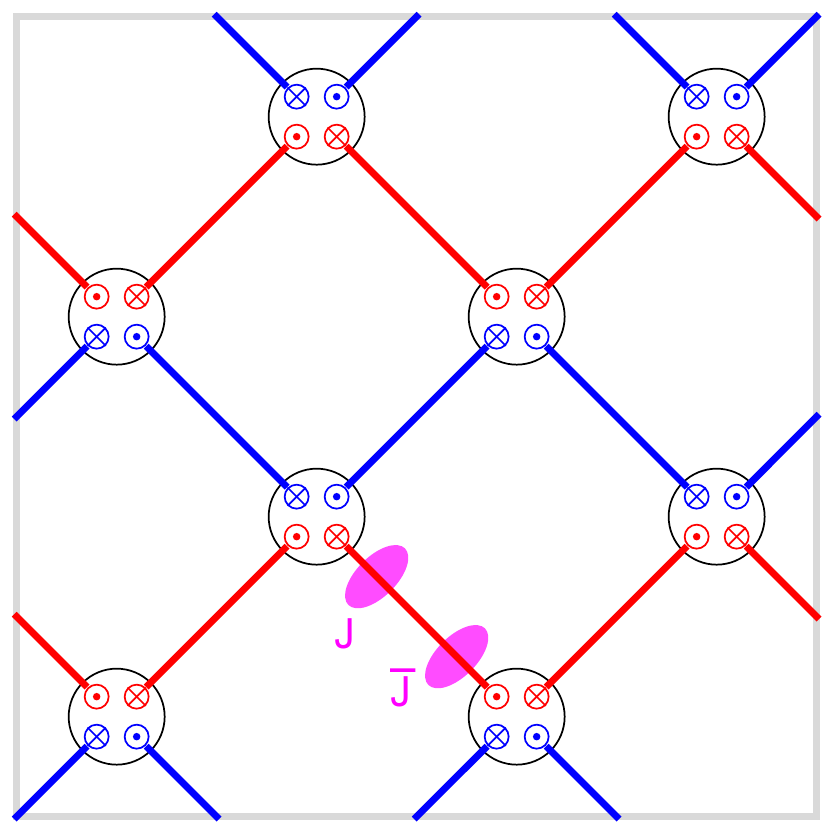} & \includegraphics[clip,width=0.2\textwidth]{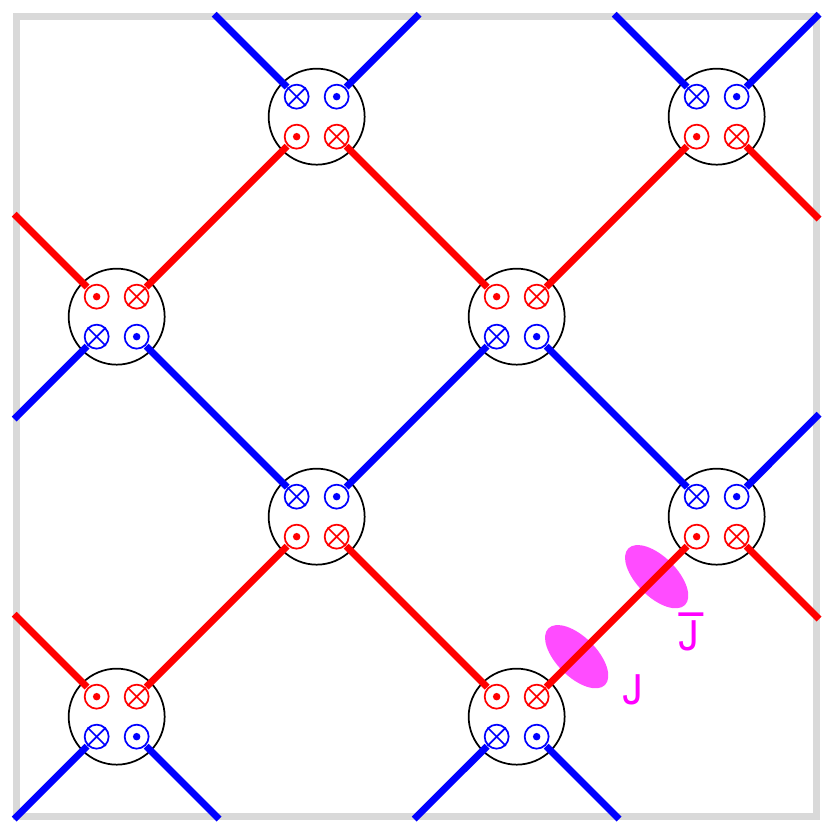} \\ \hline
$10$ & \includegraphics[clip,width=0.2\textwidth]{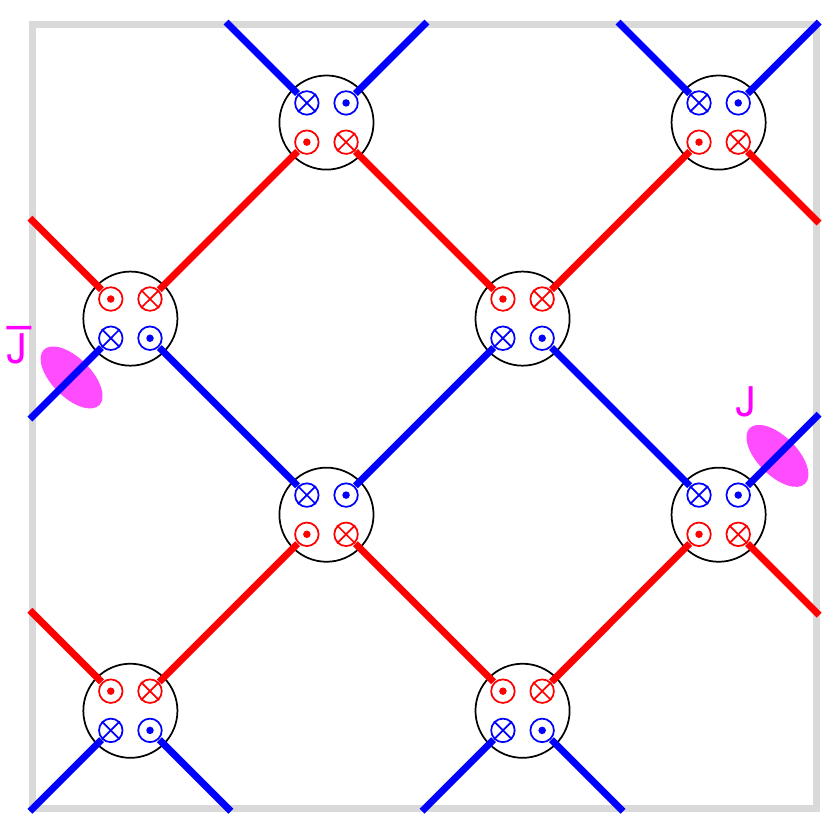} & \includegraphics[clip,width=0.2\textwidth]{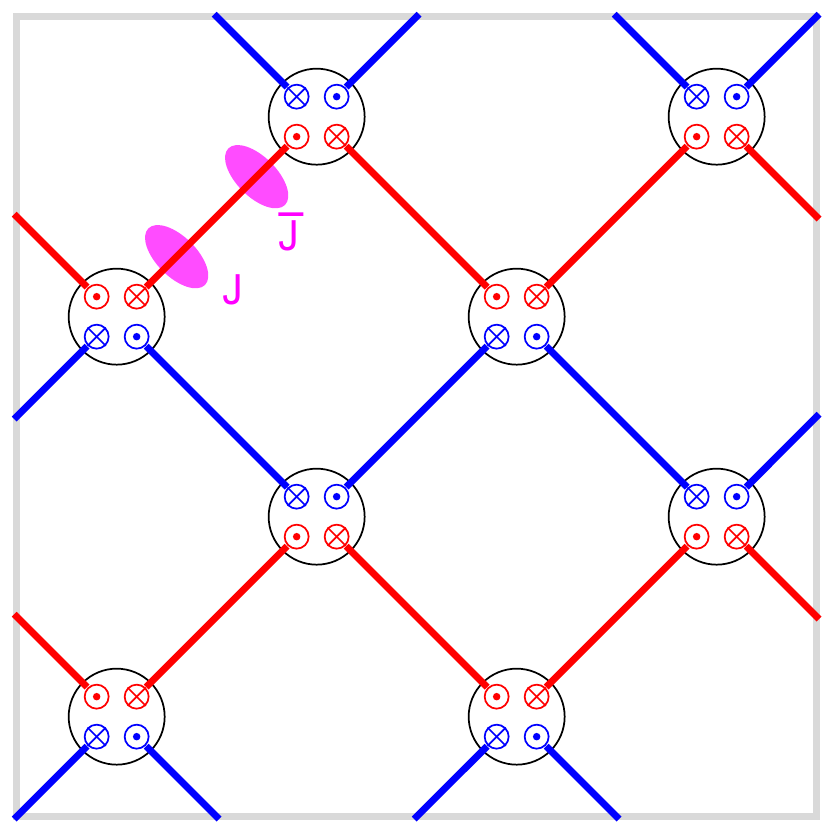} &
\includegraphics[clip,width=0.2\textwidth]{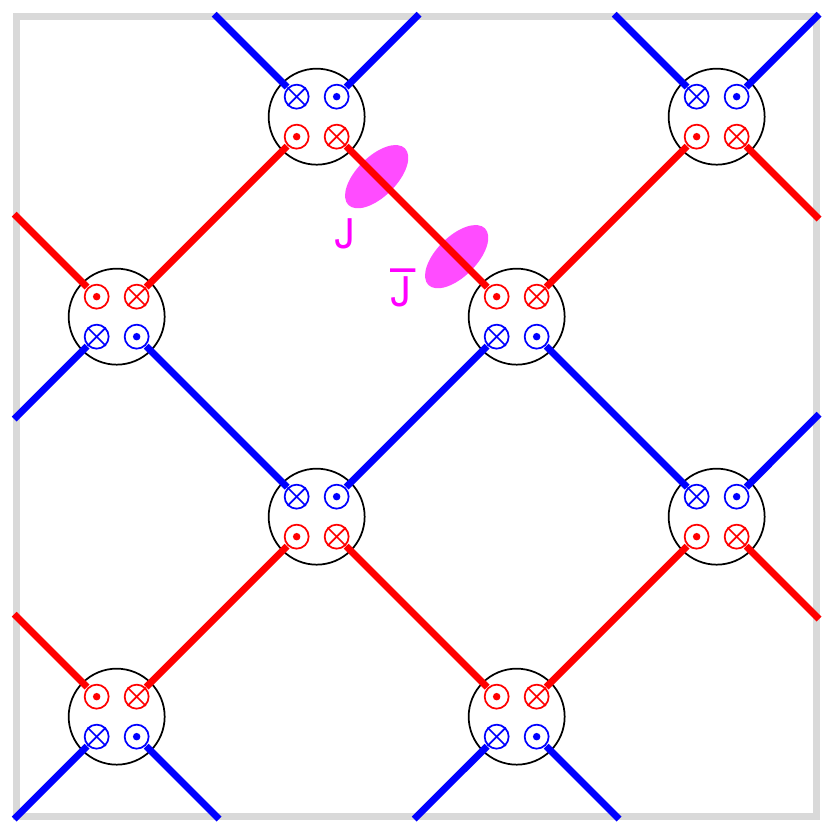} & \includegraphics[clip,width=0.2\textwidth]{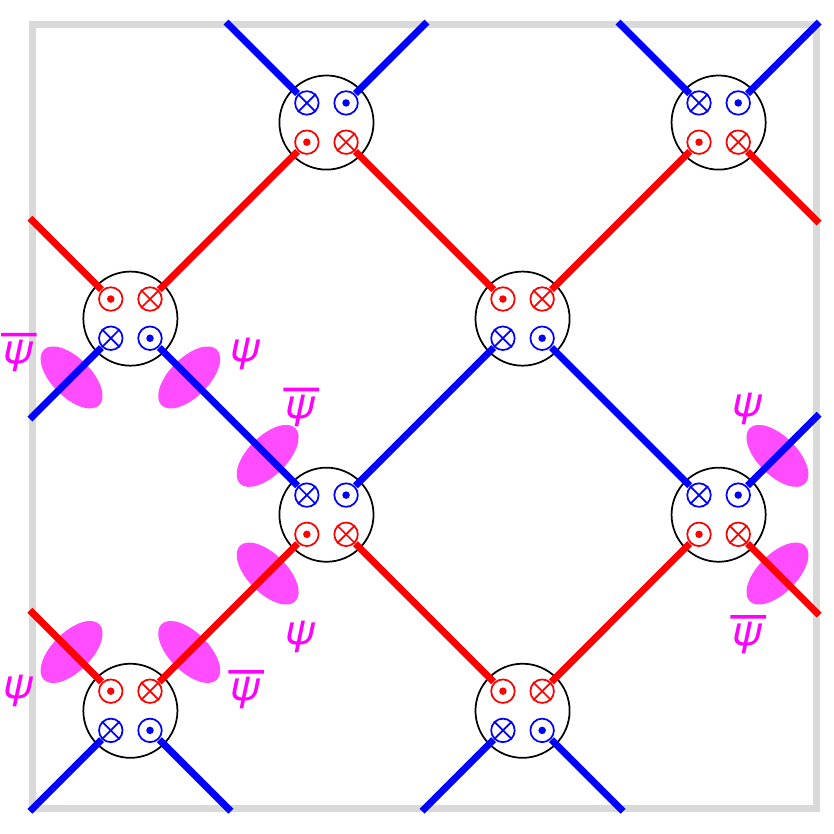} \\ \hline 
$11$ & \includegraphics[clip,width=0.2\textwidth]{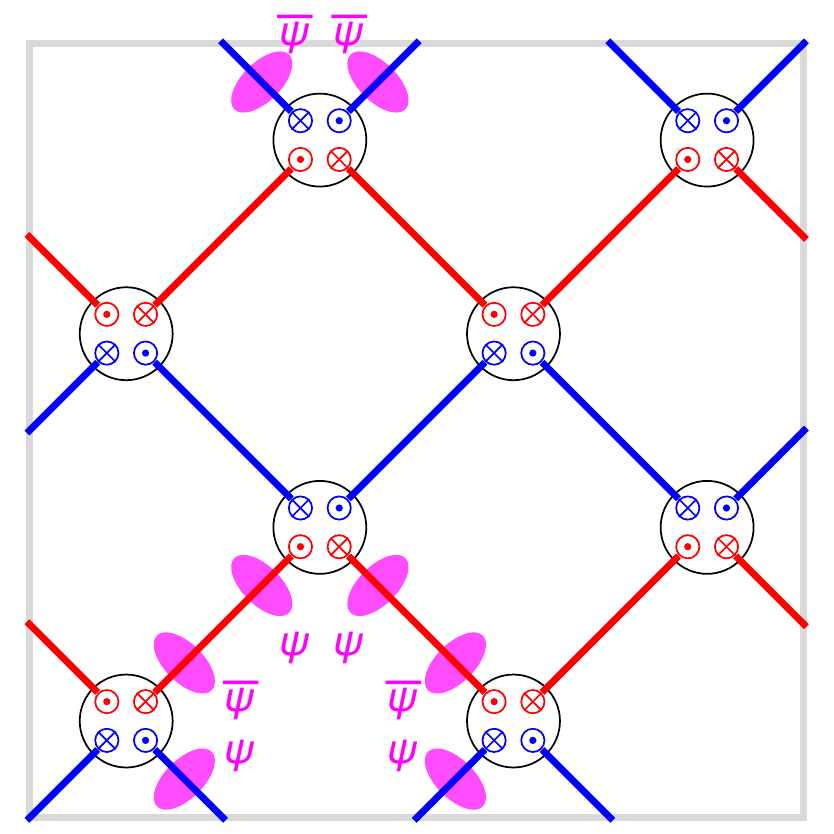} & \includegraphics[clip,width=0.2\textwidth]{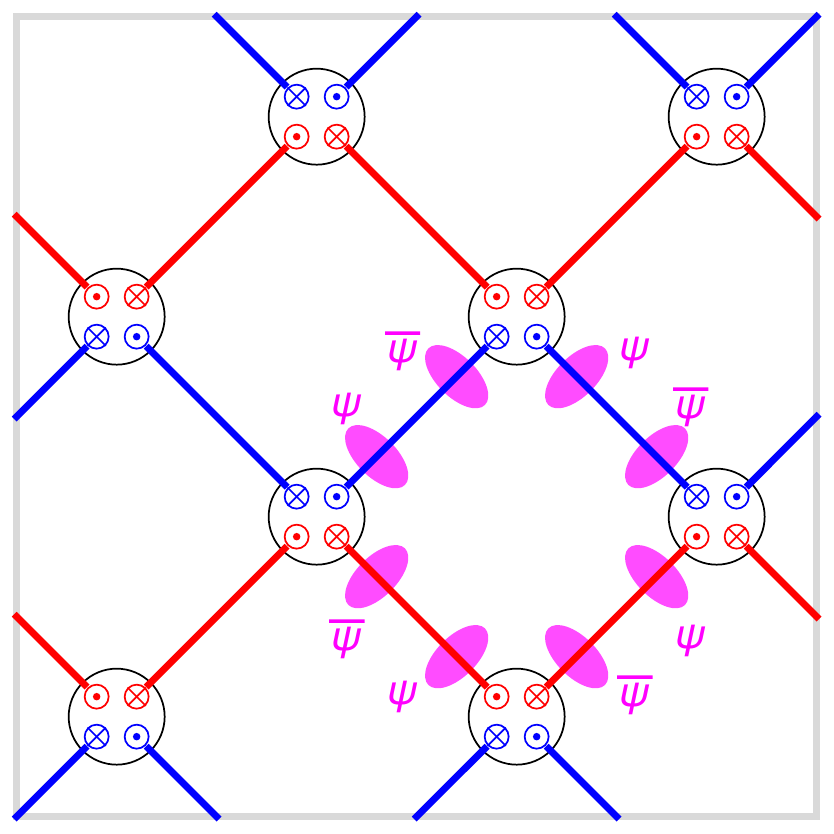} &
\includegraphics[clip,width=0.2\textwidth]{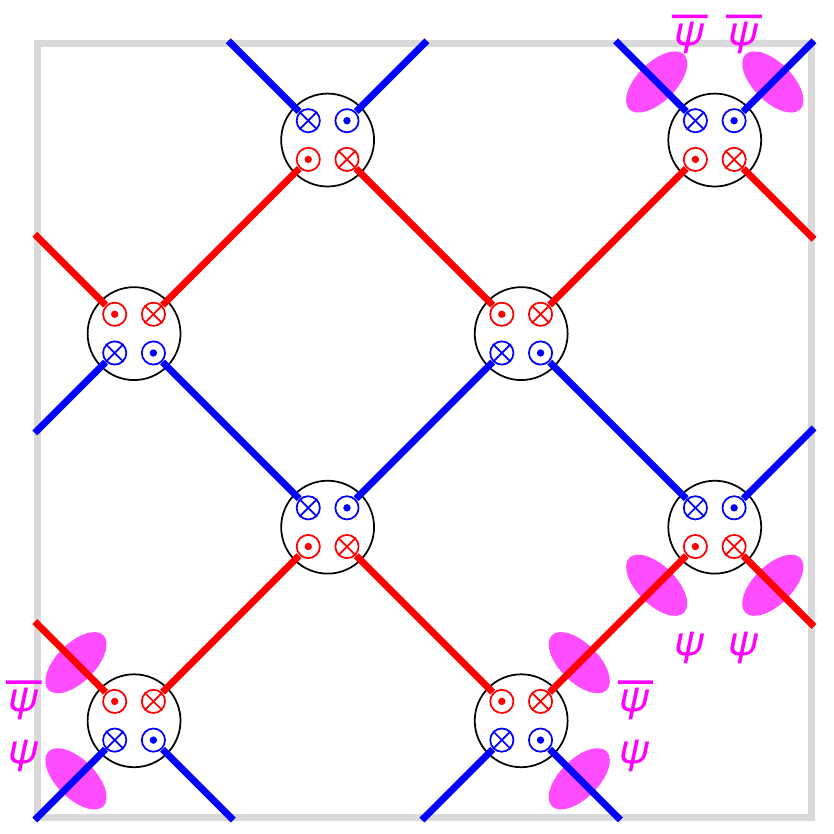} & \includegraphics[clip,width=0.2\textwidth]{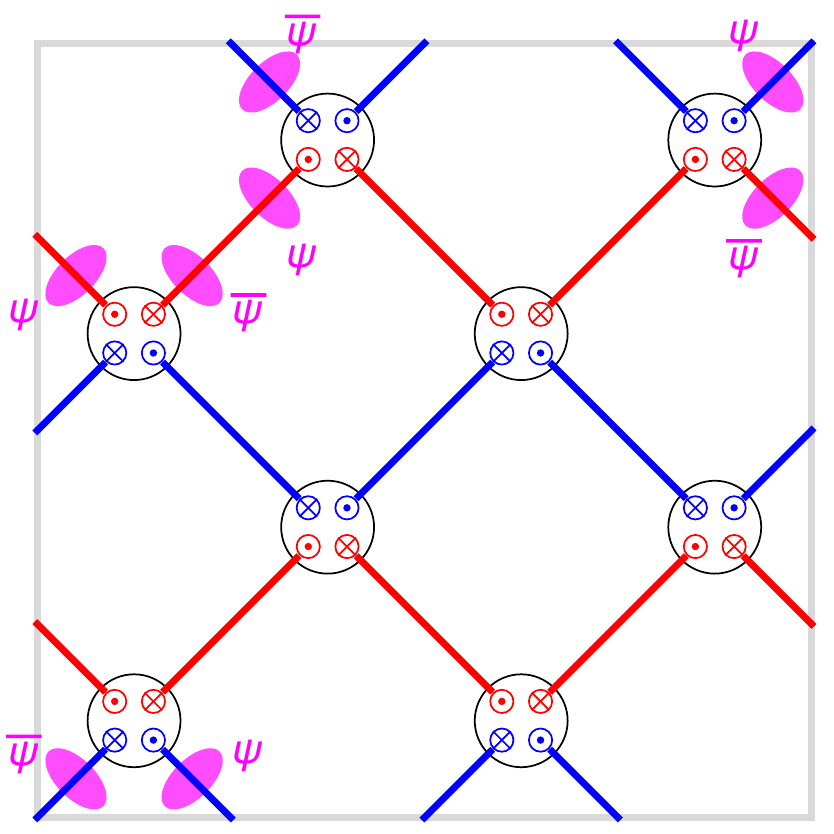} \\ \hline
$12$ & \includegraphics[clip,width=0.2\textwidth]{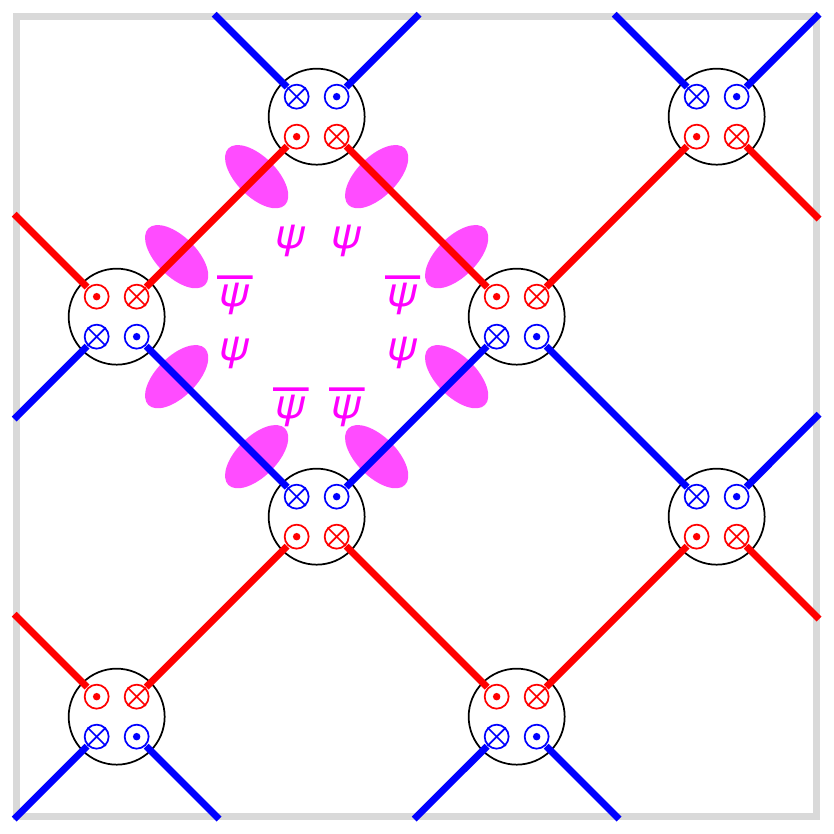} & \includegraphics[clip,width=0.2\textwidth]{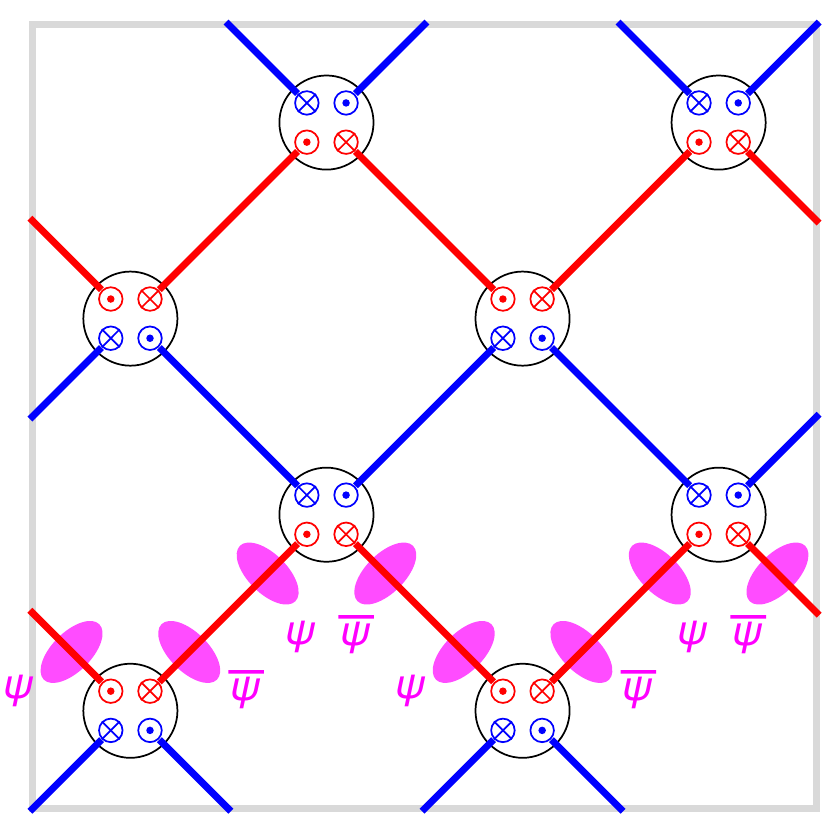} &
\includegraphics[clip,width=0.2\textwidth]{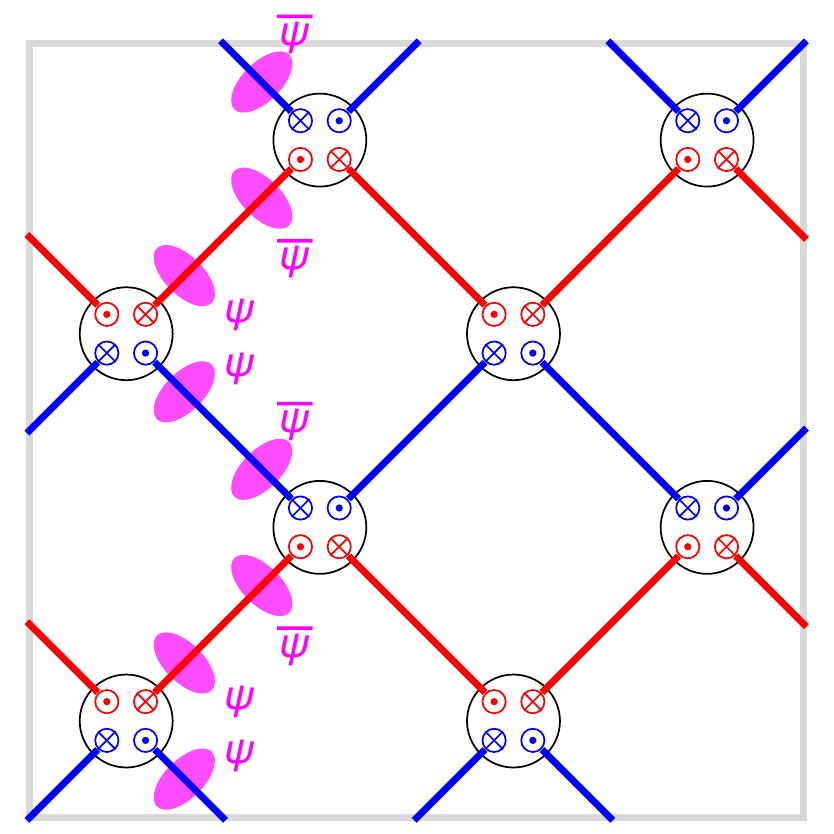} & \includegraphics[clip,width=0.2\textwidth]{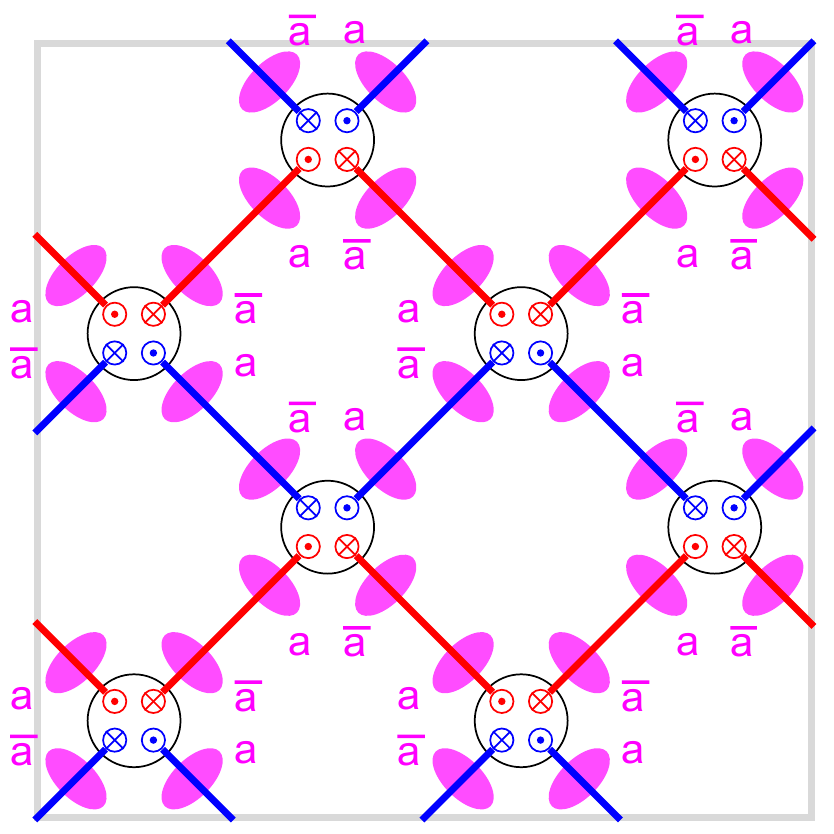} \\ \hline \hline
\end{tabular}
\end{table}

In the main text, we define the operators nontrivially acting on the ground-state manifold by 
\begin{align}
\begin{split}
X_1 &= X'_{4L_yL_z}, \\
X_2 &= X'_{4L_yL_z-2}, \\
X_3 &= X'_{4L_yL_z-1}, \\
X_I &= X'_{2L_yL_z+I-4} \ \ \ (4 \leq I \leq 2L_yL_z+1)
\end{split}
\end{align}
and $Z_I$ correspondingly. 
An important observation is that these operators $X_I$ consist of the link fields only and do not involve string operators along the $x$ axis, which are exponentials of $P^l_\bfr$ or $Q^l_\bfr$. 
The operators $\{ X_4, X_5, \cdots, X_{2L_yL_z+1} \}$ create closed fermion strings on $2L_y L_z-2$ square plaquettes in the $yz$ plane. 
One can also construct the operators that create closed fermion strings in the remaining two square plaquettes, 
\begin{align}
X_{2L_yL_z+2} &= 2\ttheta_{(L_y-1,L_z-1),\bfdelta_y+\bfdelta_z} +2\ttheta_{(L_y-1,L_z-1),-\bfdelta_y+\bfdelta_z} +2\ttheta_{(L_y-1,1),\bfdelta_y-\bfdelta_z} +2\ttheta_{(L_y-1,1),-\bfdelta_y-\bfdelta_z},\\
X_{2L_yL_z+3} &= 2\ttheta_{(L_y-1,L_z-1),\bfdelta_y+\bfdelta_z} +2\ttheta_{(L_y-1,L_z-1),\bfdelta_y-\bfdelta_z} -2\ttheta_{(1,L_z-1),-\bfdelta_y-\bfdelta_z} -2\ttheta_{(1,L_z-1),-\bfdelta_y+\bfdelta_z},
\end{align}
from appropriate products of $X_I$ with $1 \leq I \leq 2L_yL_z+1$ that involve $X^2_1$. 
The $2L_yL_z$ operators $\calS_p \in \{ X_4, X_5, \cdots, X_{2L_yL_z+3} \}$, each of which creates a closed fermion string on the plaquette labeled by $p$, are commuting with each other and strictly local operators. 
Thus, an effective Hamiltonian that consists of the sum of all $\calS_p$'s and is defined in the Hilbert space of the $4 \cdot 2^{2L_yL_z}$-fold degenerate ground states can induce a condensation of closed fermion strings. 

\subsection{Some conformal embeddings and branching rules}

We here show the branching rules for $SU(pq)_1 \supset SU(p)_q \times SU(q)_p$ discussed in the main text. 
They can be obtained from the branching rules for the finite Lie group $SU(pq) \supset SU(p) \times SU(q)$ and by applying to them elements of the outer automorphism group of the affine Lie group \cite{Walton89,dFMS}. 
In this case, it is also known that the branching index can only be $0$ or $1$ \cite{Altschuler90}. 

\subsubsection{$SU(4)_1 \supset SU(2)_2 \times SU(2)_2$}
$SU(4)_1$ has four primary fields corresponding to the Young tableau with a single column. 
We label them by the dimensions of the representations and their conformal weights are given by 
\begin{align}
\begin{array}{l|cccc}
\textrm{Rep.} & \mathbf{1} & \mathbf{4} & \overline{\mathbf{4}} & \mathbf{6} \\ \hline
h & 0 & \frac{3}{8} & \frac{3}{8} & \frac{1}{2}
\end{array}
\end{align}
For $SU(2)_2$, there are three primary fields:
\begin{align}
\begin{array}{l|ccc}
\textrm{Rep.} & \mathbf{1} & \mathbf{2} & \mathbf{3} \\ \hline
h & 0 & \frac{3}{16} & \frac{1}{2}
\end{array}
\end{align}
The branching rule is then given by 
\begin{align}
\begin{split}
\mathbf{1} &\mapsto (\mathbf{1}_1 \otimes \mathbf{1}_2) \oplus (\mathbf{3}_1 \otimes \mathbf{3}_2), \\
\mathbf{4} &\mapsto \mathbf{2}_1 \otimes \mathbf{2}_2, \\
\overline{\mathbf{4}} &\mapsto \mathbf{2}_1 \otimes \mathbf{2}_2, \\
\mathbf{6} &\mapsto (\mathbf{1}_1 \otimes \mathbf{3}_2) \oplus (\mathbf{3}_1 \otimes \mathbf{1}_2).
\end{split}
\end{align}

\subsubsection{$SU(8)_1 \supset SU(2)_4 \times SU(4)_2$}
For $SU(8)_1$, there are eight primary fields,
\begin{align}
\begin{array}{l|cccccccc}
\textrm{Rep.} & \mathbf{1} & \mathbf{8} & \overline{\mathbf{8}} & \mathbf{28} & \overline{\mathbf{28}} & \mathbf{56} & \overline{\mathbf{56}} & \mathbf{70} \\ \hline
h & 0 & \frac{7}{16} & \frac{7}{16} & \frac{3}{4} & \frac{3}{4} & \frac{15}{16} & \frac{15}{16} & 1 
\end{array}
\end{align}
For $SU(2)_4$, there are five primary fields,
\begin{align}
\begin{array}{l|ccccc}
\textrm{Rep.} & \mathbf{1} & \mathbf{2} & \mathbf{3} & \mathbf{4} & \mathbf{5} \\ \hline
h & 0 & \frac{1}{8} & \frac{1}{3} & \frac{5}{8} & 1
\end{array}
\end{align}
For $SU(4)_2$, there are ten primary fields,
\begin{align}
\begin{array}{l|cccccccccc}
\textrm{Rep.} & \mathbf{1} & \mathbf{4} & \overline{\mathbf{4}} & \mathbf{6} & \mathbf{10} & \overline{\mathbf{10}} & \mathbf{15} & \mathbf{20}^a & \mathbf{20}^b & \overline{\mathbf{20}}^b \\ \hline
h & 0 & \frac{5}{16} & \frac{5}{16} & \frac{5}{12} & \frac{3}{4} & \frac{3}{4} & \frac{2}{3} & 1 & \frac{13}{16} & \frac{13}{16}
\end{array}
\end{align}
where $\mathbf{20}^a$ denote the self-conjugate representation with four boxes and two columns in the Young tableau, while $\mathbf{20}^b$ denote the representation with five boxes and two columns and is conjugate to that with three boxes.
The branching rule is given by 
\begin{align}
\begin{split}
\mathbf{1} &\mapsto (\mathbf{1}_1 \otimes \mathbf{1}_2) \oplus (\mathbf{3}_1 \otimes \mathbf{15}_2) \oplus (\mathbf{5}_1 \otimes \mathbf{20}^a_2), \\
\mathbf{8} &\mapsto (\mathbf{2}_1 \otimes \mathbf{4}_2) \oplus (\mathbf{4}_1 \otimes \mathbf{20}^b_2), \\
\overline{\mathbf{8}} &\mapsto (\mathbf{2}_1 \otimes \overline{\mathbf{4}}_2) \oplus (\mathbf{4}_1 \otimes \overline{\mathbf{20}}^b_2), \\
\mathbf{28} &\mapsto (\mathbf{3}_1 \otimes \mathbf{6}_2) \oplus (\mathbf{1}_1 \otimes \mathbf{10}_2) \oplus (\mathbf{5}_1 \otimes \overline{\mathbf{10}}_2), \\
\overline{\mathbf{28}} &\mapsto (\mathbf{3}_1 \otimes \mathbf{6}_2) \oplus (\mathbf{1}_1 \otimes \overline{\mathbf{10}}_2) \oplus (\mathbf{5}_1 \otimes \mathbf{10}_2), \\
\mathbf{56} &\mapsto (\mathbf{4}_1 \otimes \overline{\mathbf{4}}_2) \oplus (\mathbf{2}_1 \otimes \overline{\mathbf{20}}^b), \\
\overline{\mathbf{56}} &\mapsto (\mathbf{4}_1 \otimes \mathbf{4}_2) \oplus (\mathbf{2}_1 \otimes \mathbf{20}^b_2), \\
\mathbf{70} &\mapsto (\mathbf{1}_1 \otimes \mathbf{20}^a_2) \oplus (\mathbf{5}_1 \otimes \mathbf{1}_2) \oplus (\mathbf{3}_1 \otimes \mathbf{15}_2).
\end{split}
\end{align}

\subsubsection{$SU(9)_1 \supset SU(3)_3 \times SU(3)_3$}

For $SU(9)_1$, there are nine primary fields:
\begin{align}
\begin{array}{l|ccccccccc}
\textrm{Rep.} & \mathbf{1} & \mathbf{9} & \overline{\mathbf{9}} & \mathbf{36} & \overline{\mathbf{36}} & \mathbf{84} & \overline{\mathbf{84}} & \mathbf{126} & \overline{\mathbf{126}} \\ \hline
h & 0 & \frac{4}{9} & \frac{4}{9} & \frac{7}{9} & \frac{7}{9} & 1 & 1 & \frac{10}{9} & \frac{10}{9} 
\end{array}
\end{align}
For $SU(3)_3$, there are ten primary fields:
\begin{align}
\begin{array}{l|cccccccccc}
\textrm{Rep.} & \mathbf{1} & \mathbf{3} & \overline{\mathbf{3}} & \mathbf{6} & \overline{\mathbf{6}} & \mathbf{8} & \mathbf{10} & \overline{\mathbf{10}} & \mathbf{15} & \overline{\mathbf{15}} \\ \hline
h & 0 & \frac{2}{9} & \frac{2}{9} & \frac{5}{9} & \frac{5}{9} & \frac{1}{2} & 1 & 1 & \frac{8}{9} & \frac{8}{9}
\end{array}
\end{align}
The branching rule is given by 
\begin{align}
\begin{split}
\mathbf{1} &\mapsto (\mathbf{1}_1 \otimes \mathbf{1}_2) \oplus (\mathbf{8}_1 \otimes \mathbf{8}_2) \oplus (\mathbf{10}_1 \otimes \overline{\mathbf{10}}_2) \oplus (\overline{\mathbf{10}}_1 \otimes \mathbf{10}_2), \\
\mathbf{9} &\mapsto (\mathbf{3}_1 \otimes \mathbf{3}_2) \oplus (\overline{\mathbf{6}}_1 \otimes \mathbf{15}_2) \oplus (\mathbf{15}_1 \otimes \overline{\mathbf{6}}_2), \\
\overline{\mathbf{9}} &\mapsto (\overline{\mathbf{3}}_1 \otimes \overline{\mathbf{3}}_2) \oplus (\mathbf{6}_1 \otimes \overline{\mathbf{15}}_2) \oplus (\overline{\mathbf{15}}_1 \otimes \mathbf{6}_2), \\
\mathbf{36} &\mapsto (\overline{\mathbf{3}}_1 \otimes \mathbf{6}_2) \oplus (\mathbf{6}_1 \otimes \overline{\mathbf{3}}_2) \oplus (\overline{\mathbf{15}}_1 \otimes \overline{\mathbf{15}}_2), \\
\overline{\mathbf{36}} &\mapsto (\mathbf{3}_1 \otimes \overline{\mathbf{6}}_2) \oplus (\overline{\mathbf{6}}_1 \otimes \mathbf{3}_2) \oplus (\mathbf{15}_1 \otimes \mathbf{15}_2), \\
\mathbf{84} &\mapsto (\mathbf{1}_1 \otimes \mathbf{10}_2) \oplus (\mathbf{10}_1 \otimes \mathbf{1}_2) \oplus (\mathbf{8}_1 \otimes \mathbf{8}_2) \oplus (\overline{\mathbf{10}}_1 \otimes \overline{\mathbf{10}}_2), \\
\overline{\mathbf{84}} &\mapsto (\mathbf{1}_1 \otimes \overline{\mathbf{10}}_2) \oplus (\overline{\mathbf{10}}_1 \otimes \mathbf{1}_2) \oplus (\mathbf{8}_1 \otimes \mathbf{8}_2) \oplus (\mathbf{10}_1 \otimes \mathbf{10}_2), \\
\mathbf{126} &\mapsto (\mathbf{15}_1 \otimes \mathbf{3}_2) \oplus (\mathbf{3}_1 \otimes \mathbf{15}_2) \oplus (\overline{\mathbf{6}}_1 \otimes \overline{\mathbf{6}}_2), \\
\overline{\mathbf{126}} &\mapsto (\overline{\mathbf{15}}_1 \otimes \overline{\mathbf{3}}_2) \oplus (\overline{\mathbf{3}}_1 \otimes \overline{\mathbf{15}}_2) \oplus (\mathbf{6}_1 \otimes \mathbf{6}_2).
\end{split}
\end{align}

\subsubsection{$SU(6)_1 \supset SU(2)_3 \times SU(3)_2$}

For $SU(6)_1$, there are six primary fields:
\begin{align}
\begin{array}{l|cccccc}
\textrm{Rep.} & \mathbf{1} & \mathbf{6} & \overline{\mathbf{6}} & \mathbf{15} & \overline{\mathbf{15}} & \mathbf{20} \\ \hline
h & 0 & \frac{5}{12} & \frac{5}{12} & \frac{2}{3} & \frac{2}{3} & \frac{3}{4} 
\end{array}
\end{align}
For $SU(2)_3$, there are four primary fields:
\begin{align}
\begin{array}{l|cccc}
\textrm{Rep.} & \mathbf{1} & \mathbf{2} & \mathbf{3} & \mathbf{4} \\ \hline
h & 0 & \frac{3}{20} & \frac{2}{5} & \frac{3}{4}
\end{array}
\end{align}
For $SU(3)_2$, there are six primary fields:
\begin{align}
\begin{array}{l|cccccc}
\textrm{Rep.} & \mathbf{1} & \mathbf{3} & \overline{\mathbf{3}} & \mathbf{6} & \overline{\mathbf{6}} & \mathbf{8} \\ \hline
h & 0 & \frac{4}{15} & \frac{4}{15} & \frac{2}{3} & \frac{2}{3} & \frac{3}{5}
\end{array}
\end{align}
The branching rule is given by 
\begin{align}
\begin{split}
\mathbf{1} &\mapsto (\mathbf{1}_1 \otimes \mathbf{1}_2) \oplus (\mathbf{3}_1 \otimes \mathbf{8}_2), \\
\mathbf{6} &\mapsto (\mathbf{2}_1 \otimes \mathbf{3}_2) \oplus (\mathbf{4}_1 \otimes \overline{\mathbf{6}}_2), \\
\overline{\mathbf{6}} &\mapsto (\mathbf{2}_1 \otimes \overline{\mathbf{3}}_2) \oplus (\mathbf{4}_1 \otimes\mathbf{6}_2), \\
\mathbf{15} &\mapsto (\mathbf{1}_1 \otimes \mathbf{6}_2) \oplus (\mathbf{3}_1 \otimes \overline{\mathbf{3}}_2), \\
\overline{\mathbf{15}} &\mapsto (\mathbf{1}_1 \otimes \overline{\mathbf{6}}_2) \oplus (\mathbf{3}_1 \otimes \mathbf{3}_2), \\
\mathbf{20} &\mapsto (\mathbf{4}_1 \otimes \mathbf{1}_2) \oplus (\mathbf{2}_1 \otimes \mathbf{8}_2).
\end{split}
\end{align}

\end{widetext}

\end{document}